\documentclass[iop]{emulateapj}
\usepackage{epsfig}
\usepackage{amsmath}
\usepackage{natbib}
\newcommand{\kepler}{{\it Kepler}}
\newcommand{\el}{$\ell$}
\newcommand{\nheartbeat}{$17$}
\newcommand{\thor}{KIC~4248941}
\newcommand{\alan}{KIC~3547874}
\newcommand{\carl}{KIC~8719324}
\newcommand{\ardi}{KIC~3230227}
\newcommand{\zeko}{KIC~4936180}

\newcommand{\hank}{KIC~9790355}
\newcommand{\lars}{KIC~4372379}
\newcommand{\matt}{KIC~5034333}
\newcommand{\gene}{KIC~7622059}

\newcommand{\logg}{$\log g$}
\newcommand{\teff}{$T_{\rm eff}$}

\begin{document}

\title{A Class of Eccentric Binaries with Dynamic Tidal Distortions discovered with Kepler}
\author{Susan E. Thompson\altaffilmark{1,2},  Mark Everett\altaffilmark{3}, Fergal Mullally\altaffilmark{2,4}, Thomas Barclay\altaffilmark{4,5}, Steve B. Howell\altaffilmark{4}, Martin Still\altaffilmark{4,5}, Jason Rowe\altaffilmark{2,4}, Jessie L. Christiansen\altaffilmark{2,4}, Donald W. Kurtz\altaffilmark{6}, Kelly Hambleton\altaffilmark{6}, Joseph D. Twicken\altaffilmark{2,4}, Khadeejah A. Ibrahim\altaffilmark{7,4} and Bruce D. Clarke\altaffilmark{2,4}}

\altaffiltext{1}{{\it a.k.a.} Susan E. Mullally, NASA Ames Research Center, Moffett Field, CA 94035, USA, susan.e.thompson@nasa.gov}
\altaffiltext{2}{SETI Institute, 189 Bernardo Ave Suite 100, Mountain View, CA 94043, USA}
\altaffiltext{3}{National Optical Astronomy Observatory, Tucson, AZ 85719, USA}
\altaffiltext{4}{NASA Ames Research Center, Moffett Field, CA 94035, USA}
\altaffiltext{5}{Bay Area Environmental Research Institute, 560 Third St West, Sonoma, CA, 95476, USA}
\altaffiltext{6}{Jeremiah Horrocks Institute, University of Central Lancashire, Preston PR1\,2HE, UK}
\altaffiltext{7}{Orbital Sciences Corporation, 17143 Flight Systems Drive, Mojave, CA 93501}

\shorttitle{Eccentric Binaries with Dynamic Tidal Distortions}
\shortauthors{Thompson et al.}

\begin{abstract}
We have discovered a class of eccentric binary systems within the \kepler\ data archive that have dynamic tidal distortions and tidally-induced pulsations.  Each has a uniquely shaped light curve that is characterized by periodic brightening or variability at time scales of 4-20 days, frequently accompanied by shorter period oscillations.  We can explain the dominant features of the entire class with orbitally-varying tidal forces that occur in close, eccentric binary systems. The large variety of light curve shapes arises from viewing systems at different angles. This hypothesis is supported by spectroscopic radial velocity measurements for five systems, each showing evidence of being in an eccentric binary system. Prior to the discovery of these 17 new systems, only four stars, where KOI-54 is the best example, were known to have evidence of these dynamic tides and tidally-induced oscillations. We perform preliminary fits to the light curves and radial velocity data, present the overall properties of this class and discuss the work required to accurately model these systems. 

\end{abstract}

\section{Introduction}
\kepler, in its search for extrasolar planets, obtains high-precision time series photometry of 115.6 square degrees of sky \citep{Koch2010}. With this new combination of photometric quality and uninterrupted coverage of more than 160\,000 stars, \kepler\ is discovering variability never before identified in stars. Here we present light curves that demonstrate a periodic brightening or variability on time scales of days, frequently accompanied by harmonic oscillations at shorter periods. 

These light curves are reminiscent of the tidally distorted light curves theorized by \citet{Kumar1995} for eccentric orbits.  KOI-54 (KIC~8112007), a highly eccentric binary system composed of two A-type stars observed by \kepler\ \citep{Welsh2011}, was the first system discovered whose light curve is predominantly characterized by these dynamic distortions.  Binary tidal forces also induce pulsations at integer harmonics of the orbital period \citep{Kumar1995, Willems2003}; making KOI-54 the best example of pulsations driven in this way. See \citet{Handler2002}, \citet{DeCat2000} and \citet{Maceroni2009} for other examples of systems with evidence of this type of driving. The discovery of KOI-54 has provided critical observational data to constrain the theories of dynamic tidal forces on stellar binaries \citep{Fuller2011,Burkart2012}, but more examples of these stars are required to generally test these theories.  

Despite having unique light curves, we demonstrate that the \nheartbeat\ systems described in this paper are all tidally interacting binary systems in eccentric orbits.  In \S\ref{s:lightcurves} we describe the \kepler\ photometric time series of these systems and propose a simple model to describe their dominant features in \S\ref{s:model}.  We fit our simple model to these systems and estimate their orbital parameters in \S\ref{s:fittingkumar}, demonstrating a technique proposed by \citet{Kumar1995}.  In \S\ref{s:spectra} we present spectroscopy for each system including radial velocity measurements of five systems that confirm they are in eccentric binary systems. Finally we discuss the properties of this class of stars, the effects of the tidal interactions, and the remaining work to accurately model these systems in \S\ref{s:discussion}.

\section{Kepler Light Curves}
\label{s:lightcurves}
\begin{figure*}
\includegraphics[scale=0.9,angle=270]{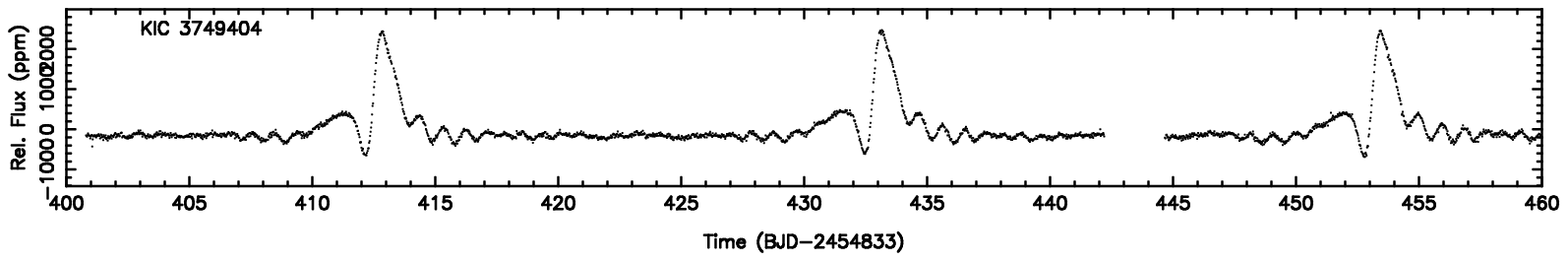}\\
\includegraphics[scale=0.9,angle=270]{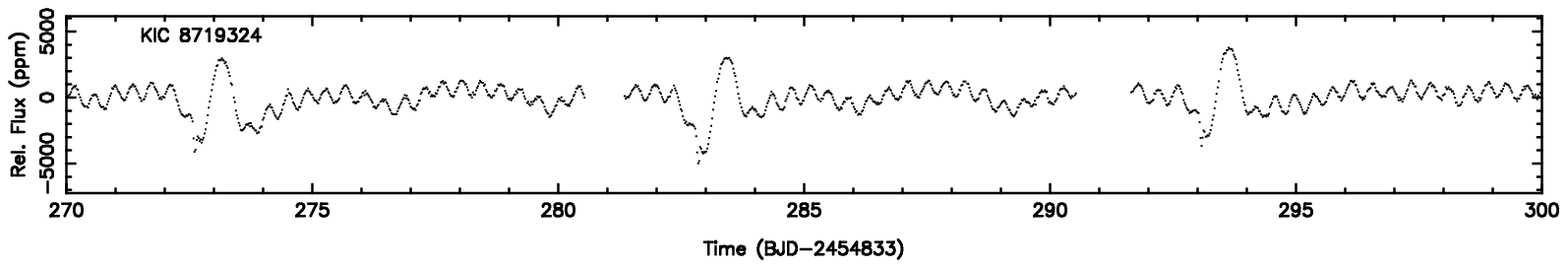}\\
\includegraphics[scale=0.9,angle=270]{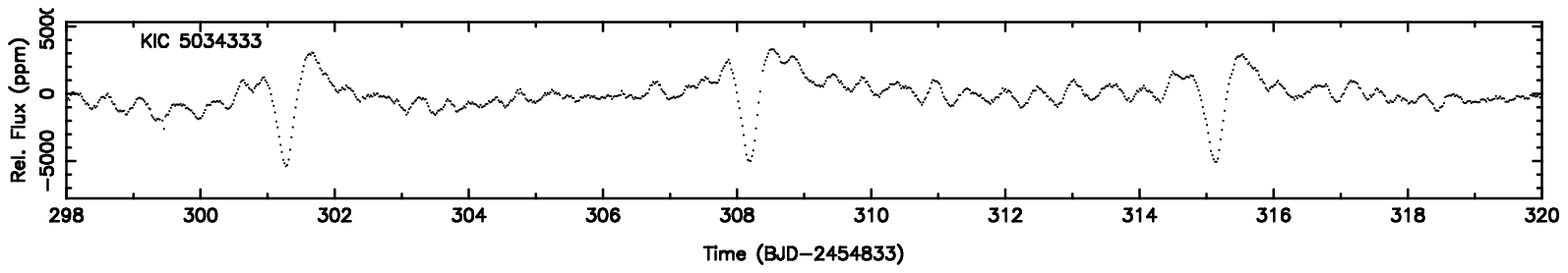}\\
\includegraphics[scale=0.9,angle=270]{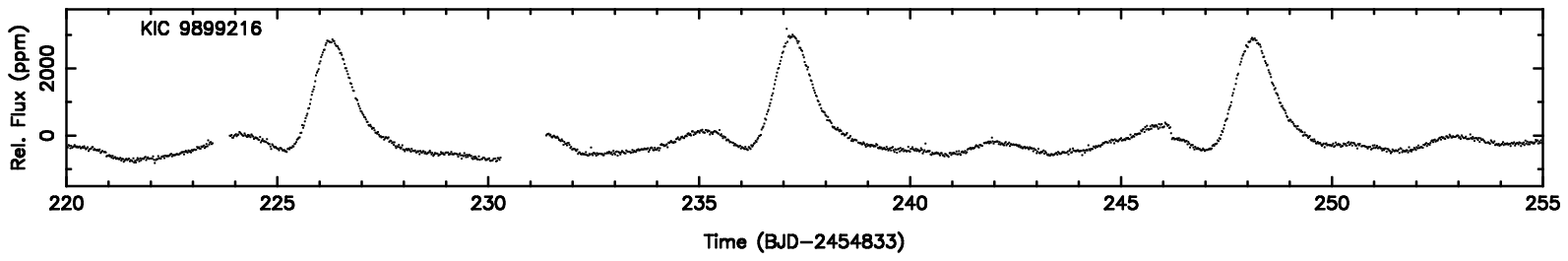}\\

\caption{\label{rawlc}Example light curves of heartbeat stars found in the \kepler\ data. }
\end{figure*}

We have discovered \nheartbeat\ objects in the \kepler\ data archive that demonstrate unique periodic variability.  We refer to these objects as ``heartbeat'' stars because the shape of the variability for several are reminiscent of an echocardiogram.  We show example light curves in Figure~\ref{rawlc} that exemplify the sudden and periodic variation in brightness. Notice that the heartbeat profile comes in many varieties and appears to be correlated with other periodic oscillations in the light curve of the star.  

 For each system we analyzed a subset of the long cadence (one exposure taken every 29.4\,min) light curves observed during \kepler\ quarters 0 through 7 (MJD spanning 54\,953.03 to 55\,552.06). Each quarter is one fourth of the \kepler\ heliocentric orbital period of 372\,d, except for the quarters~0 and~1 (with lengths of approximately 10\,d and 33\,d respectively). Each subset contains enough data to constrain the heartbeat period to better than two minutes. See Table~\ref{table} for a list of long cadence quarters used to study each object.  
 
 We used the light curves created by the \kepler\ simple aperture photometry module \citep{Twicken2010,Fraquelli2011}. These light curves are the result of performing simple aperture photometry on the calibrated pixels in the optimal aperture \citep{Bryson2010}. For more details on the calibration steps performed by the \kepler\ pipeline see \citet{Jenkins2010}, \citet{Caldwell2010}, and \citet{KDPH}. To remove the long period effects that result from differential velocity abberation and thermal effects from the spacecraft \citep{VanCleve2011}, we removed a piece-wise polynomial fit from each quarter using the program {\sc wqed} \citep{Thompson2009}.  We did not adjust the lightcurves for contamination from other stars. The contamination is small (typically $<5$ per~cent) for these stars, as reported by the \kepler\ archive \citep{Fraquelli2011}. The contamination will have a small effect on the amplitude of the features we see in the light curves, but will not affect the shapes.

\begin{figure*}
\includegraphics[scale=0.8,angle=270]{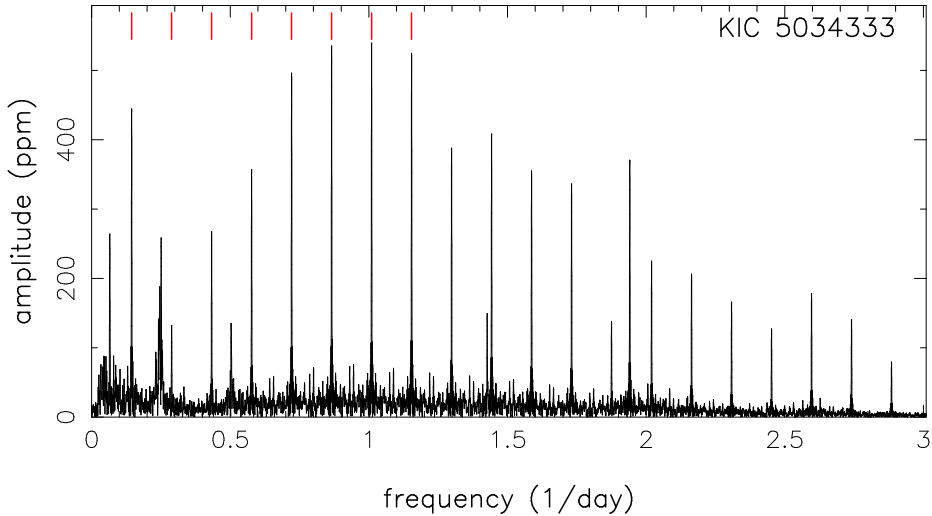}
\caption{\label{ft}The amplitude spectrum of the Fourier transform of \matt. Eight evenly spaced harmonics are labeled as red lines at the top of the figure. To determine the heartbeat period, we fit the light curve at the fundamental frequency and this series of harmonics. }
\end{figure*}

We measured the heartbeat period by performing a nonlinear least-squares fit to the ``comb" of equally spaced frequencies found in the power spectrum of each light curve. This comb results from the periodic, yet sudden, changes in the brightness and from any harmonic pulsations present on the star.  We fitted a sine wave at the orbital frequency found in the power spectrum as well as at a minimum of six harmonics of the heartbeat frequency to establish the heartbeat period \citep[using Period04,][]{Period04} (See Figure~\ref{ft}). In performing the fit, the amplitude and phase of each harmonic were allowed to vary, while the frequency was fixed to an integer multiple of the orbital frequency.  The fitted orbital periods are reported in Table~\ref{table}. At this stage we do not attempt to determine what fraction of the fitted amplitudes is due to the heartbeat shape and what fraction is due to stellar pulsation; it is not required to establish the period of the heartbeat.

\begin{figure*}[ht]

\includegraphics[scale=.6,angle=270]{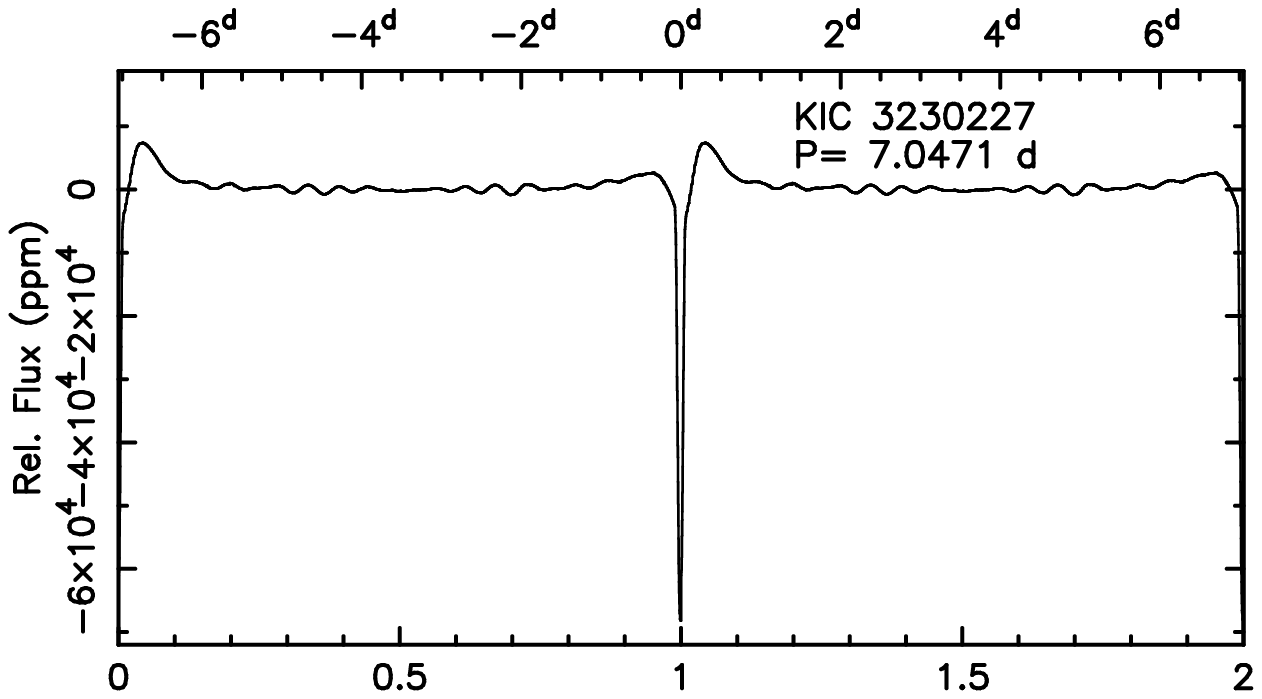}
\includegraphics[scale=.6,angle=270]{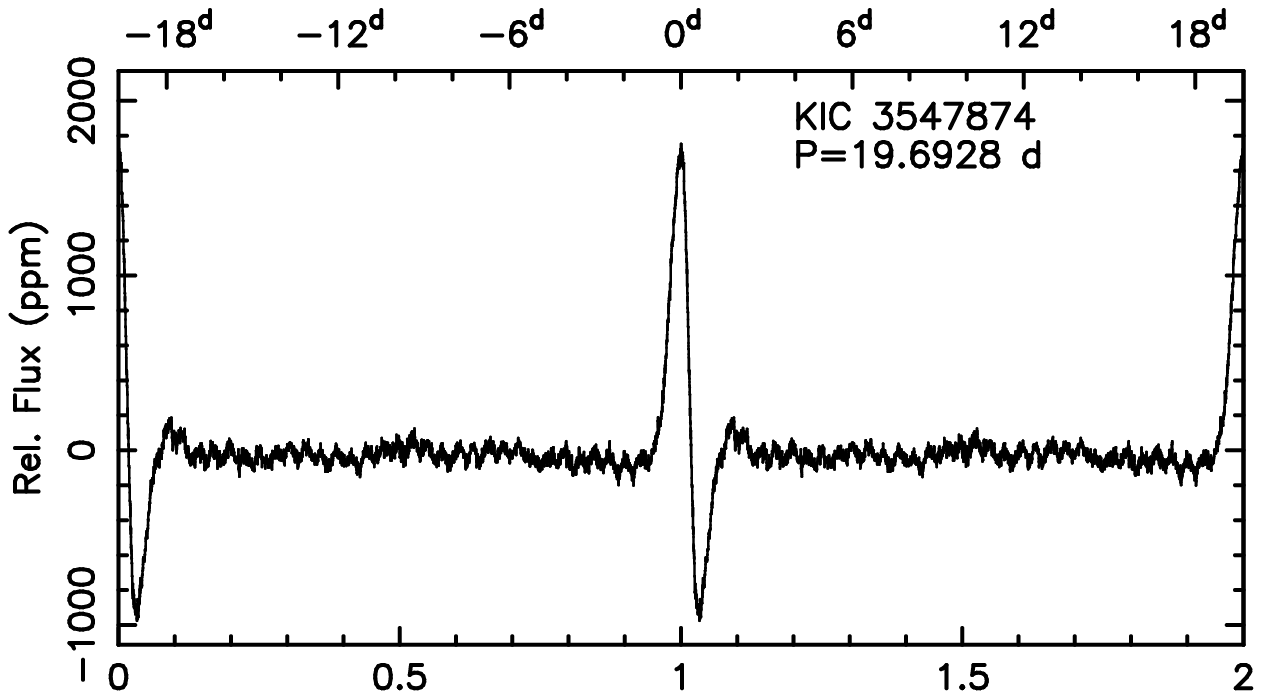}\\
\includegraphics[scale=.6,angle=270]{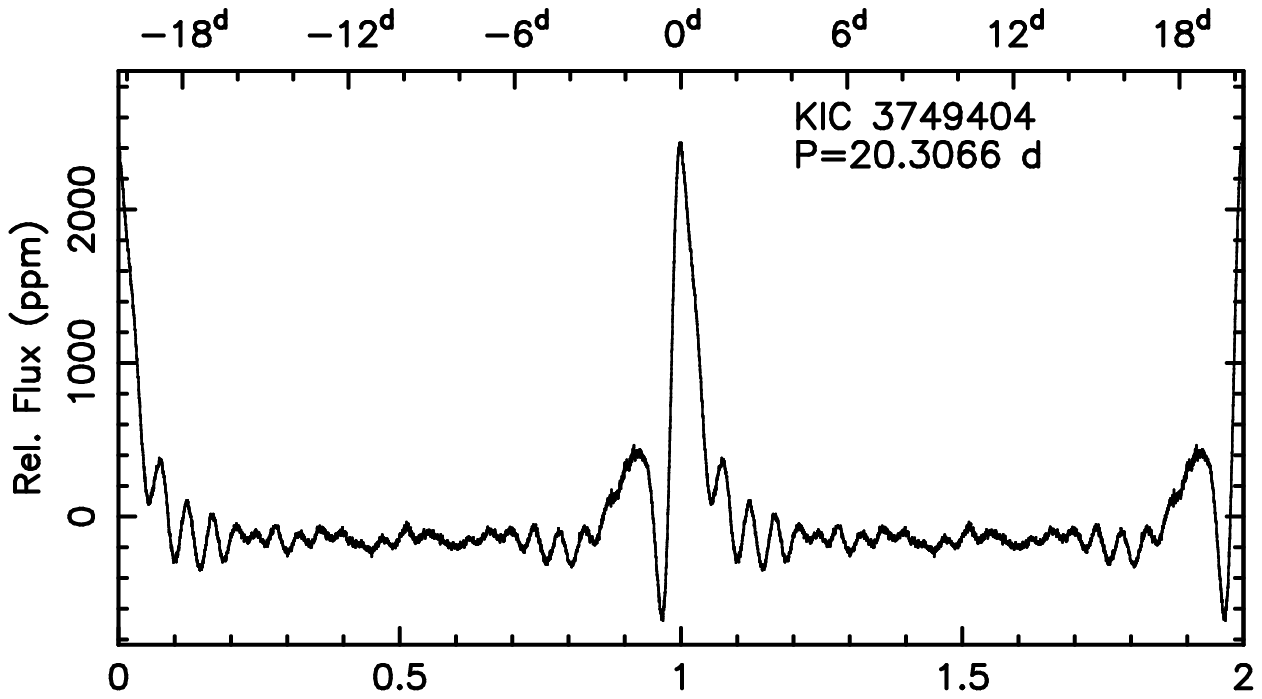}
\includegraphics[scale=.6,angle=270]{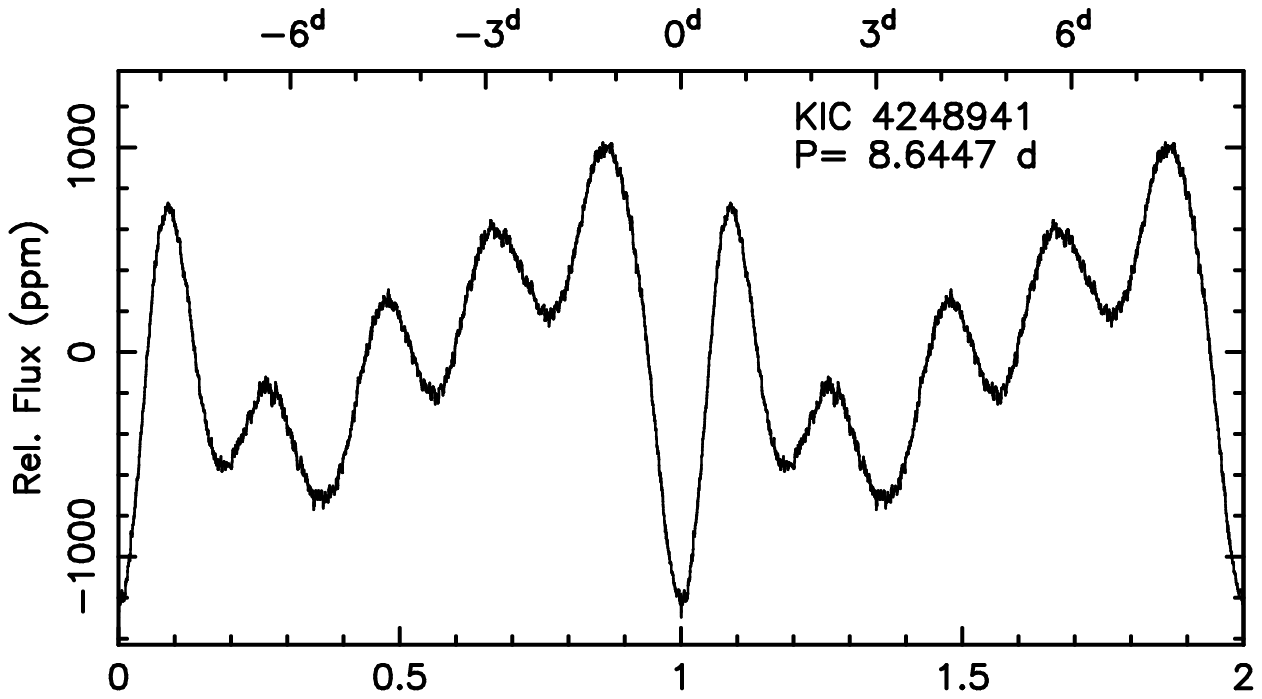}\\
\includegraphics[scale=.6,angle=270]{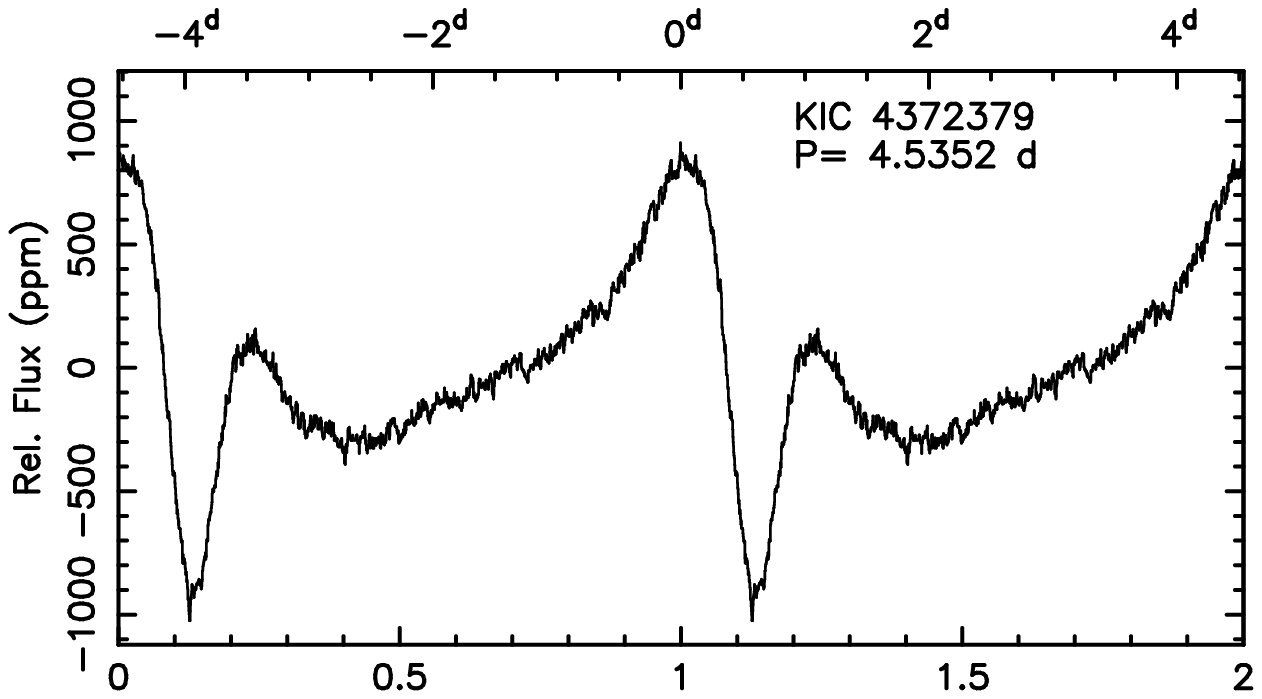}
\includegraphics[scale=.6,angle=270]{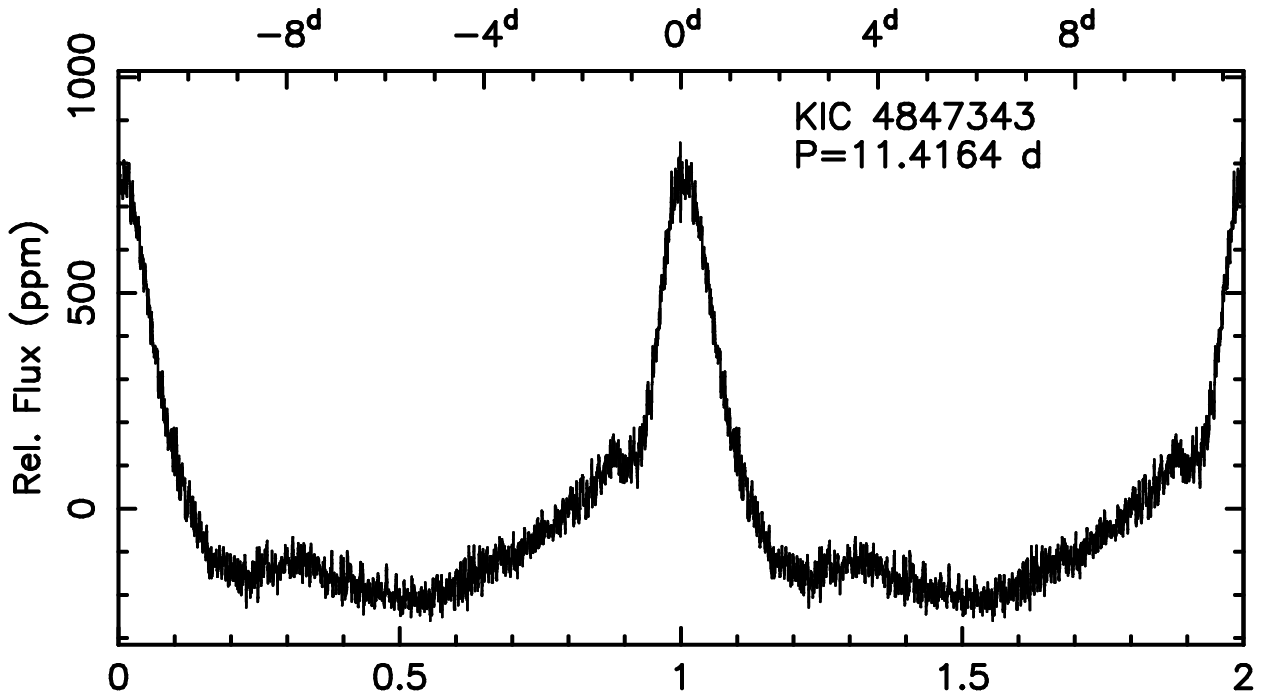}\\
\includegraphics[scale=.6,angle=270]{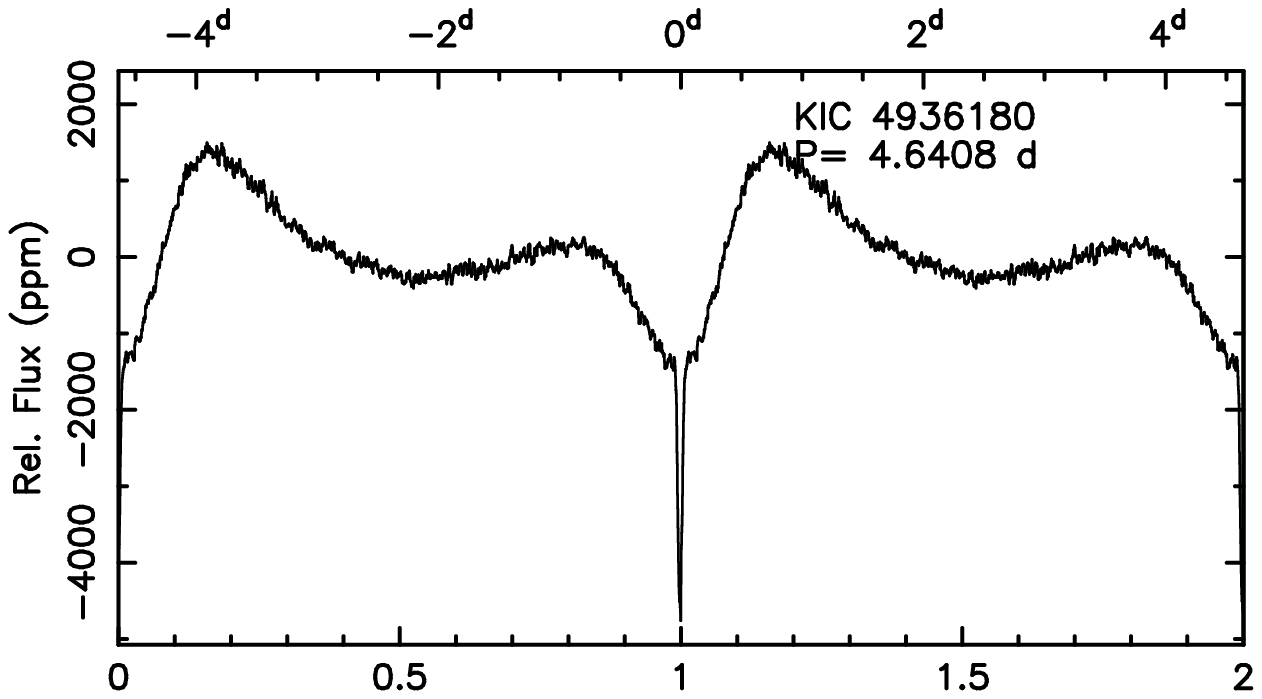}
\includegraphics[scale=.6,angle=270]{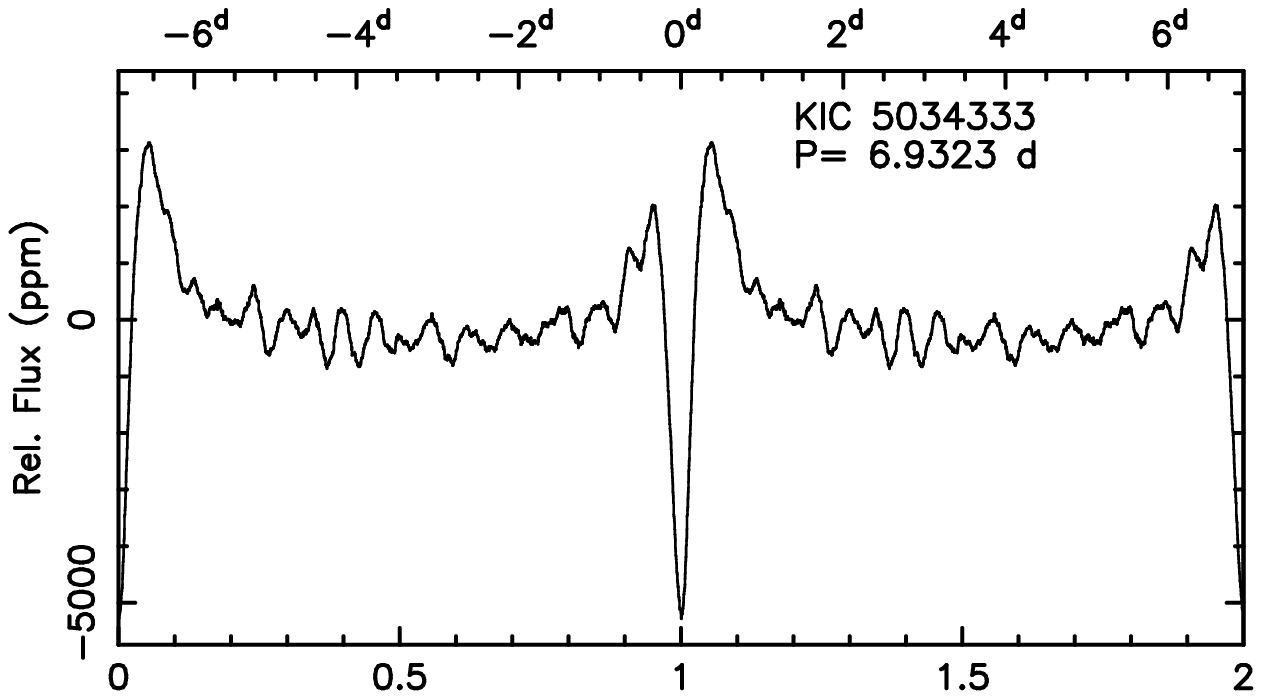}\\
\includegraphics[scale=.6,angle=270]{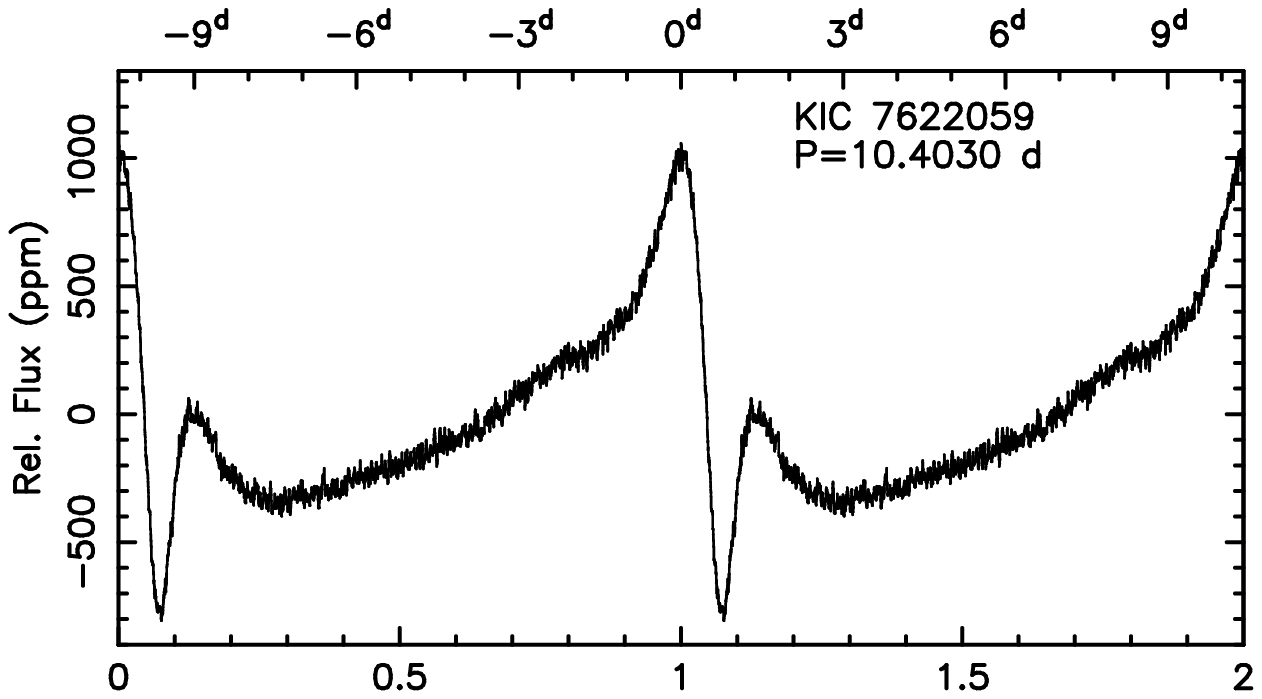}
\includegraphics[scale=.6,angle=270]{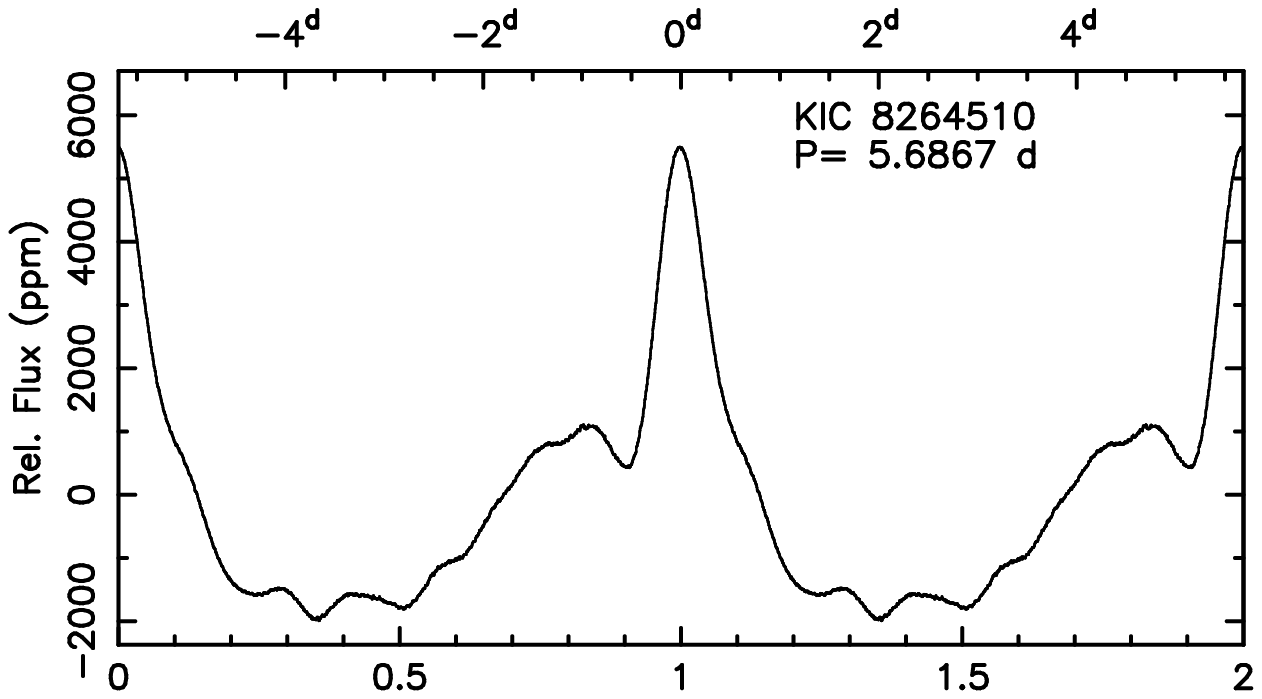}\\
\caption{\label{folded}Folded light curves plotted as relative flux against orbital phase; the zero phase relates to the brightest or dimmest event in the folded light curve. Phase in fractional units is labeled along the bottom axis and time in days is labeled along the top axis.  The KIC number and period are labeled on each plot. We smooth each folded light curve by performing a 15-min running average. }
\end{figure*}

\begin{figure*}
\includegraphics[scale=.6,angle=270]{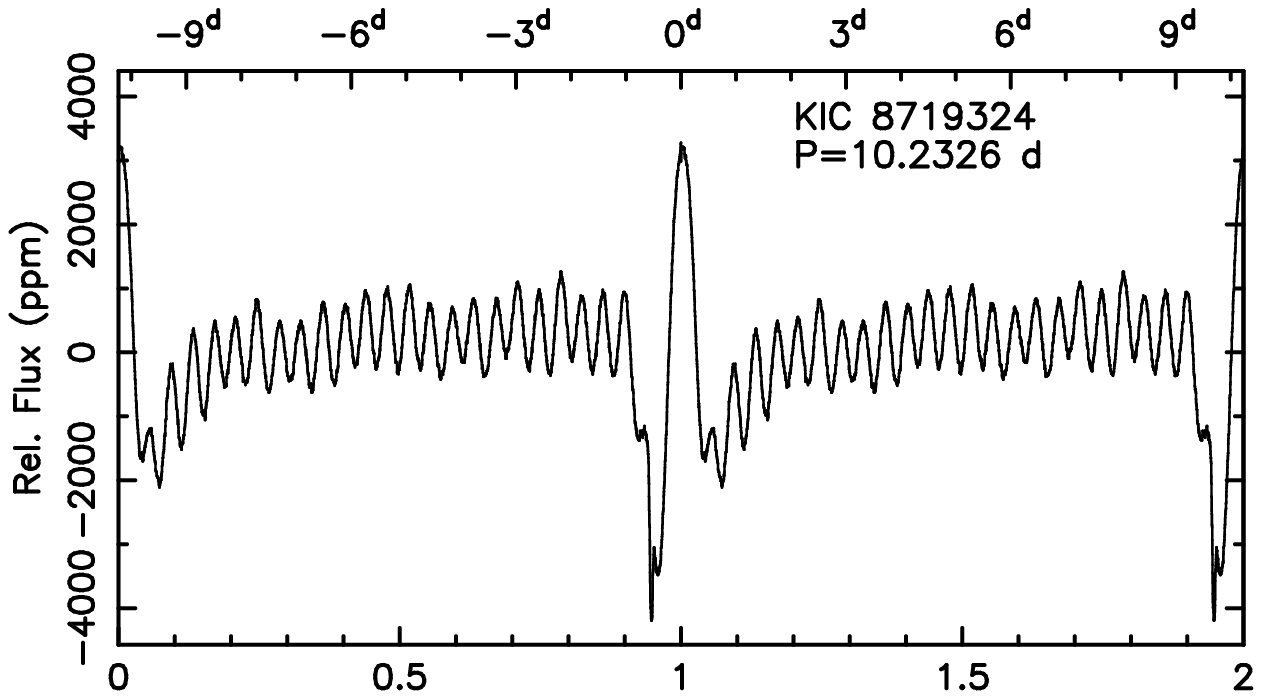}
\includegraphics[scale=.6,angle=270]{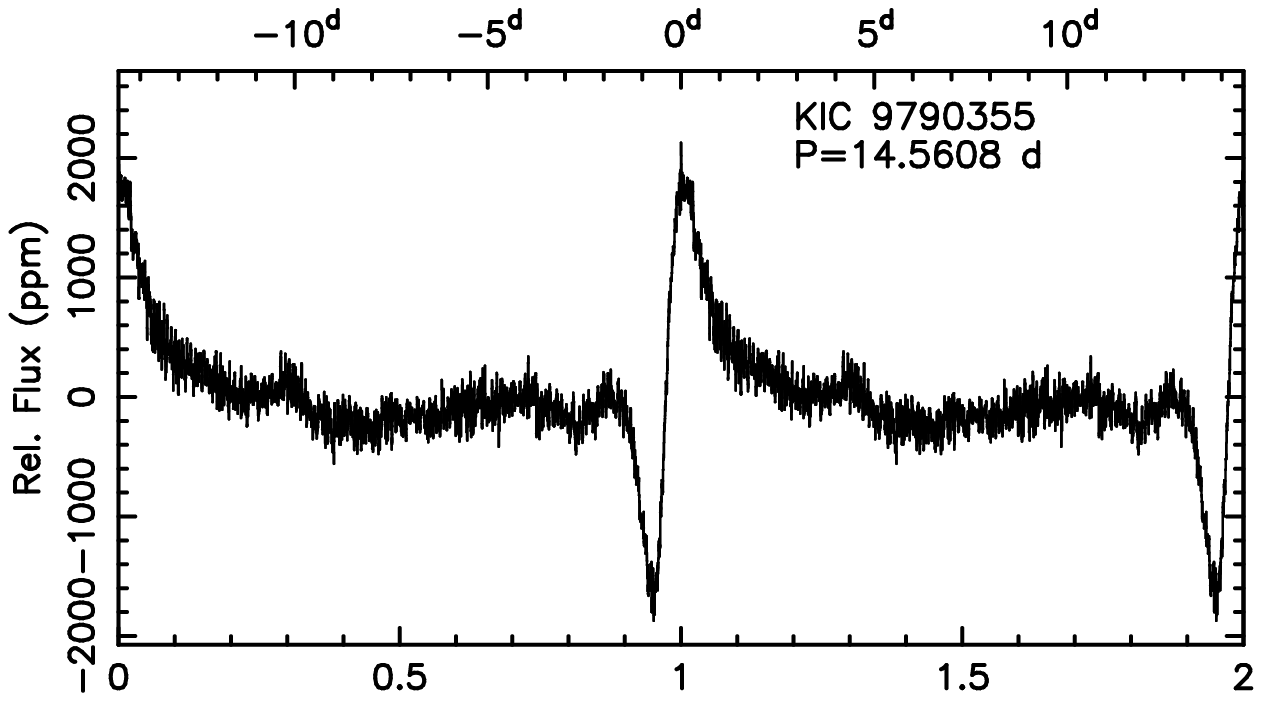}\\
\includegraphics[scale=.6,angle=270]{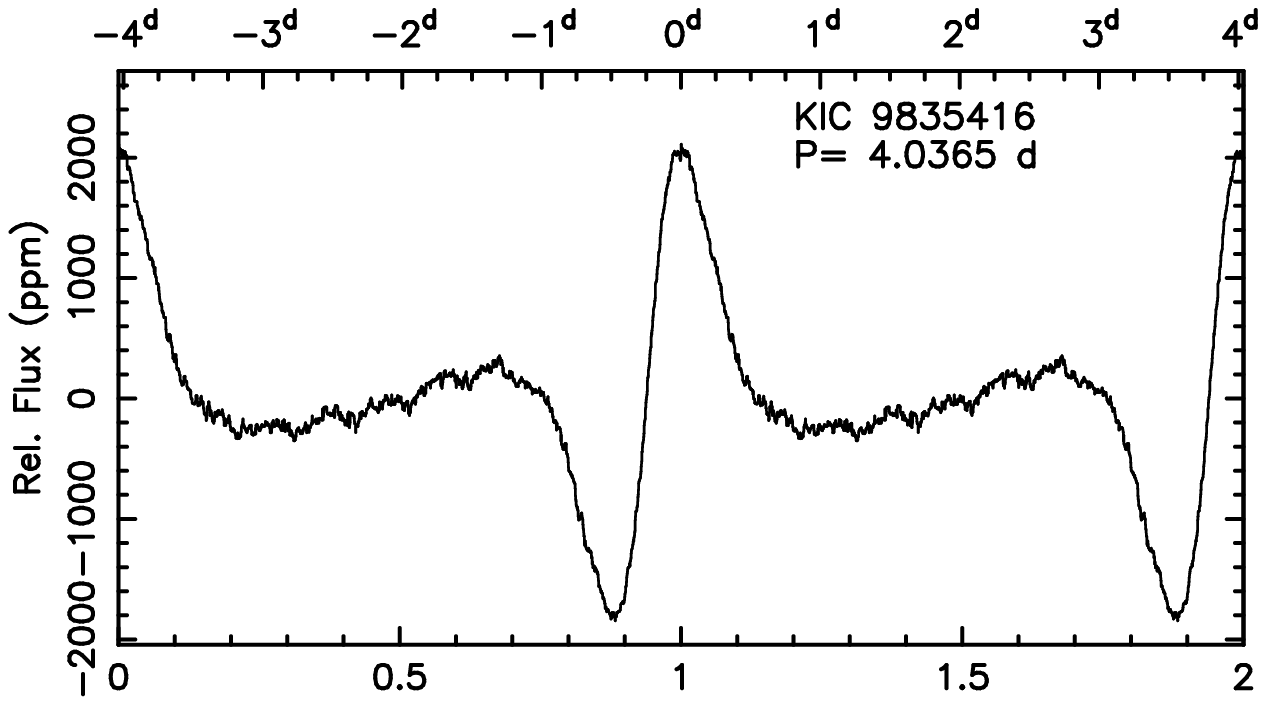}
\includegraphics[scale=.6,angle=270]{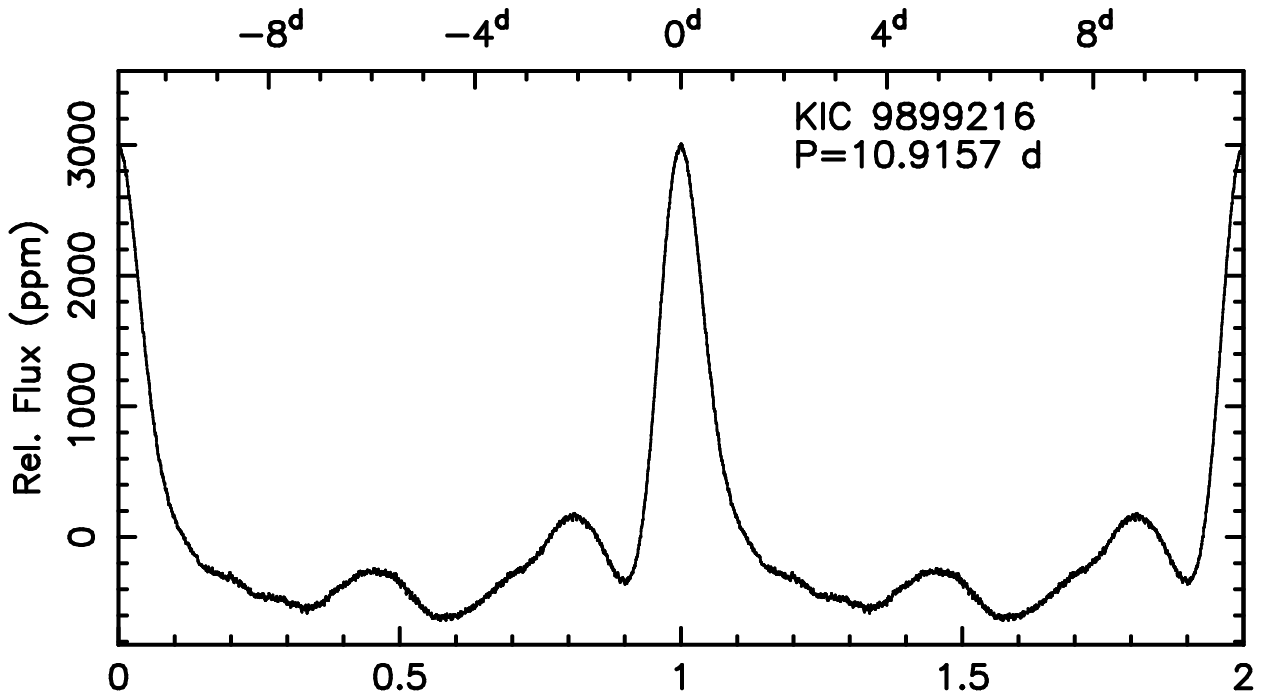}\\
\includegraphics[scale=.6,angle=270]{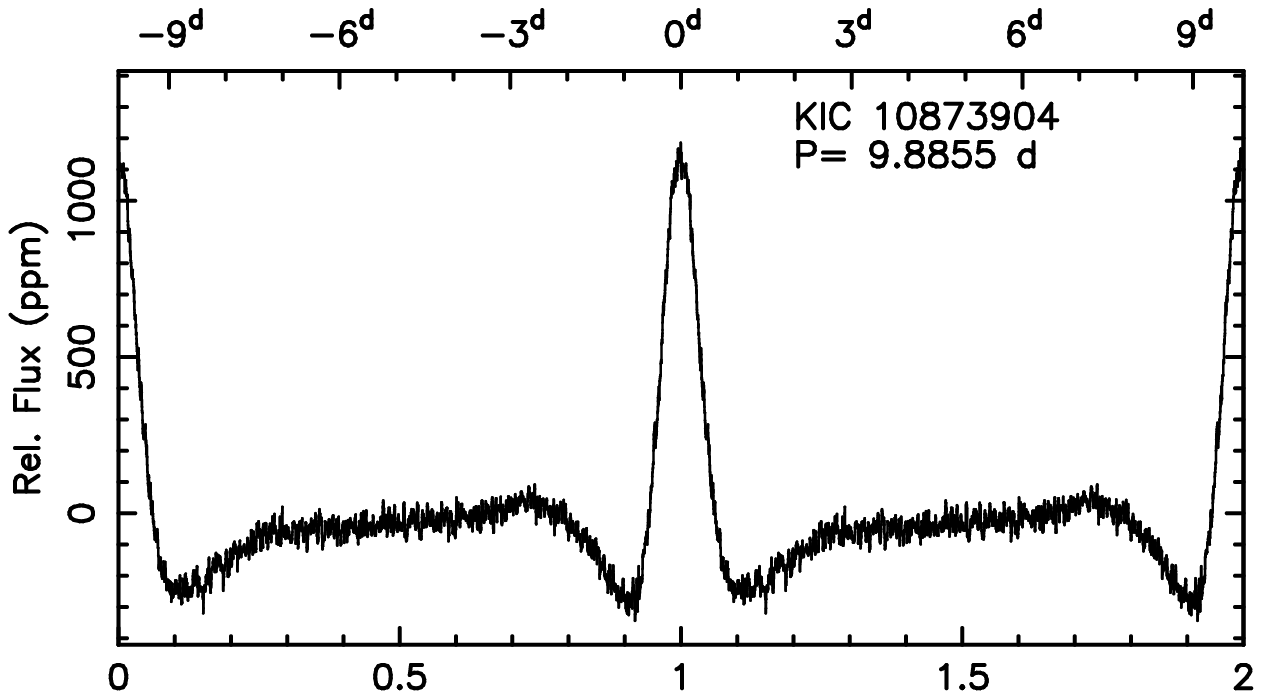}
\includegraphics[scale=.6,angle=270]{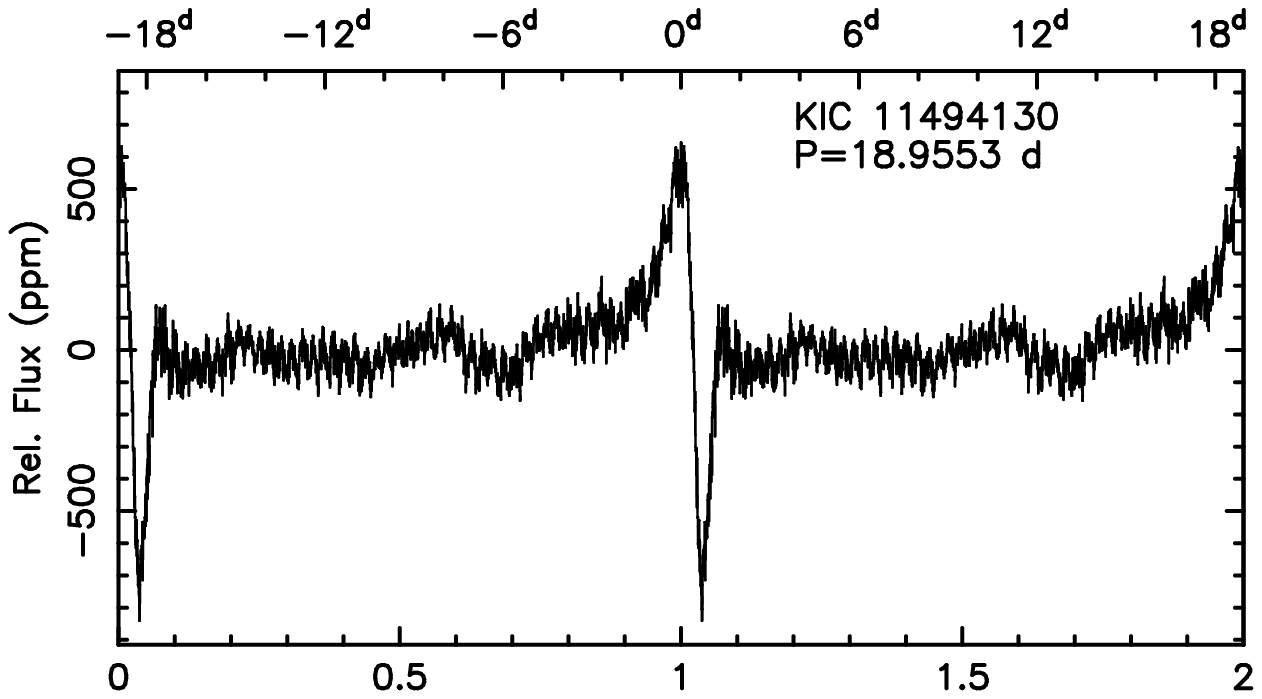}\\
\includegraphics[scale=.6,angle=270]{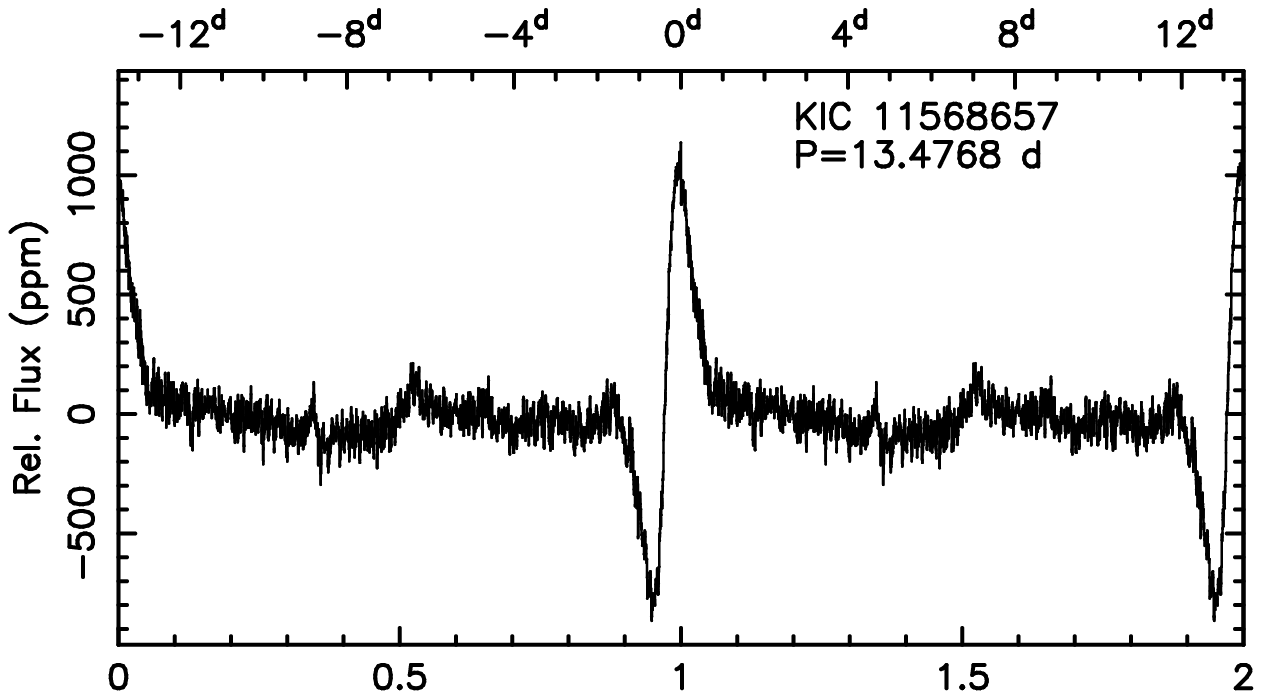}
\caption{\label{folded2} Same as Figure~\ref{folded} for the remaining heartbeat stars.}
\end{figure*}

We folded each light curve at the measured period and then smoothed by performing a running average spanning 15~minutes (see Figures~\ref{folded} and~\ref{folded2}).  We set the zero phase to either the maximum or the minimum of the hearbeat feature, depending on the shape of the feature. Because of the quality and length of the \kepler\ data, almost all the features shown in the folded light curves are real.  However, because these are folded and smoothed curves, any variations that are not harmonics of the heartbeat period are lost.  

Notice that each folded light curve has a unique shape, quite different from KOI-54. Some dim before they brighten, others dim after they brighten, and others show distinct `W' or 'M' shapes.  Besides the heartbeat, stellar oscillations at harmonics of the heartbeat period are also apparent. For example, \ardi , identified by \citet{Uytterhoeven2011} as an eclipsing $\delta$\,Sct star, has short period oscillations evident in the folded light curve and thus harmonics of the orbital frequency. Similarly, Figures~\ref{folded} and \ref{folded2} reveal that most of the sample have pulsations at harmonics of the heartbeat frequency.   

The presence of harmonic oscillations is a clue to the nature of these systems.  For KOI-54, \citet{Welsh2011} show that short period modes that are harmonics of the orbital frequency dominate the light curve once the effects of the tidal distortion and external heating have been removed. Another hint to the nature of these objects comes from \ardi, \zeko , and \carl. They all show evidence of a grazing eclipse. Therefore, it is likely that the heartbeat shape is also caused by a binary companion. While several of the heartbeat systems identified here are also in the \kepler\ eclipsing binary star catalog \citep{slawson2011,prsa2011}, all of those in the catalog (except \ardi) show no evidence of an eclipse.

In the next section we explain how this varied collection of light curve shapes can be created with tidal distortions in an eccentric binary system.

\section{Tidal Distortions in an Eccentric Orbit}
\label{s:model}

Tidal forces change the shape of a star causing brightness variations \citep{BrownKilic2011,Welsh2010,Morris1985}.  For circular orbits, this results in a non-spherical star producing brightness variations at twice the orbital period, known as ellipsoidal variations.  For eccentric binaries, the tidal force is phase dependent and the largest variations in shape and brightness occur at periastron.  As a result, instead of seeing the double-wave in the flux variations, as in the case of the circular orbit, the flux changes drastically when the binary system nears periastron.  

As a star passes through periastron, in the static tide limit, the star's shape adjusts to the instantaneous tidal force. The shape of the star's distortion can be approximated by summing over axisymmetric and sectoral spherical degree two (\el=2) modes.  When observed at different inclinations, the relative contributions of the \el=2, $m$=~$-2$,~0,~2 modes, integrated over the observable sphere, will change.  Additionally, the observed phase of these modes will depend on the angle of periastron for the binary system.  As a result the shape of the flux variations due to the tidal distortions will depend on inclination, angle of periastron, and eccentricity \citep{Kumar1995}. The amplitude of these variations will depend on factors that influence the strength of the tidal forces, such as the masses of the objects and the orbital separation at peristron.

\begin{figure*}

\includegraphics{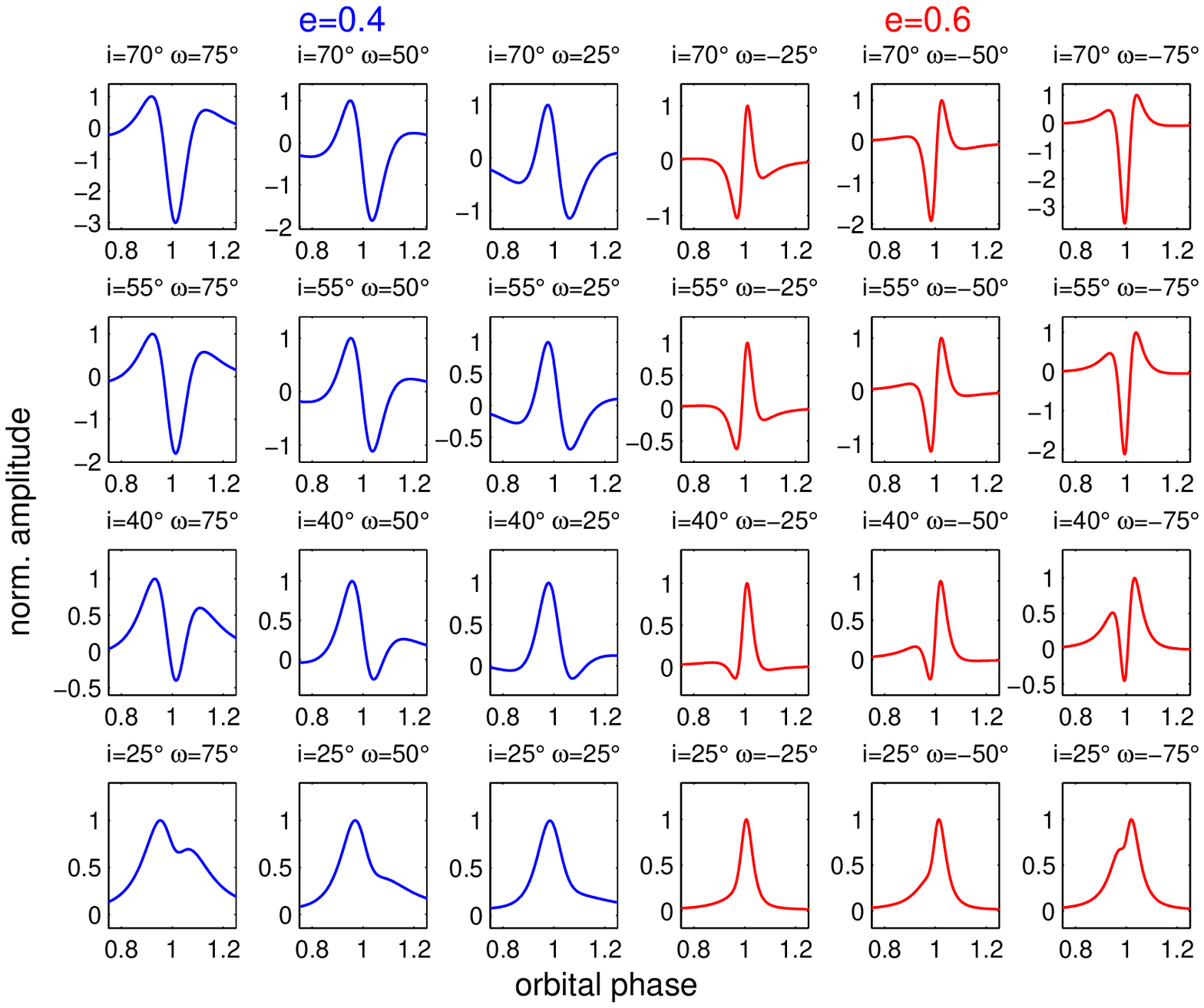}
\caption{\label{kumargrid}Synthetic light curves generated from the analytic model of \citet{Kumar1995} for tidal distortions in an eccentric binary.  The inclination ($i$) increases vertically and periastron angle ($\omega$) increase horizontally. The relative flux of each curve has been normalized by the largest change in brightness. Since the grid is symmetric around $\omega=0$, we present the model at two eccentricities, 0.6 and 0.4 (red and blue lines respectively).}
\end{figure*}

The tidal brightness variations were derived by \citet{Kumar1995} while predicting the tidal oscillations induced in a B-type star orbiting a neutron star.  Even though the effect was too subtle to observe at the time, they did describe an analytic model whose shape depends on the orbital inclination, eccentricity and angle of periastron of the binary orbit. We calculate a grid of solutions to this model for different inclinations and periastron angles and plot each solution in Figure~\ref{kumargrid} for two different eccentricities. 

A side-by-side comparison of the grid in Figure~\ref{kumargrid} to the folded light curves in Figures~\ref{folded} and~\ref{folded2} shows that this fairly simple model can reproduce many of the features observed in these systems. For instance, the shape of the KOI-54 brightenings occur for systems at low orbital inclinations while larger inclinations can cause the light curve to first increase in brightness and then decrease (\alan ), or vice versa (\hank ), depending on the angle of periastron.  Eccentricity has the effect of changing the duration of the heartbeat event; heartbeats in more eccentric orbits span a shorter phase interval.

Tidal distortion is not the only physical effect that can change the observed brightness of an eccentric binary system. Irradiation, tidally-induced pulsations \citep{Kumar1995,Willems2002}, Doppler boosting \citep{Bloemen2011,vanKerkwijk2010,Loeb2003}, and eclipses must also be considered in order to obtain the best fit to the light curves. Fitting an extensive set of physical effects to each system, as done for KOI-54 \citep{Welsh2011,Burkart2012}, is complex and beyond the scope of this discovery paper. We do, however, show that the tidal distortions alone can fit the dominant features of most of the light curves. By doing so we also approximate the orbital parameters of these systems. 

\subsection{Fitting Eccentric Tidal Distortions}
\label{s:fittingkumar}

We use the analytic model of \citet[][equation~44]{Kumar1995} for the tidal oscillations induced in eccentric binaries to obtain the orbital parameters for each of the heartbeat stars. Our fit to the fractional flux ($\frac{\delta F}{F}$) contains six parameters: eccentricity ($e$), orbital inclination ($i$), angle of periastron ($\omega$), an amplitude scaling factor ($S$), a fractional flux offset ($C$), and time of periastron. The period and the semi-major axis ($a$) were fixed to one when performing the fit to the folded light curves. $R(t)$ describes the distance between the two stars as a function of time and $\phi(t)$ gives the phase of the ellipse as a function of time. 

\begin{equation}
\frac{\delta F}{F} = S\cdot \frac{1-3\sin^2(i) \sin^2(\phi(t)-\omega )}{(R(t)/a)^{3}} + C
\end{equation}

 Inclination is defined such that the plane of the orbit is perpendicular to the line of sight at an inclination of zero. Because the stellar distortion is symmetric along the axis of the elongation, the angle of periastron is reported from $-90^{\circ}$ to $+90^{\circ}$; $\omega=0$ when looking down the minor axis of the ellipse and increases in the direction of the orbiting primary. The fit was performed using a Levenberg-Marquardt chi-square minimization.

\begin{figure*}

\includegraphics[scale=.31,angle=0]{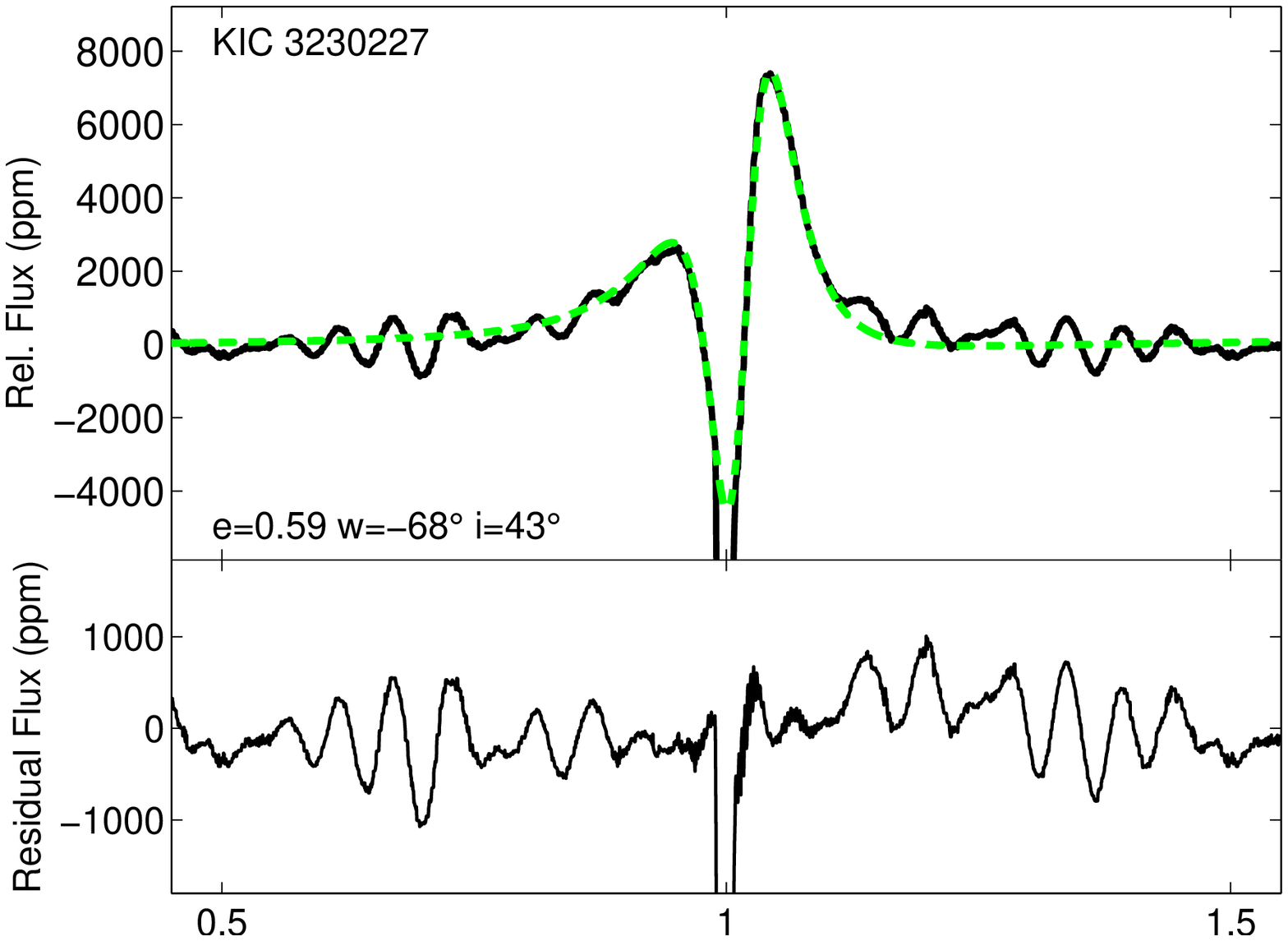}
\includegraphics[scale=.31,angle=0]{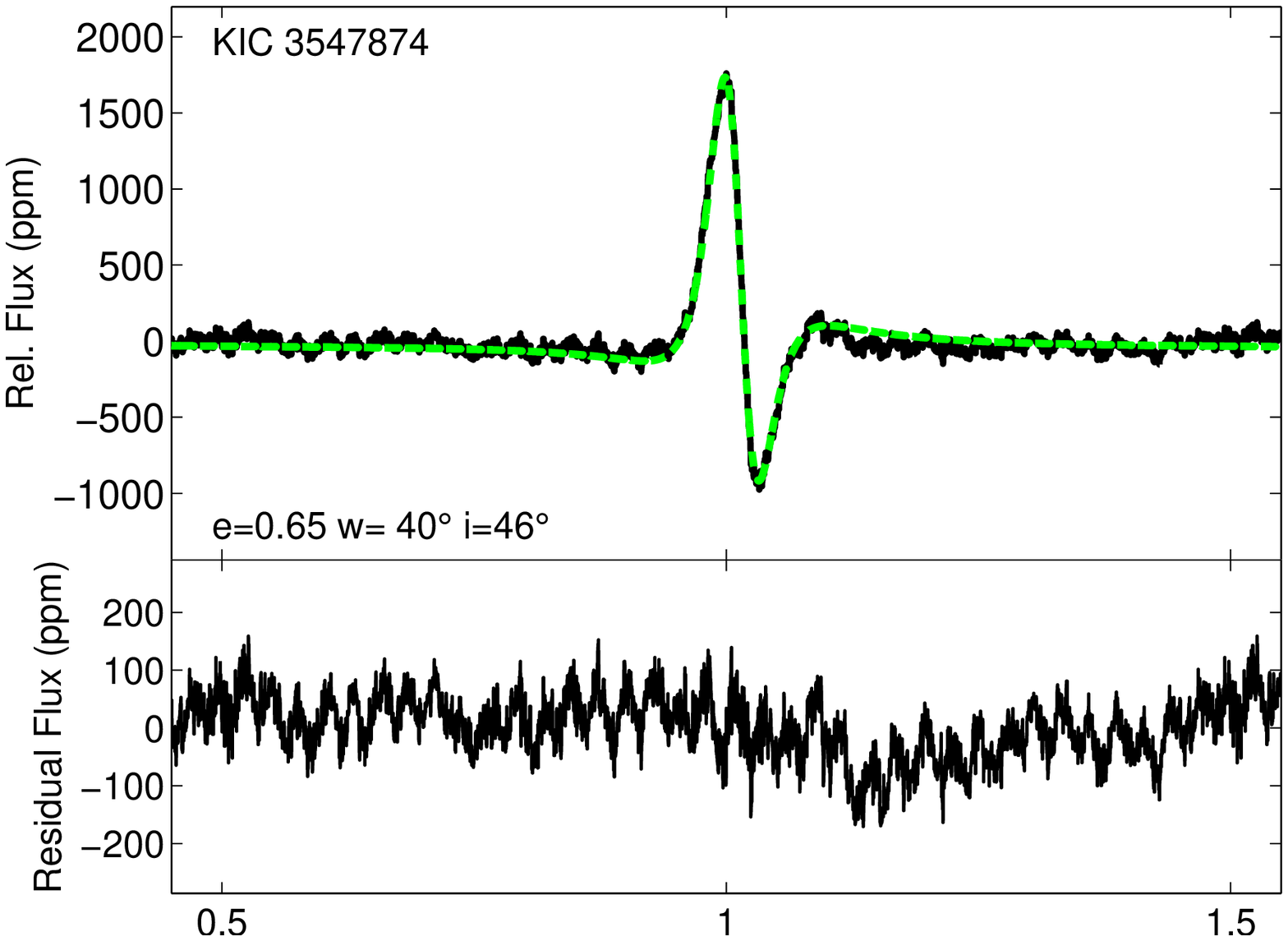}
\includegraphics[scale=.31,angle=0]{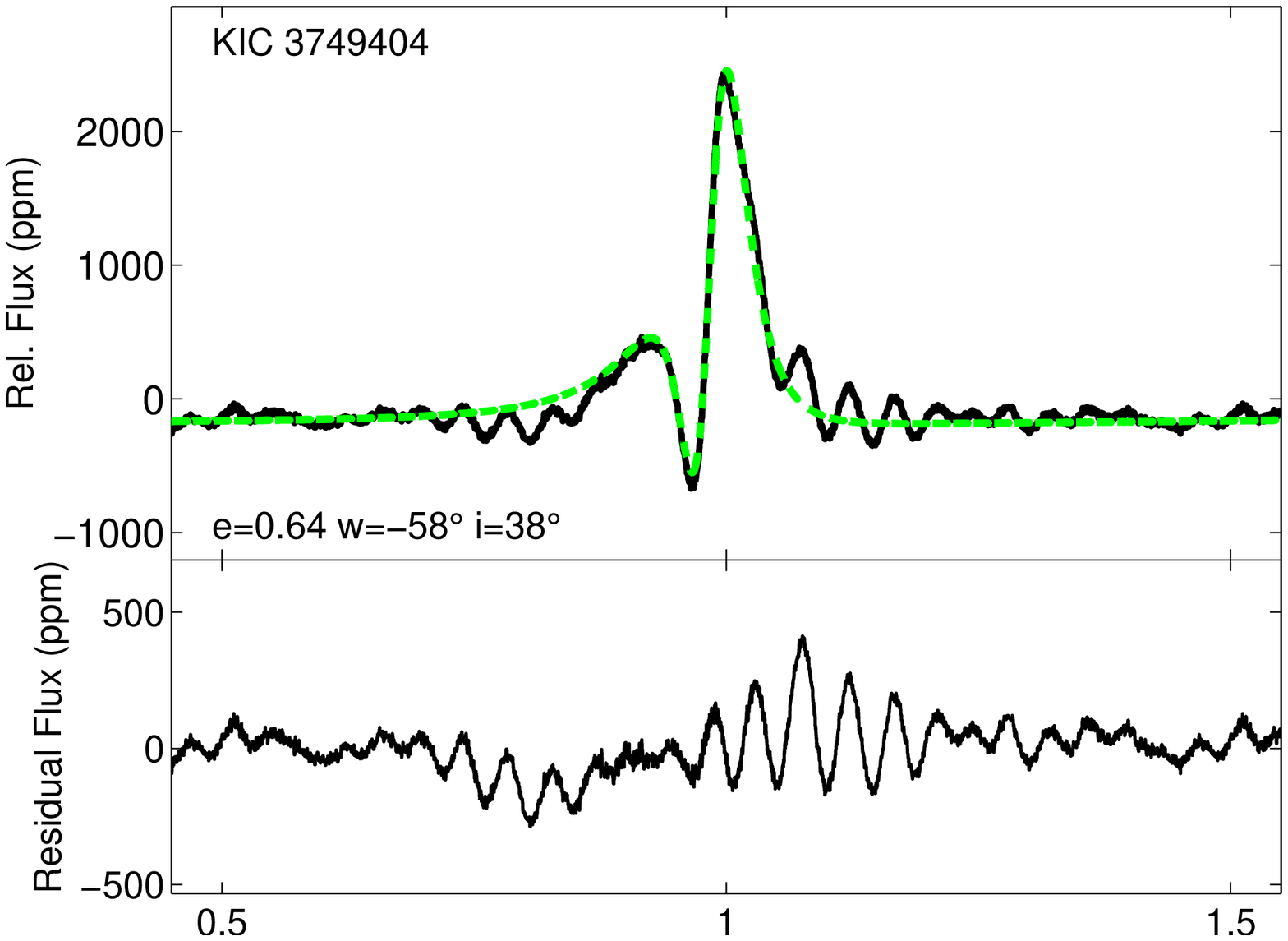}\\
\includegraphics[scale=.31,angle=0]{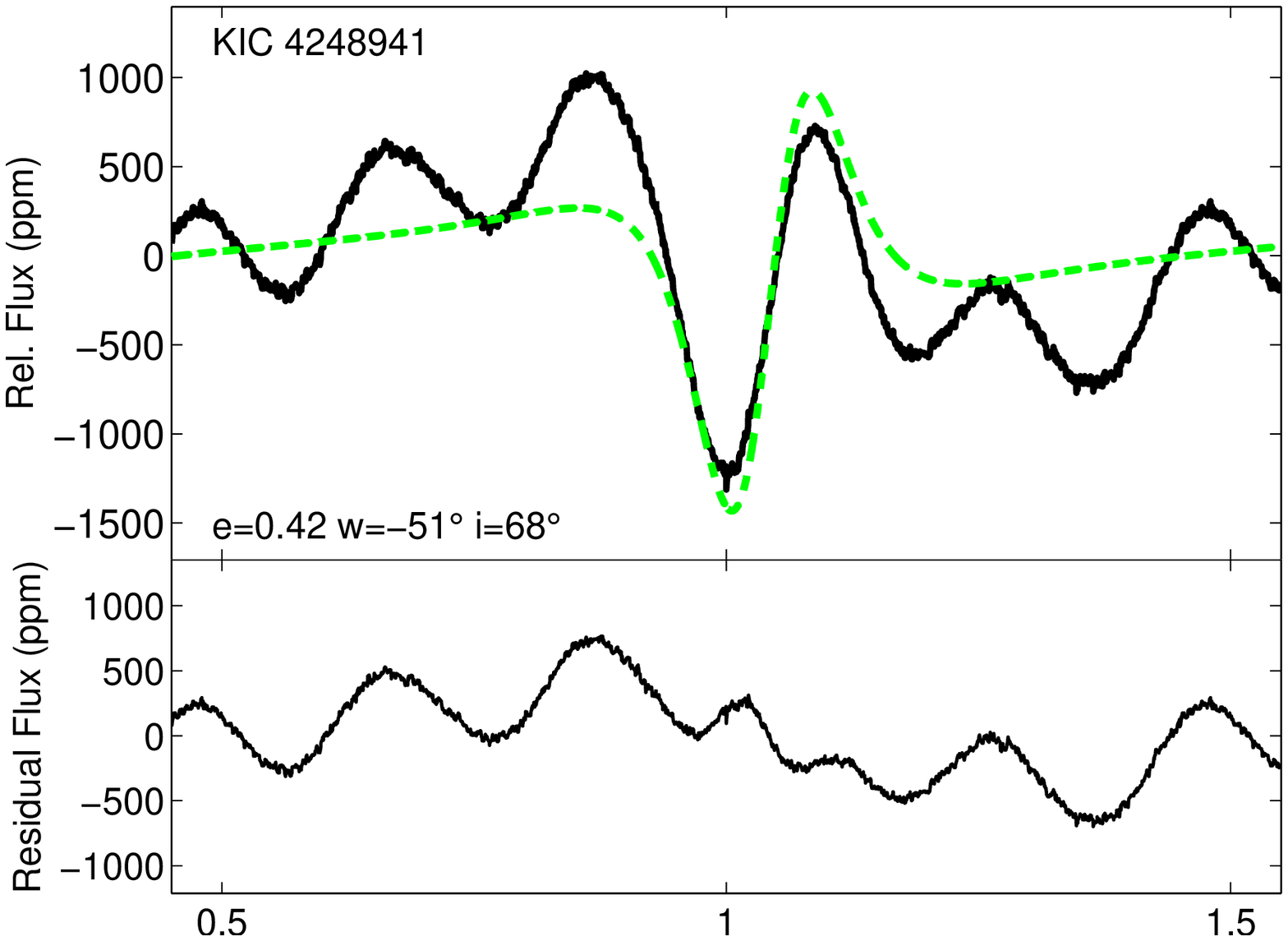}
\includegraphics[scale=.31,angle=0]{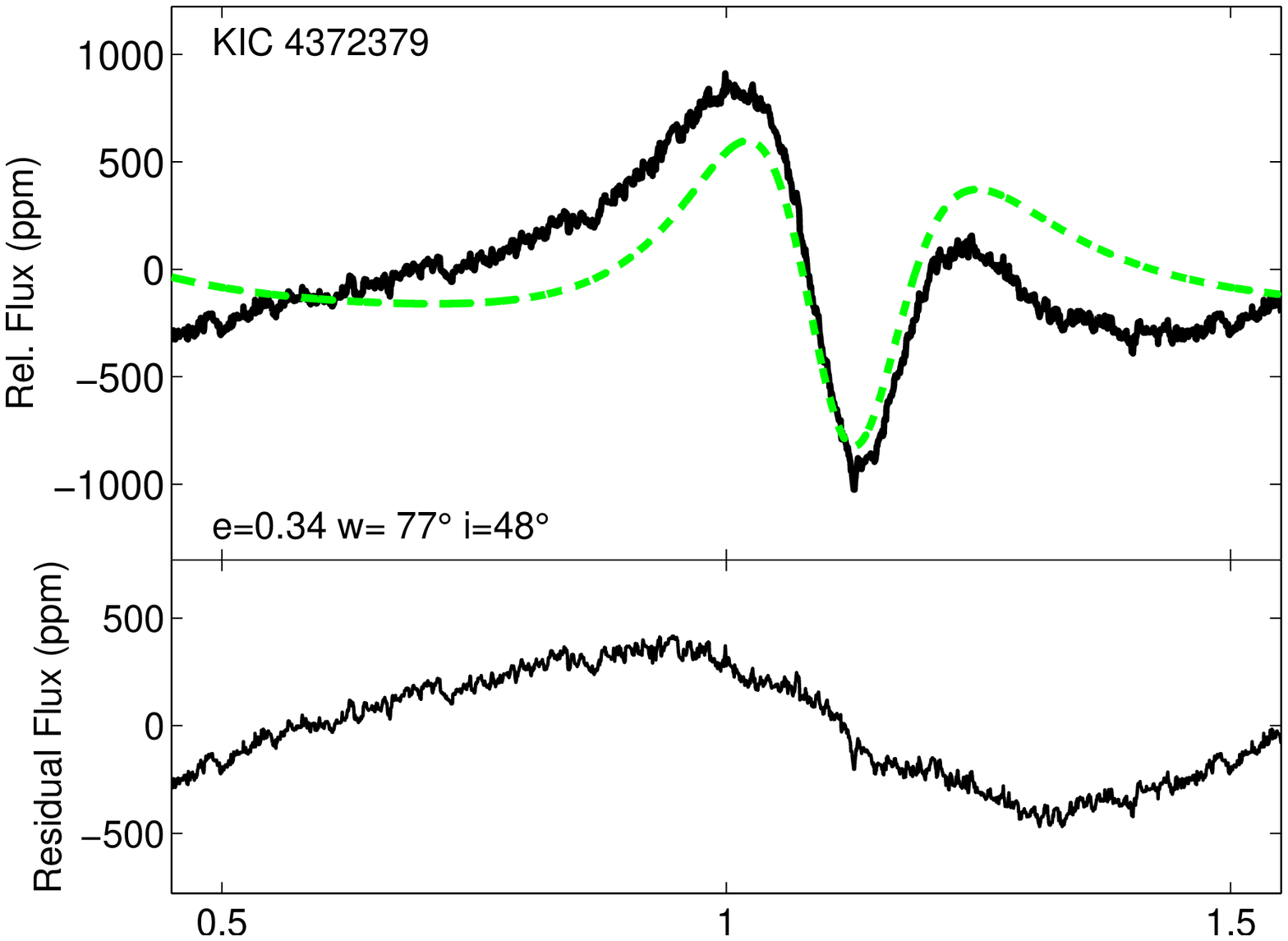}
\includegraphics[scale=.31,angle=0]{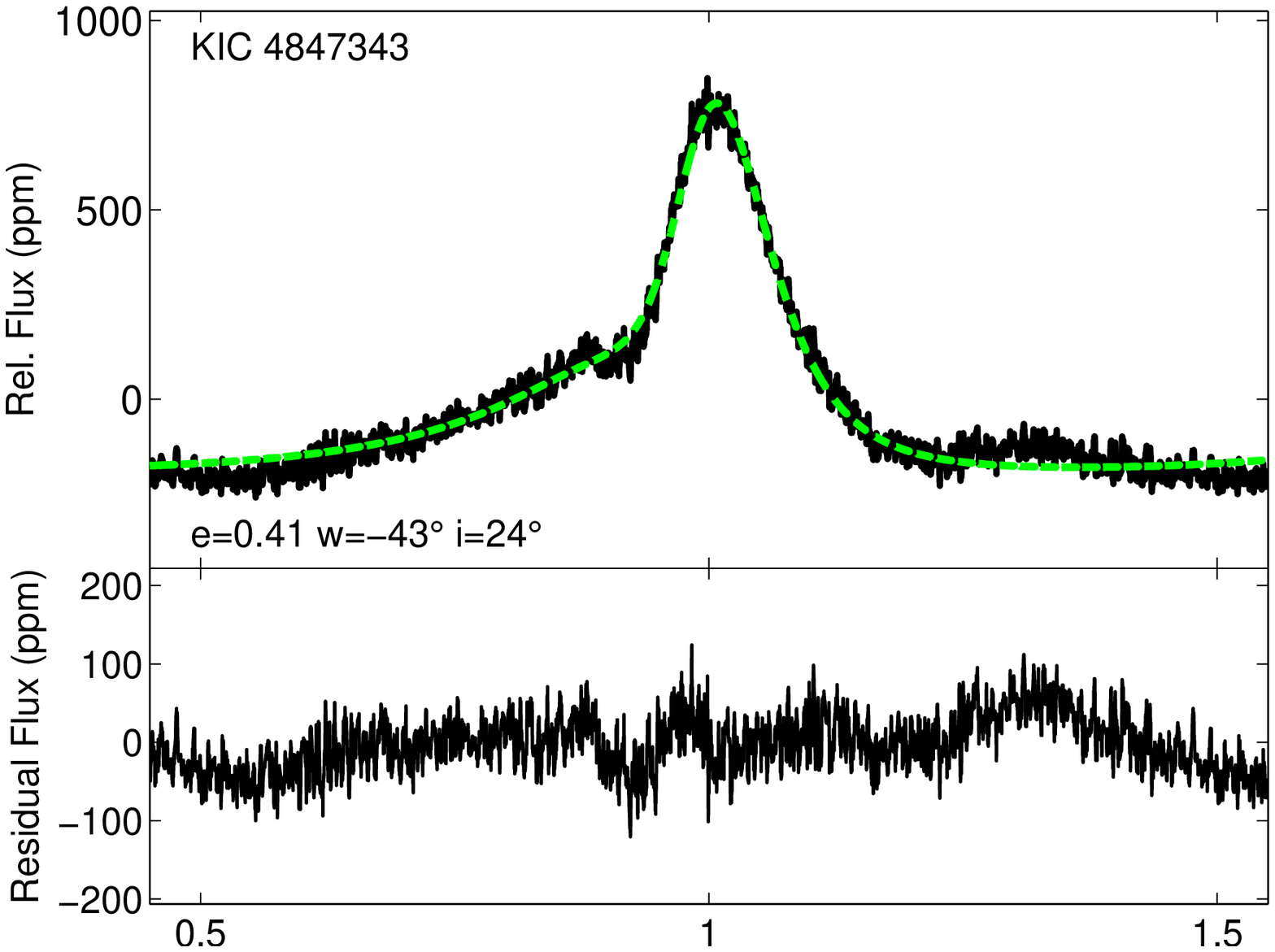}\\
\includegraphics[scale=.31,angle=0]{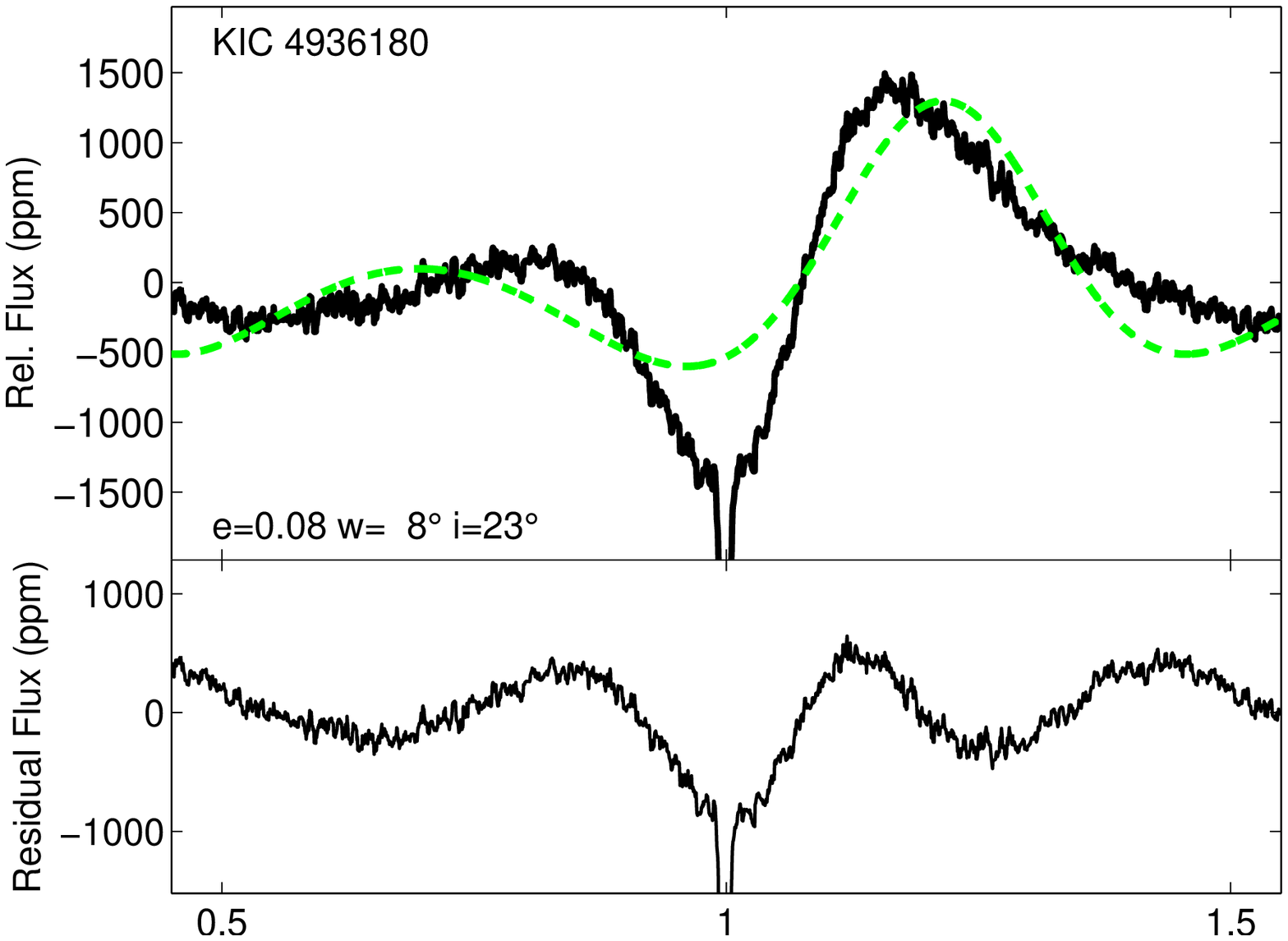}
\includegraphics[scale=.31,angle=0]{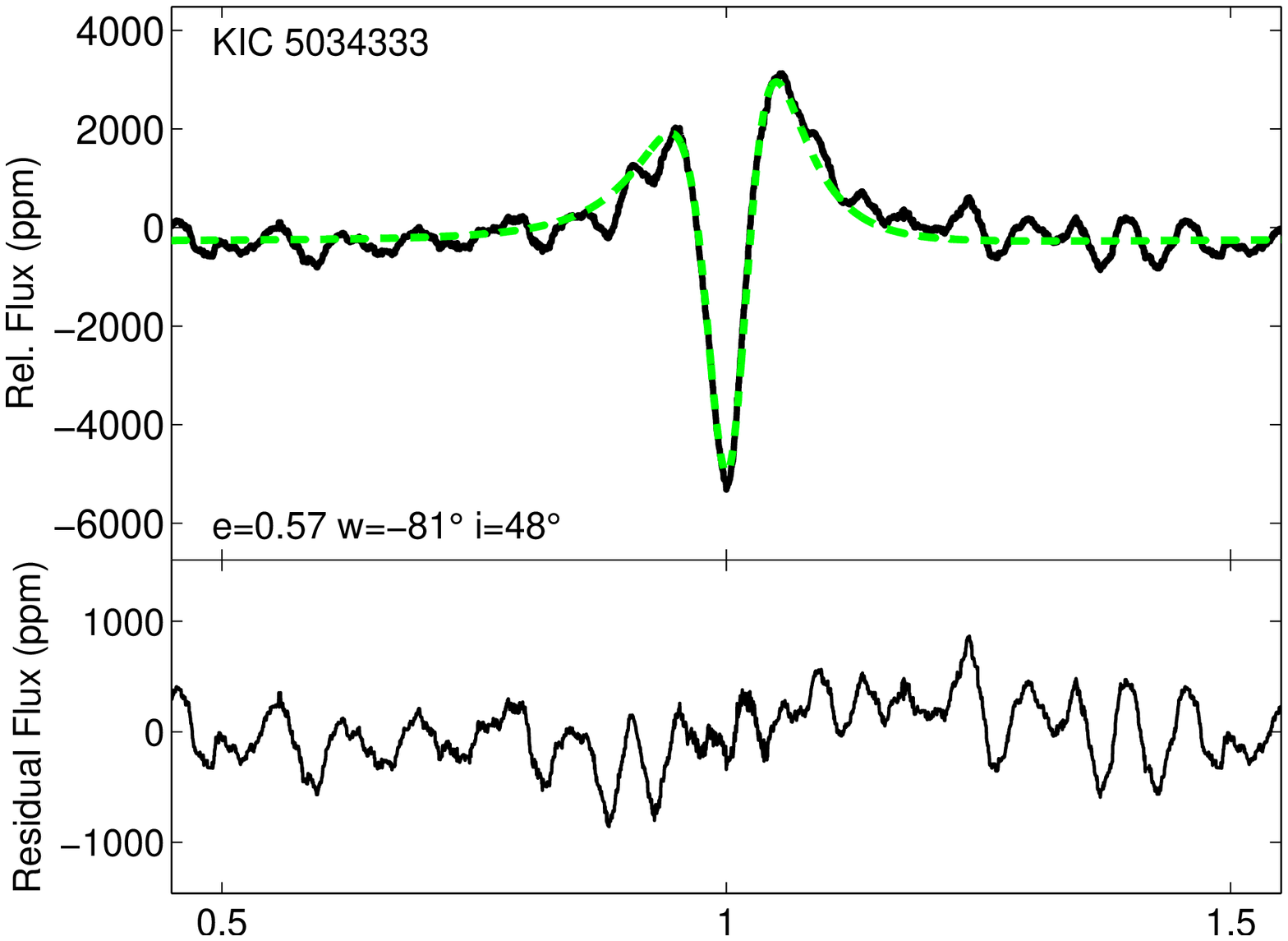}
\includegraphics[scale=.31,angle=0]{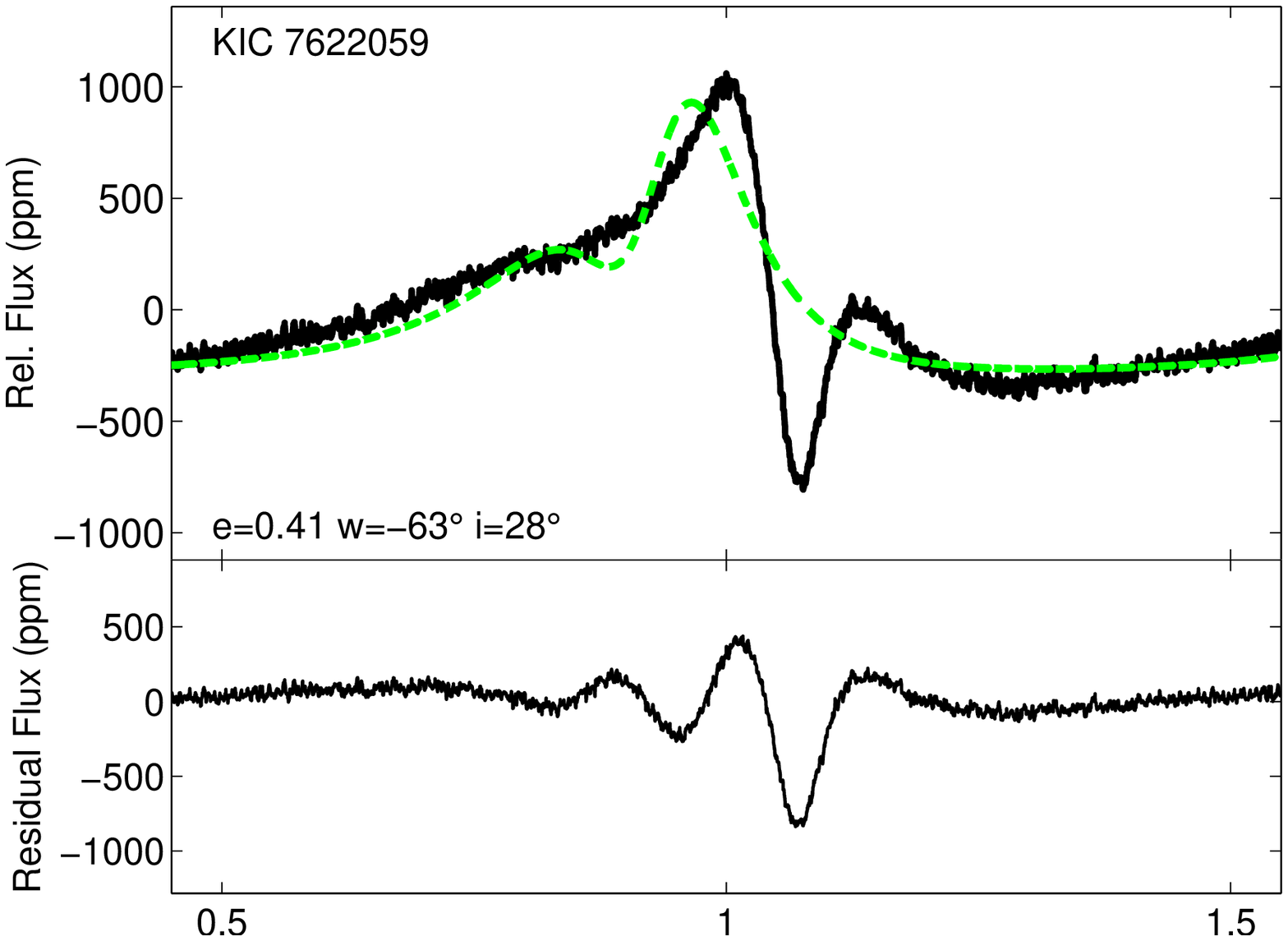}\\
\includegraphics[scale=.31,angle=0]{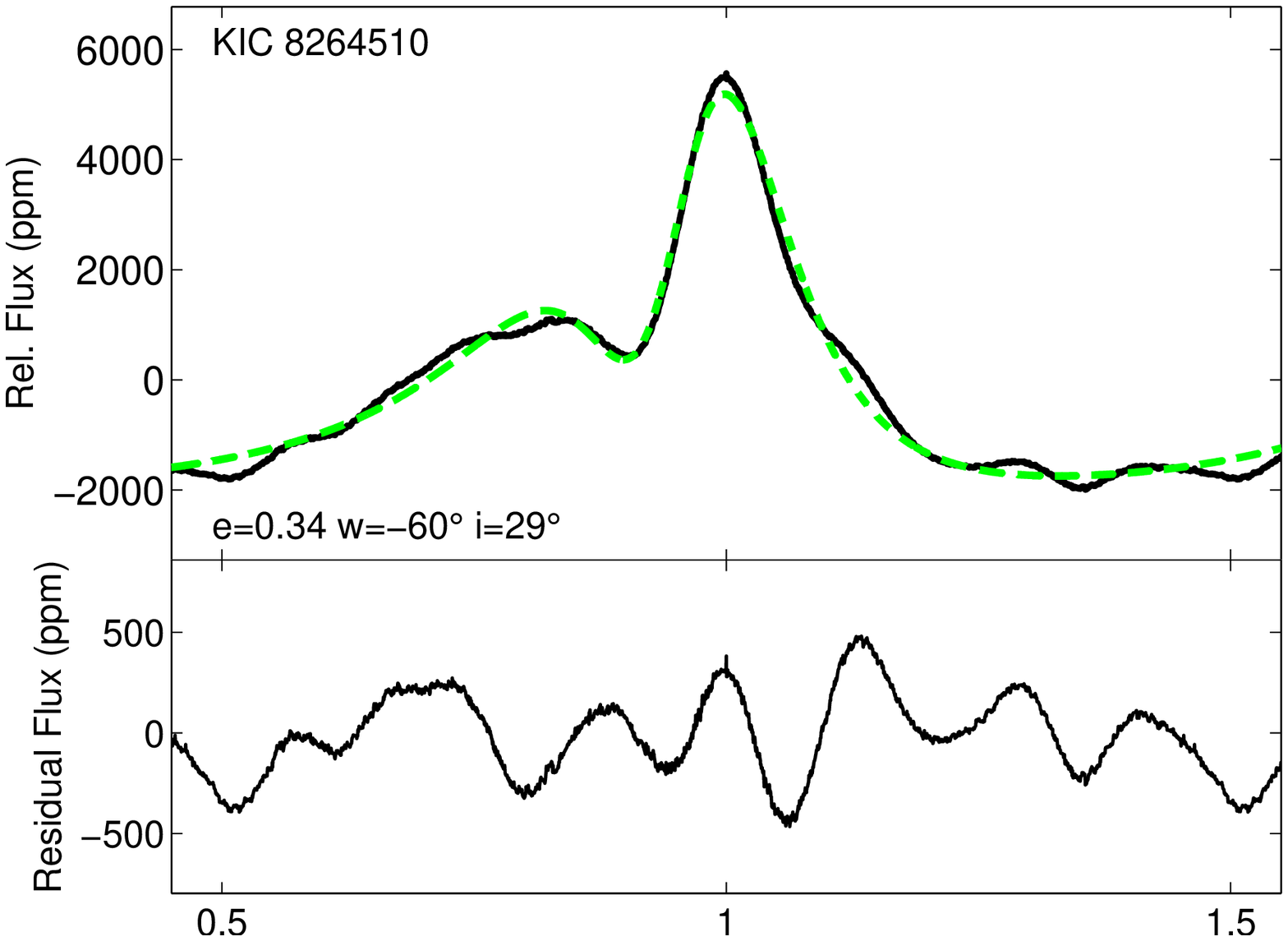}
\includegraphics[scale=.31,angle=0]{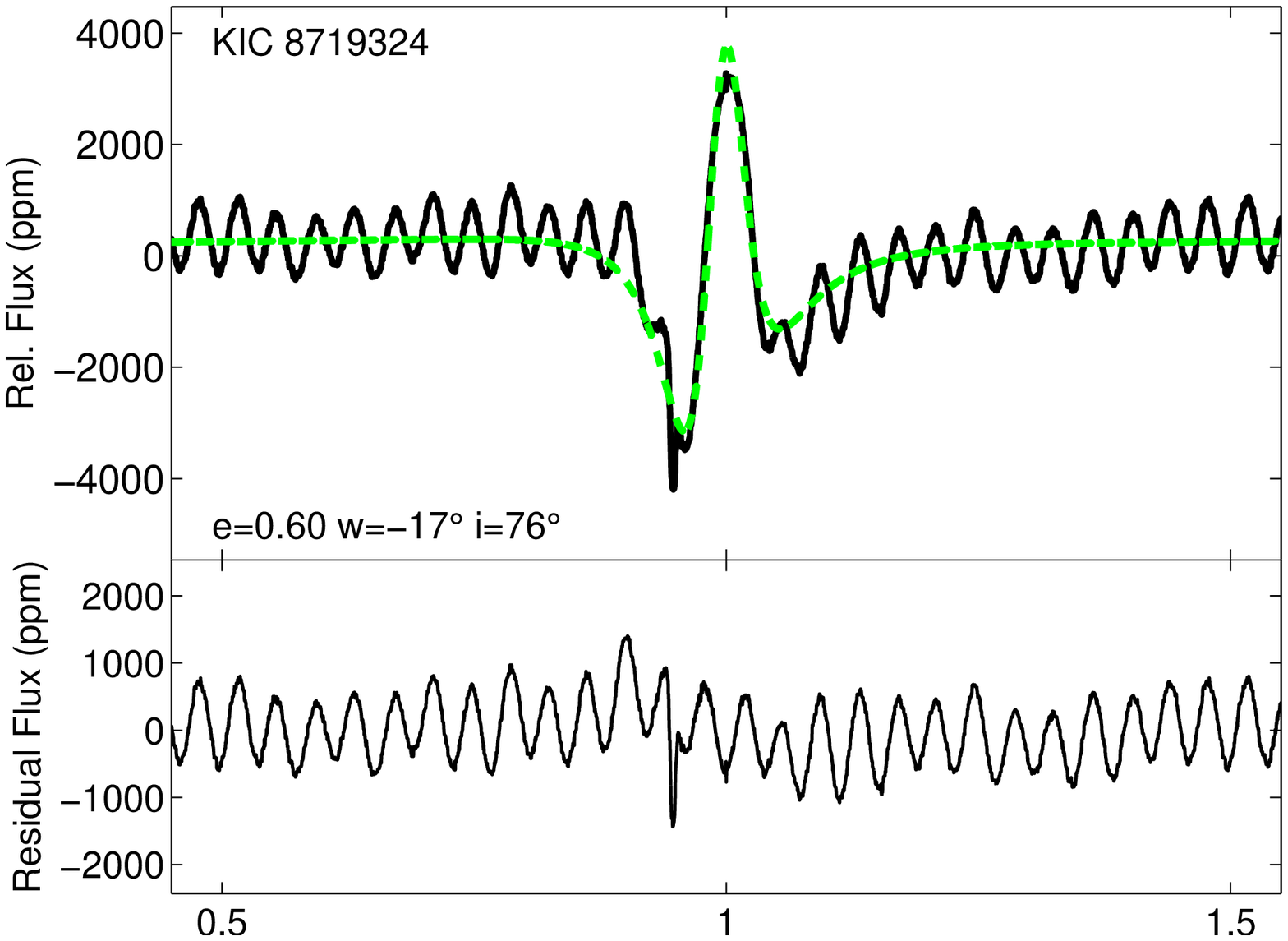}
\includegraphics[scale=.31,angle=0]{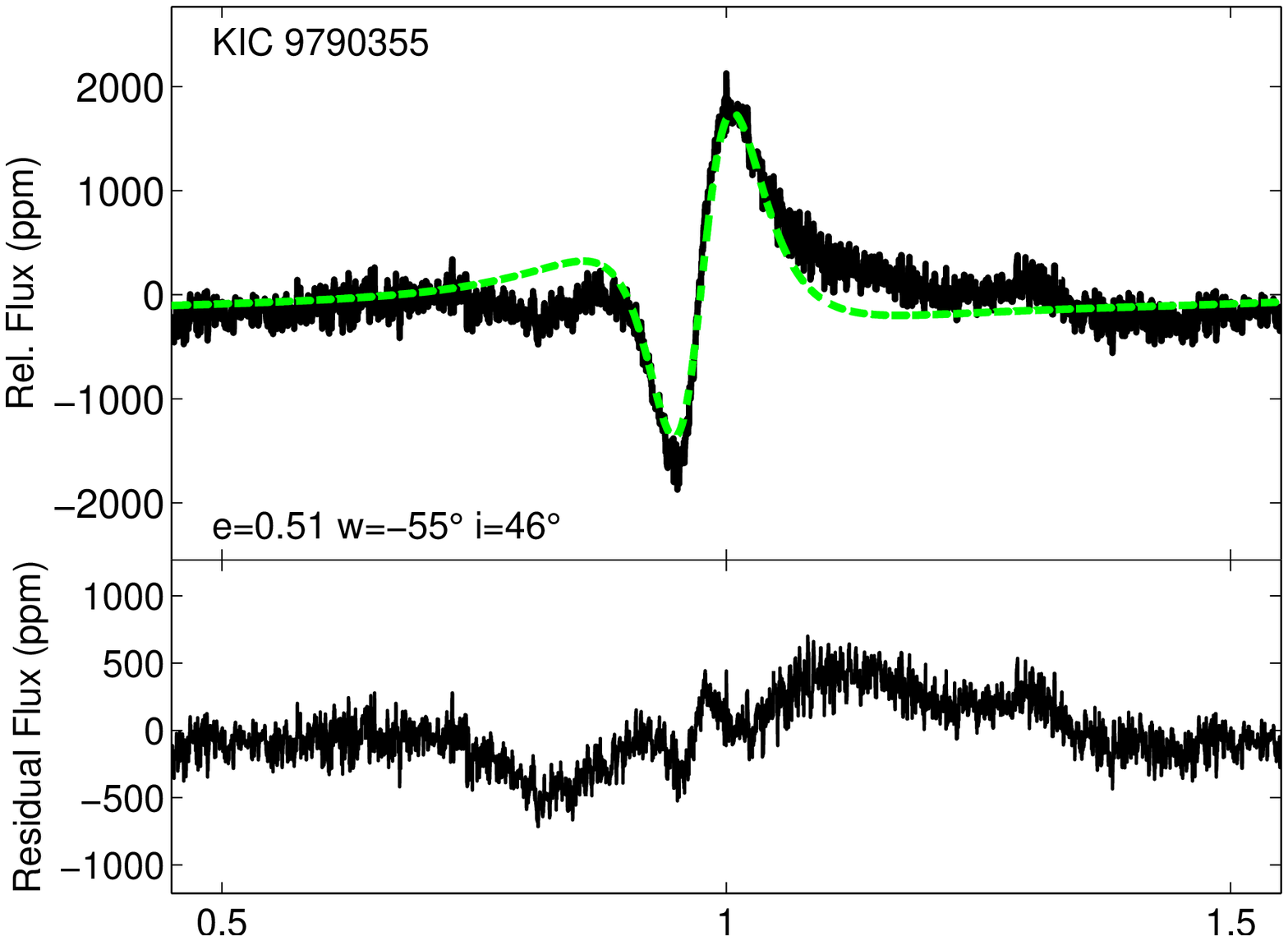}\\
\caption{\label{kumarfits} Best fits of the analytic model of \citet{Kumar1995} (dashed green line) to folded light curves (solid black). The eccentricity, inclination, and angle of periastron ($e$, $i$, and $\omega$ respectively) established from the fit are noted at the bottom of each plot. The lower panel for each set contains the residuals of the fit. Our model excludes many known second-order physical effects, leaving large variations in the residuals (see \S\ref{s:fittingkumar}).  Regions with apparent eclipses were removed from \zeko\ and \ardi\ before fitting. See Figure~\ref{kumarfits2} for fits to the remaining stars.}
\end{figure*}

\begin{figure*}

\includegraphics[scale=.31,angle=0]{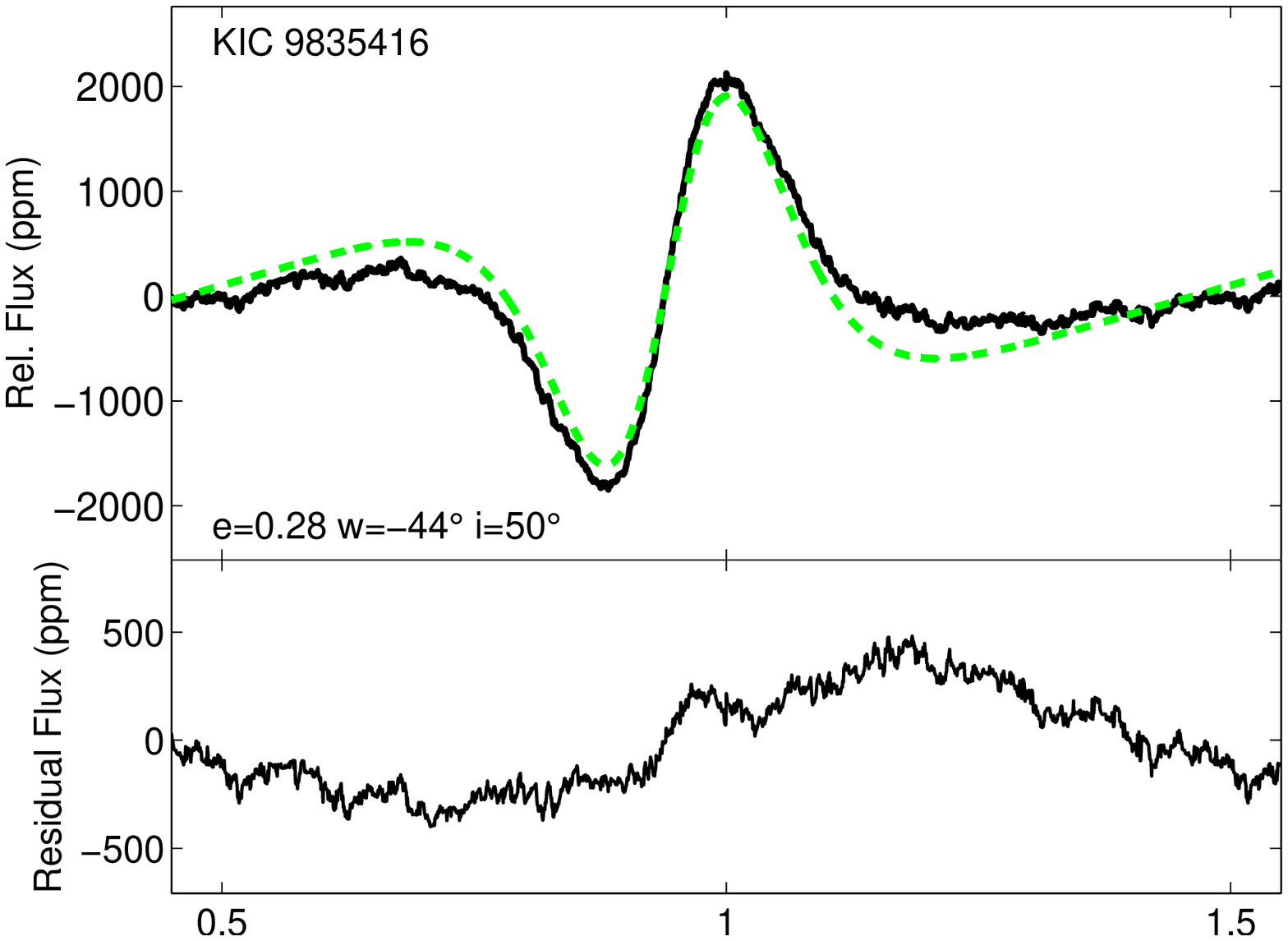}
\includegraphics[scale=.31,angle=0]{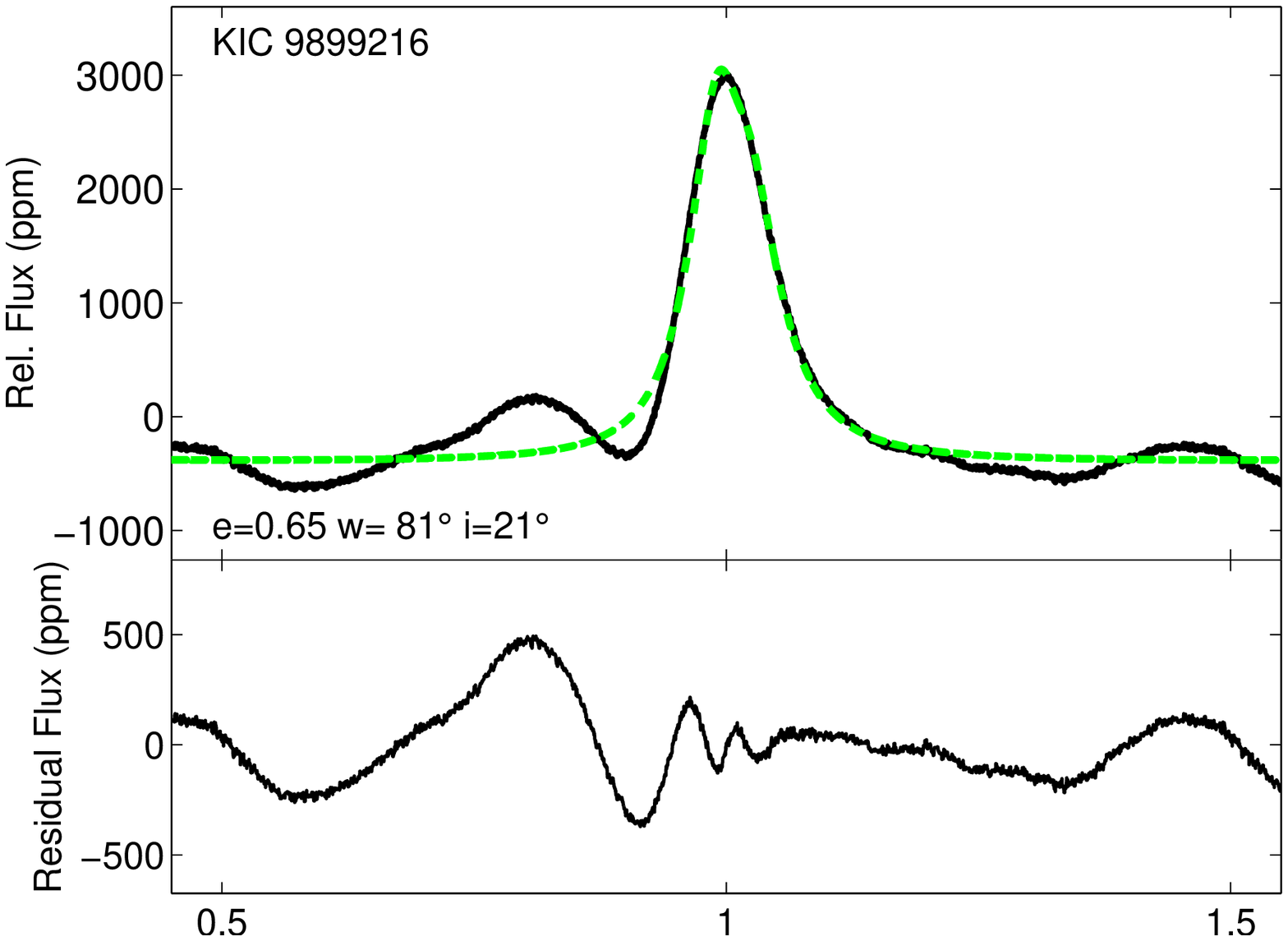}
\includegraphics[scale=.31,angle=0]{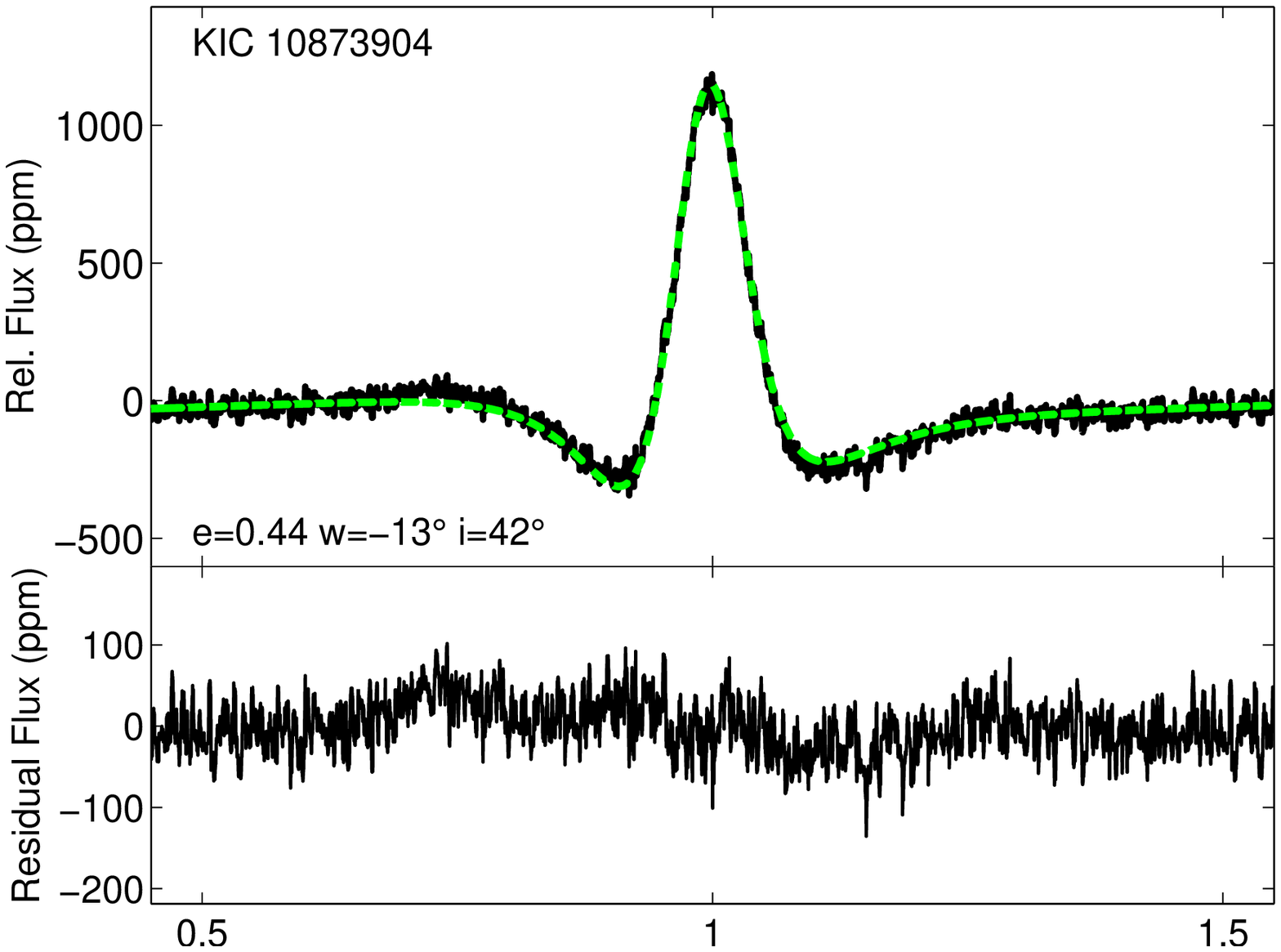}\\
\includegraphics[scale=.31,angle=0]{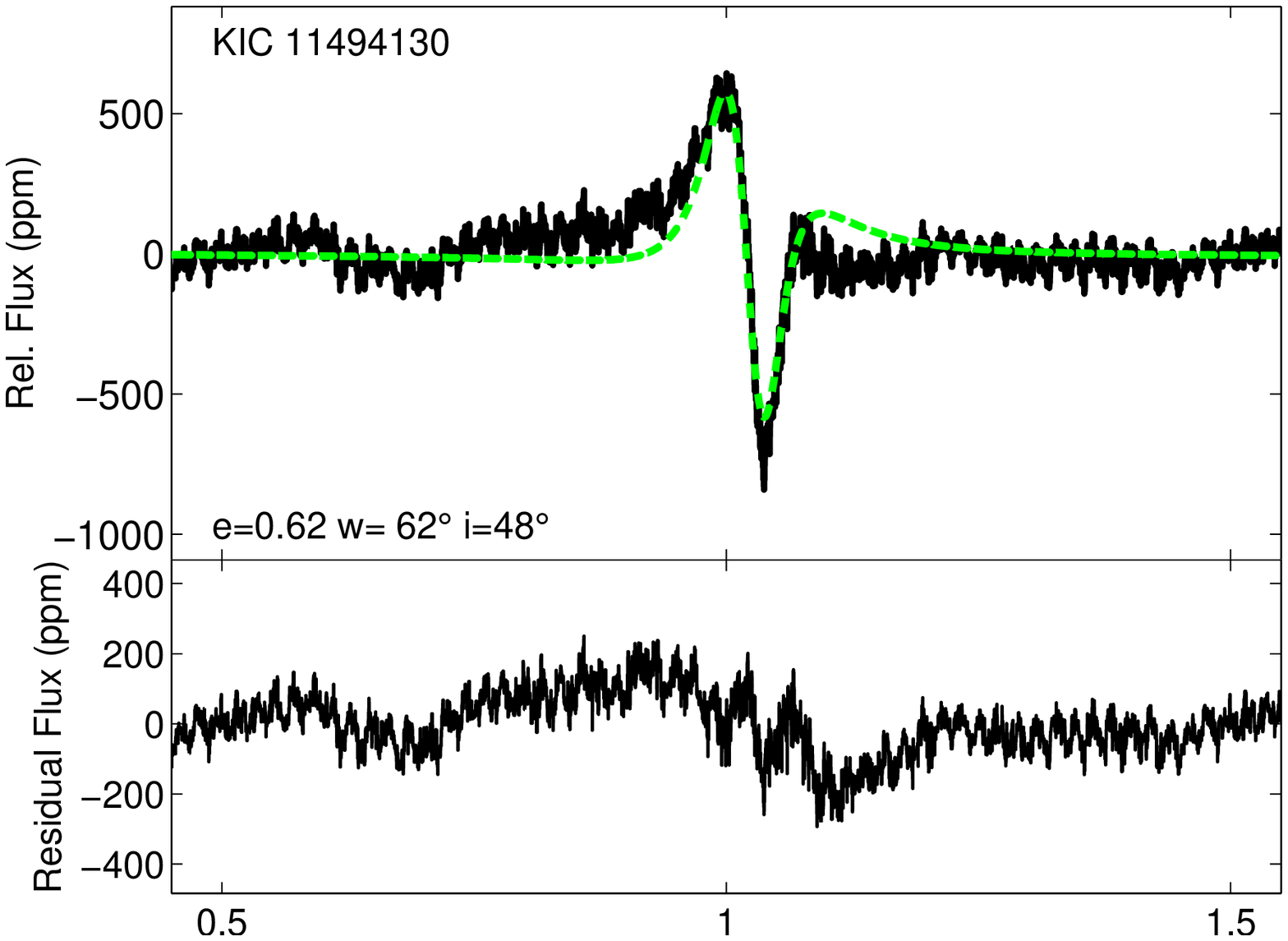}
\includegraphics[scale=.31,angle=0]{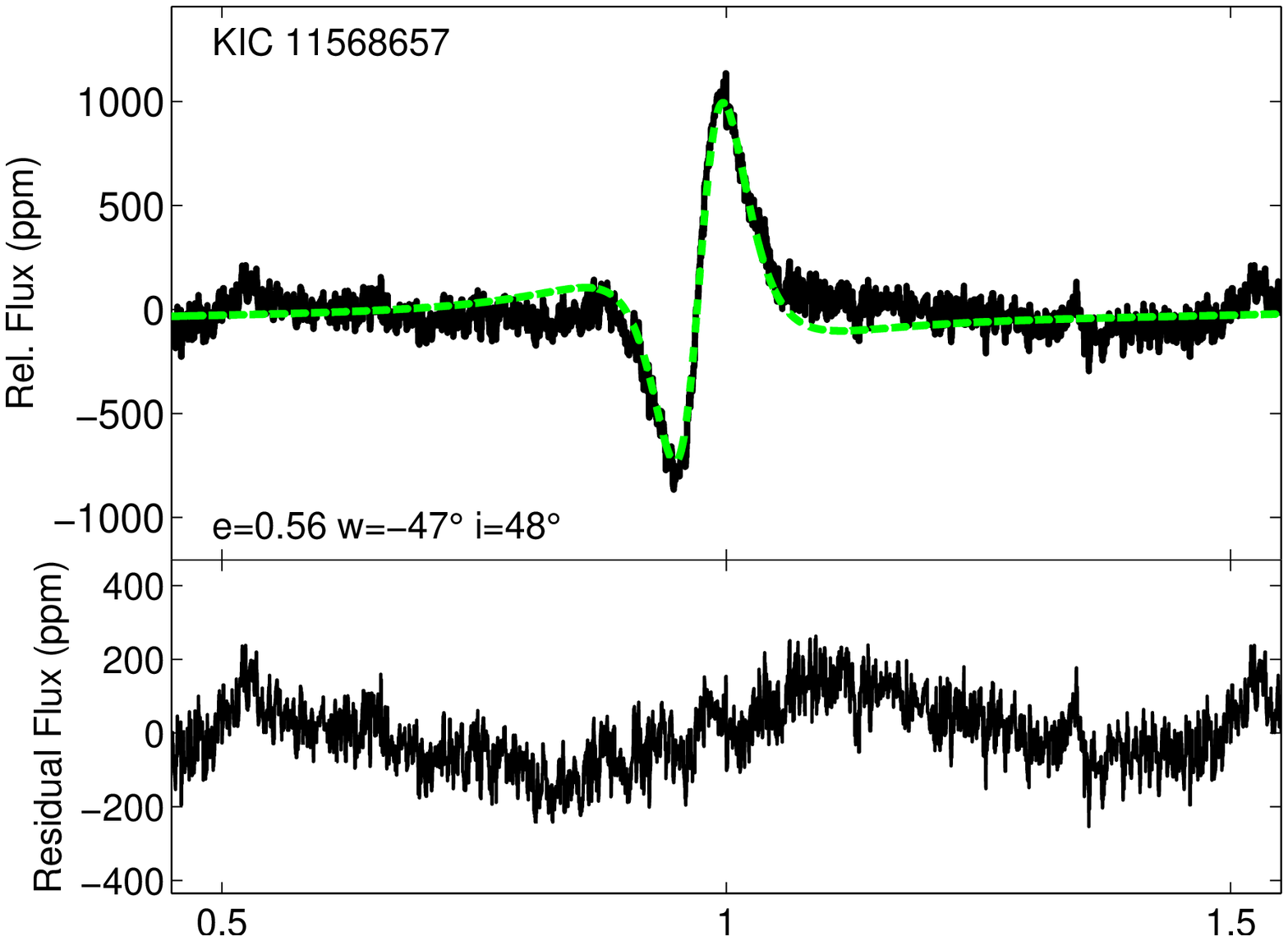}
\includegraphics[scale=.31,angle=0]{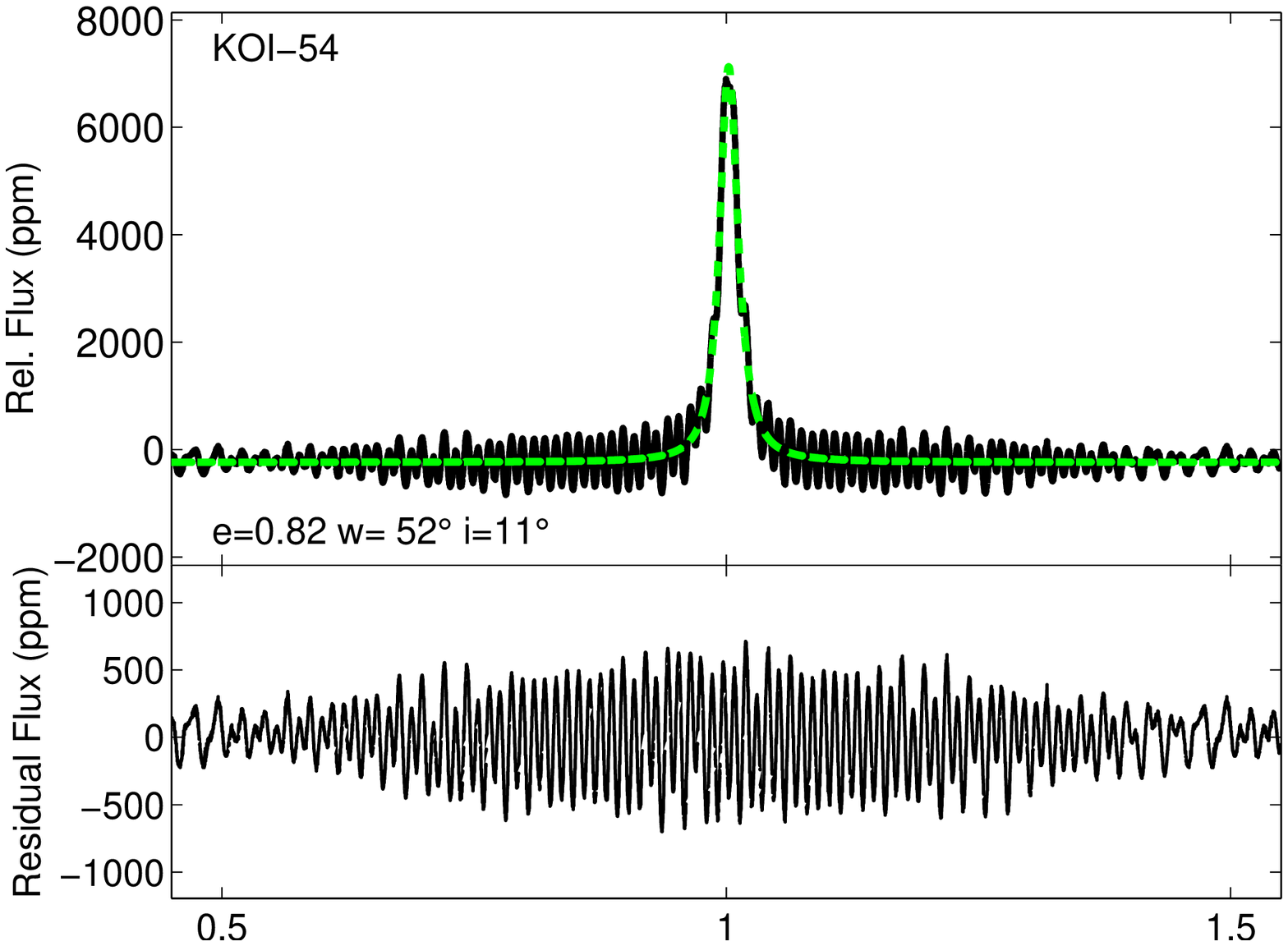}\\
\caption{\label{kumarfits2} Same as Figure~\ref{kumarfits}.  The last panel shows KOI-54 fit by the same model for comparision. Our model excludes the irradiative effect included by \citet{Welsh2011}; see \S\ref{s:fittingkumar}.}
\end{figure*}

The results of our fits can be found in Figures~\ref{kumarfits} and \ref{kumarfits2} and Table~\ref{table}.  Since we fit folded light curves with extremely small uncertainties, and did not fit all the features present in the light curve, our fits result in large reduced chi-square values with small error bars.  To compensate, the 1$\sigma$ uncertainties on the fitted parameters presented in Table~\ref{table} represent the error on the fitted parameters when the uncertainties on the folded light curve are increased so that the reduced chi-square is equal to one. The error bars vary greatly depending on the chi-square of the fit. 

We have not included several physical parameters known to affect the light curves of such highly eccentric systems. These effects will have a different impact depending on the properties of the stars in the system. In many cases the harmonic pulsations are large and may significantly alter the results of the fit. As an example, \thor\ shows large harmonic pulsations that could easily sway the results of our fits. Also, the exclusion of orbital effects such as irradiation (and to a lesser degree Doppler boosting) can prevent us from obtaining a good fit for the true orbital parameters. For instance \lars, \zeko, and \gene, all show a heartbeat shape but are not fit well by our simple tidal distortion model.  Adding the additional physical effects may account for the discrepancy.

While incomplete, we note that the \citet{Kumar1995} model can give a good first estimate of the oribtal parameters even in cases where the irradiation is significant, and has the advantage of not requiring knowledge about the properties of the stars to make this estimate.  For KOI-54, where the irradiation makes-up almost half of the change in flux, the fit to the tidal distortion model gives a reasonable estimate of the orbital parameters. According to the detailed modeling of \citet{Welsh2011}, the orbital parameters of KOI-54 are $e=$0.8315(32), $\omega =$39$^\circ$.46(51), and $i=$5$^\circ$.52(10). Our fit is presented in the last panel of Figure~\ref{kumarfits2}. While our model does not include irradiation, the pulsations, nor the information from radial velocity measurements, our eccentricity, 0.82(01), angle of periastron, 52$^\circ$(30), and inclination, 11$^\circ$(3), all match within the two sigma error bars. 

The ability to fit KOI-54 well with only the tidal distortion model is in part because of its inclination.  At low inclinations the shape of the contribution from irradiation closely resembles that of the tidal distortion. For higher inclinations this will not always be the case, and so when the irradiation is significant it can appreciably alter the shape of the heartbeat \citep[see][]{Burkart2012}. Once knowledge of the companion can be obtained, the effects of irradiation can be included accurately in the model of the light curve. 

We have three obviously eclipsing systems: \ardi, \zeko, and \carl.  We fit these systems with the inclinations of 42.79$^\circ$, 23.2$^\circ$, and 75.78$^\circ$, respectively. Our model for \zeko\ is obviously incomplete and so the fitted inclination is likely incorrect. Both \carl\ and \ardi\ have very good fits, but rather small inclinations for eclipsing systems.  It is possible to achieve an eclipse at these inclinations if the distance at periastron for these highly eccentric systems is very small ($\lessapprox3$\,R$_\sun$).  This is possible if the secondary has a rather low mass ($\lessapprox0.5$\,M$_\sun$). A more likely possibility is that these are cases where the irradiation has a significant effect on the shape of the light curve, causing us to estimate the inclination at a lower value. 

\begin{figure}[t]
\includegraphics[scale=0.56,angle=270]{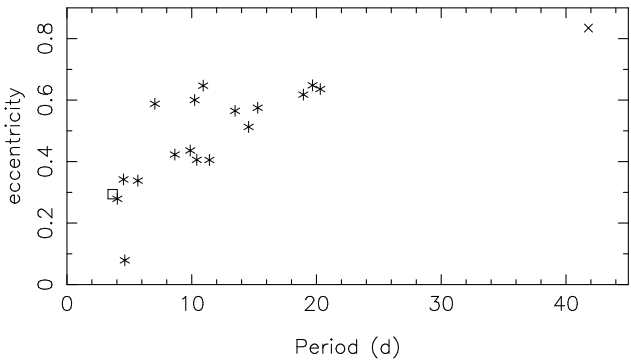}
\caption{\label{eccentricity} The best fit eccentricity measured using the photometric tidal distoriton model plotted against the orbital period for the \nheartbeat\ \kepler\ systems. KOI-54 is plotted as a `x' and HD~174884 \citet{Maceroni2009} is plotted as a square. The error bars are smaller than the size of the points.}
\end{figure}

This simple model, despite being incomplete, allows us to illustrate that tidal distortions in an eccentric binary system can explain many of the features in the light curves.  Fitting these features will allow us to estimate the orbital parameters of these systems. One trend noticed from fitting our simple model is that the eccentricity is smaller for shorter period systems; see Figure~\ref{eccentricity}.   While we are less likely to find long-period, low-eccentricity systems with observable tidally-induced variations, we would expect to detect highly eccentric, short-period systems if they existed. Since shorter period orbits undergo stronger tidal forces, they are likely to circularize more quickly, leaving a dearth of highly eccentric short period binaries.  Similar trends have been seen by \citet{Prsa2008} and \citet{Mathieu1988} for eclipsing binaries.


\begin{table*}[h]
\caption{\label{table}Eccentric Binary Systems}
\begin{tabular}{llrlllllll}
\tableline
\tableline
\multicolumn{1}{c}{KIC} & \multicolumn{1}{c}{2MASS} & $K_{\rm mag}$ & quarters & \teff & \logg & Period (d) & \multicolumn{1}{c}{$e$} & \multicolumn{1}{c}{$\omega$ ($^{\circ}$)} & \multicolumn{1}{c}{$i$ ($^{\circ}$)}\\
\tableline
3230227 & 19202701+3823595 & 9.002 & 0-3 & 8750 & 5.0 & 7.04711(87) & 0.5880(40) & -67.9(1.2) & 42.79(46)\\
3547874 & 19290789+3839565 & 10.952 & 0-6 & 6750 & 4.5 & 19.69284(59) & 0.6480(20) & 40.40(20) & 46.434(56)\\
3749404 & 19281908+3850135 & 10.575 & 0-6 & 7500 & 5.0 & 20.306630(60) & 0.6360(20) & -58.17(75) & 37.53(17)\\
4248941 & 19102476+3923516 & 12.159 & 0,2-7 & 6750 & 4.5 & 8.644718(34) & 0.423(14) & -50.5(4.3) & 68.3(5.5)\\
4372379 & 19340487+3926520 & 13.810 & 1-5,7 & 6750 & 4.5 & 4.535167(35) & 0.342(92) & 77(26) & 48(11)\\
4847343 & 19394839+3955460 & 12.837 & 0-6 & 6250 & 4.5 & 11.4162193(59) & 0.4050(30) & -42.81(62) & 24.23(16)\\
4936180 & 19402728+4004162 & 15.270 & 1-6 & 6750 & 5.0 & 4.640867(59) & 0.079(58) & 8(16) & 23.2(6.7)\\
5034333 & 19493609+4008581 & 11.520 & 1-7 & 9250 & 4.5 & 15.29434(21) & 0.5750(20) & -81.41(38) & 48.21(13)\\
7622059 & 19434079+4316543 & 13.272 & 1-7 & 6250 & 4.5 & 10.403145(43) & 0.406(91) & -63(16) & 28.0(5.5)\\
8264510 & 20035245+4411313 & 10.609 & 0-7 & 8250 & 4.0 & 5.6867311(64) & 0.3380(30) & -60.42(49) & 29.03(16)\\
8719324 & 20042742+4449328 & 11.614 & 0-6 & 7750 & 4.5 & 10.23262347(27) & 0.6000(20) & -16.61(42) & 75.78(89)\\
9790355 & 19545806+4634382 & 14.055 & 3-6 & 6000 & 3.0 & 14.56034(73) & 0.5130(70) & -55.1(2.3) & 46.33(69)\\
9835416 & 19344716+4637138 & 13.338 & 4-6 & 7250 & 5.0 & 4.03650(11) & 0.279(23) & -44.3(7.5) & 49.8(3.5)\\
9899216 & 19403879+4645023 & 10.886 & 1-6 & 7500 & 4.0 & 10.915763(20) & 0.6470(40) & 81.2(1.3) & 20.74(61)\\
10873904 & 19462402+4817456 & 13.065 & 1-6 & 6250 & 5.0 & 9.88553(23) & 0.4360(20) & -12.67(18) & 42.196(47)\\
11494130 & 18560643+4924555 & 10.988 & 0-3 & 6750 & 4.5 & 18.9555(14) & 0.6180(40) & 62.3(1.6) & 48.29(52)\\
11568657 & 19450715+4931040 & 13.246 & 0-4 & 6000 & 4.5 & 13.47697(98) & 0.5650(20) & -46.80(30) & 48.202(96)\\
\tableline
\end{tabular}

Note -- The values for \teff\ and \logg\ are the result of fits to a grid of model spectra with a spacing of 250\,K and 0.5 (cgs), respectively (see \S\ref{s:spectra}). The eccentricity ($e$), angle of periastron ($\omega$), and inclination ($i$) are the result of the fits to the tidal distortion model described in \S\ref{s:fittingkumar}. This model is likely incomplete for many of these systems; use the resulting orbital parameters with caution.  The error bars represent 1$\sigma$ uncertainties and do not include the error associated with any systematic effects.\\

\end{table*}

\section{Spectra}
\label{s:spectra}
In order to measure the stellar parameters and to confirm the prediction that these are eccentric binary systems, we obtained optical spectra of the \nheartbeat\ different heartbeat
stars at the KPNO Mayall 4-m telescope between 2011 May 14 and
September 7 (UTC) using the long-slit spectrograph RCSpec and T2KA
Camera. The KPB-22b grating was used in 2nd order with a slit width of
1.0" to produce a dispersion of $0.72$\,\AA\,${\rm pixel}^{-1}$ and resolution $R\sim4000$.  Each spectrum was observed with an exposure time between 180 and 900\,s, depending on brightness and conditions and each object exposure was preceded by a 30-s exposure of an FeAr arc lamp spectrum for wavelength calibration.  A standard star (one of BD+40\,4032, Feige~34, or Wolf\,1346) was observed each night using the same method \citep{Stone1977,Barnes1982}.  Bias exposures and quartz lamp flat fields were taken the day before each night of observations for calibration.

We reduced the spectra using standard IRAF (Image Reduction and Analysis Facility) tools.  First, the overscan bias level was subtracted from each frame.  Next, a master residual bias and normalized flat-field frame were constructed from sets of bias and quartz lamp exposures and then applied to each spectrum frame to correct for any remaining 2-D bias patterns and pixel-to-pixel variations in detector response.  The {\sc IRAF} onedspec package was used to define apertures and to extract sky-subtracted spectra.  The same apertures are also employed to extract an arc lamp spectrum for each object.  The spectra are then resampled on a linear wavelength scale using fits to the arc lamp spectra.  The reduced standards were then fit to the flux standards provided with the {\sc IRAF} onedspec package to produce a sensitivity curve in order to correct each science spectrum.

These spectra suffer from degraded focus outside of the range
$3800-4700$\,\AA\ where the fluxes are difficult to properly calibrate.
Consequently, we have restricted our spectra fits to this limited wavelength range.

\begin{figure*}[ht]

\includegraphics[scale=.42,angle=270]{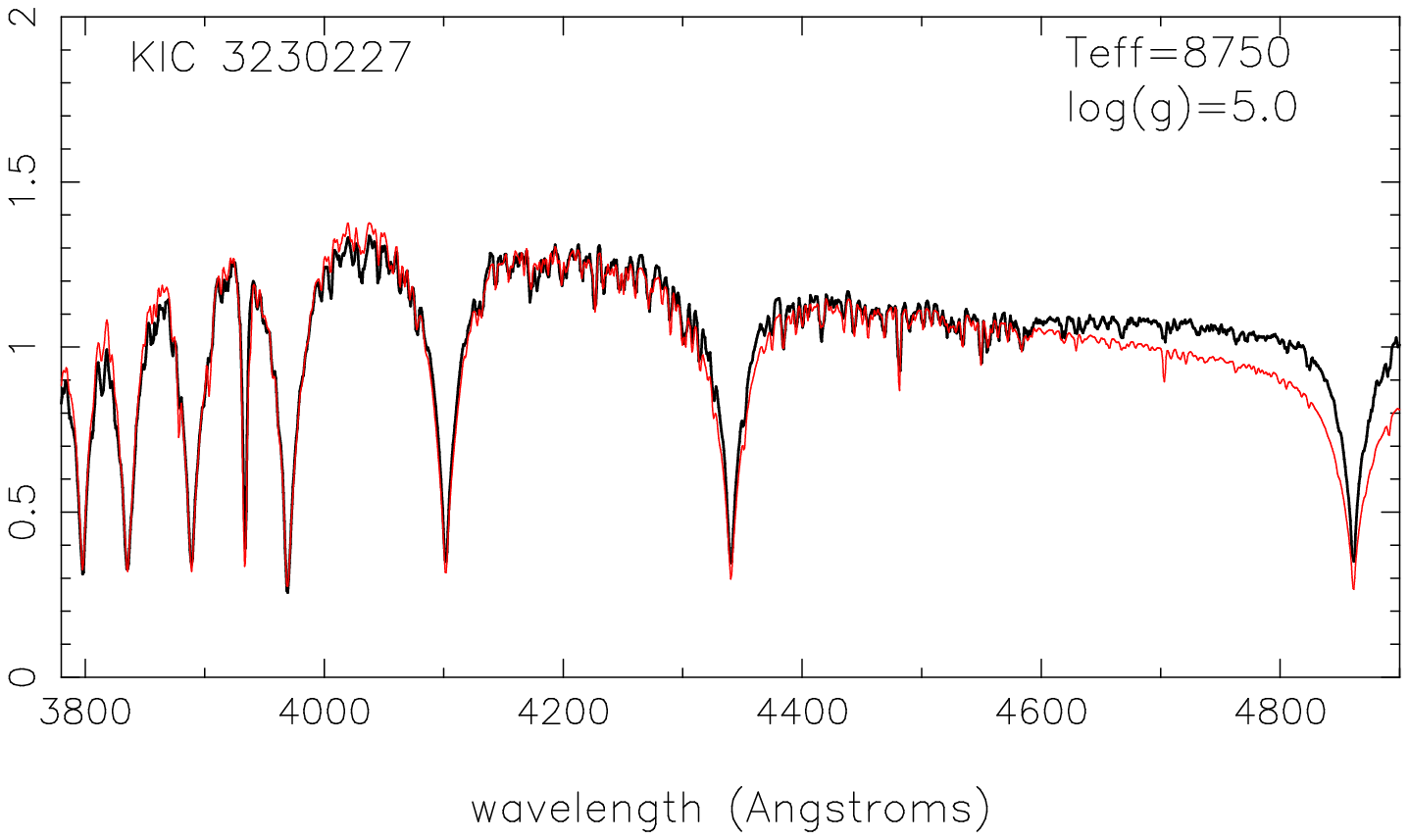}
\includegraphics[scale=.42,angle=270]{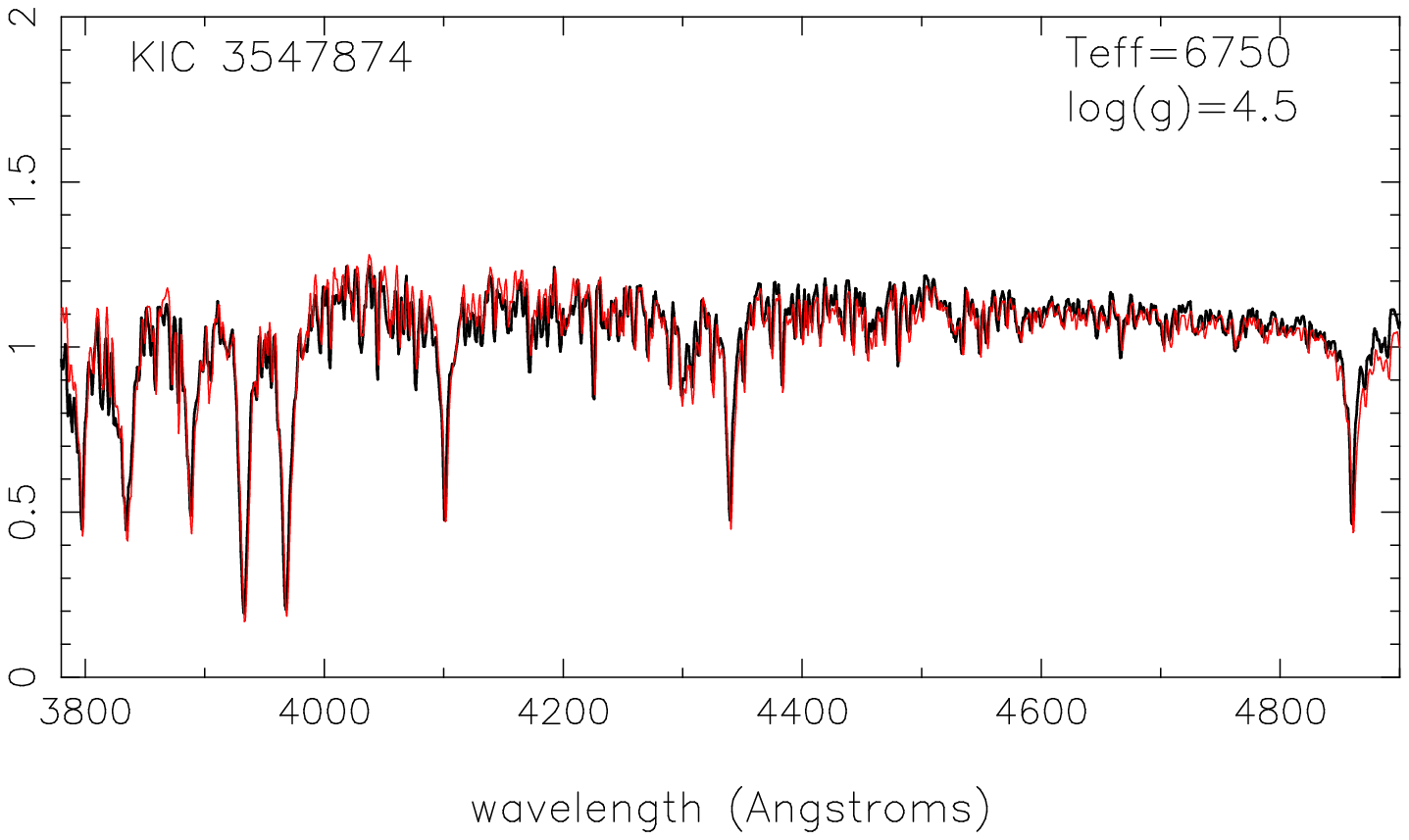}
\includegraphics[scale=.42,angle=270]{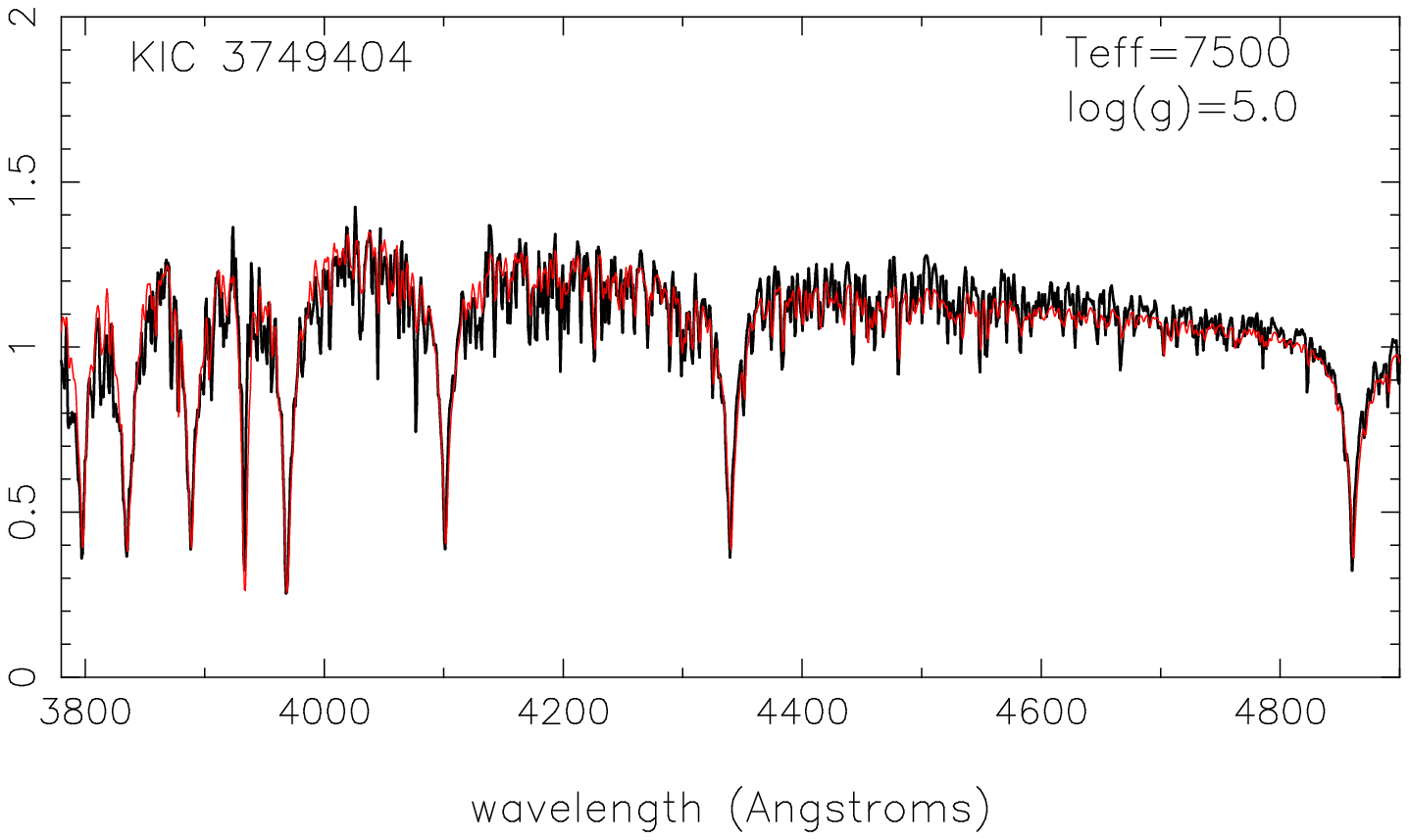}
\includegraphics[scale=.42,angle=270]{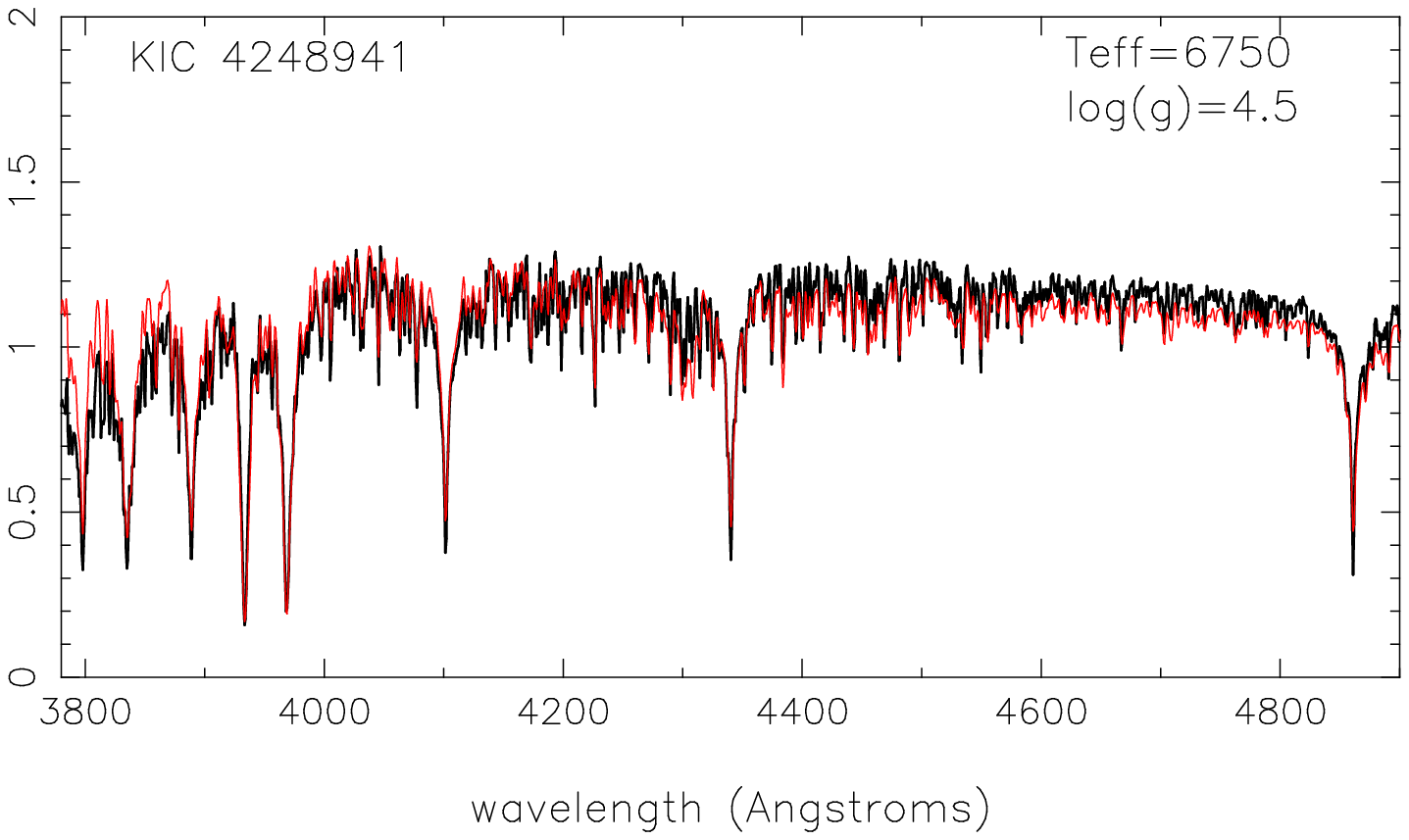}
\includegraphics[scale=.42,angle=270]{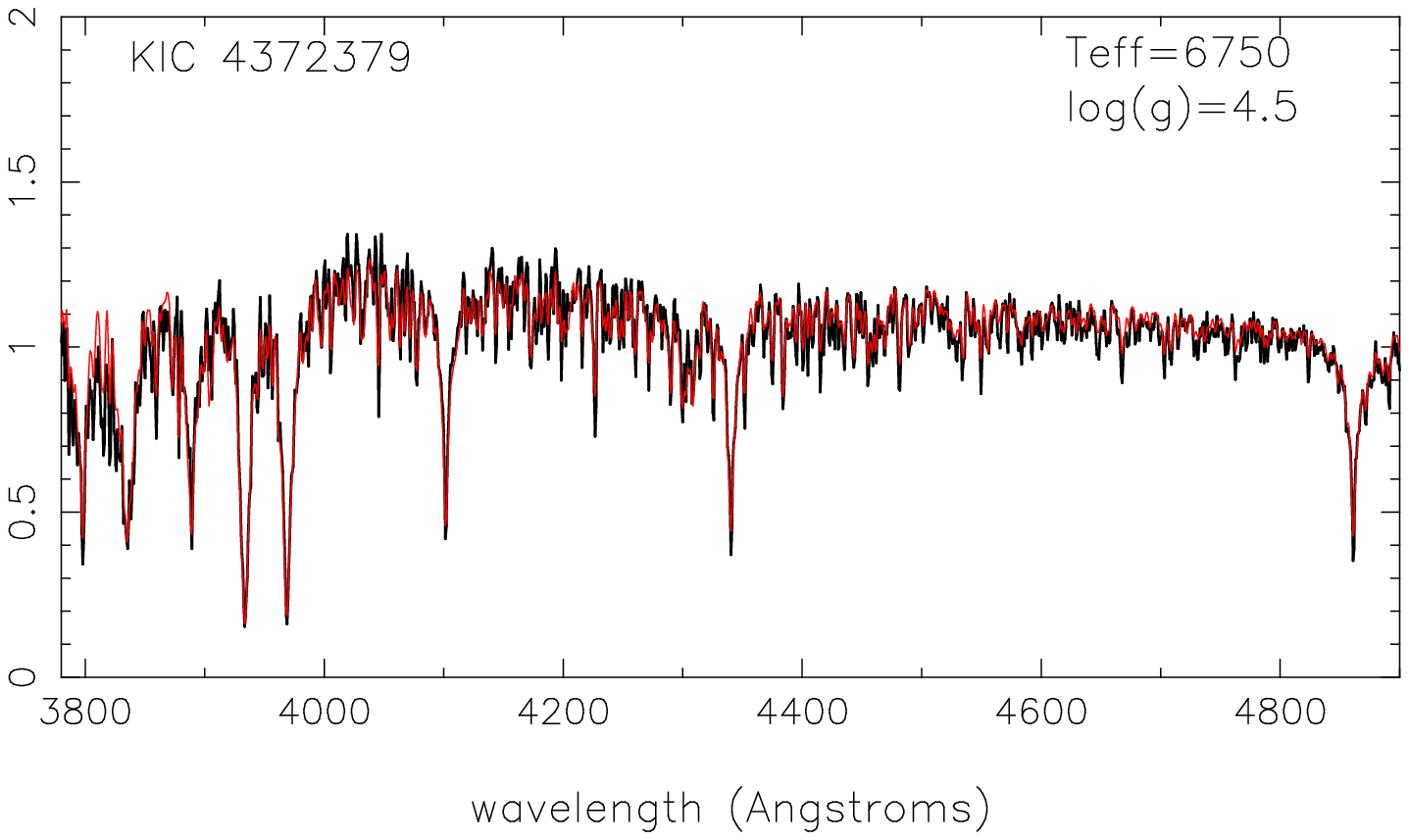}
\includegraphics[scale=.42,angle=270]{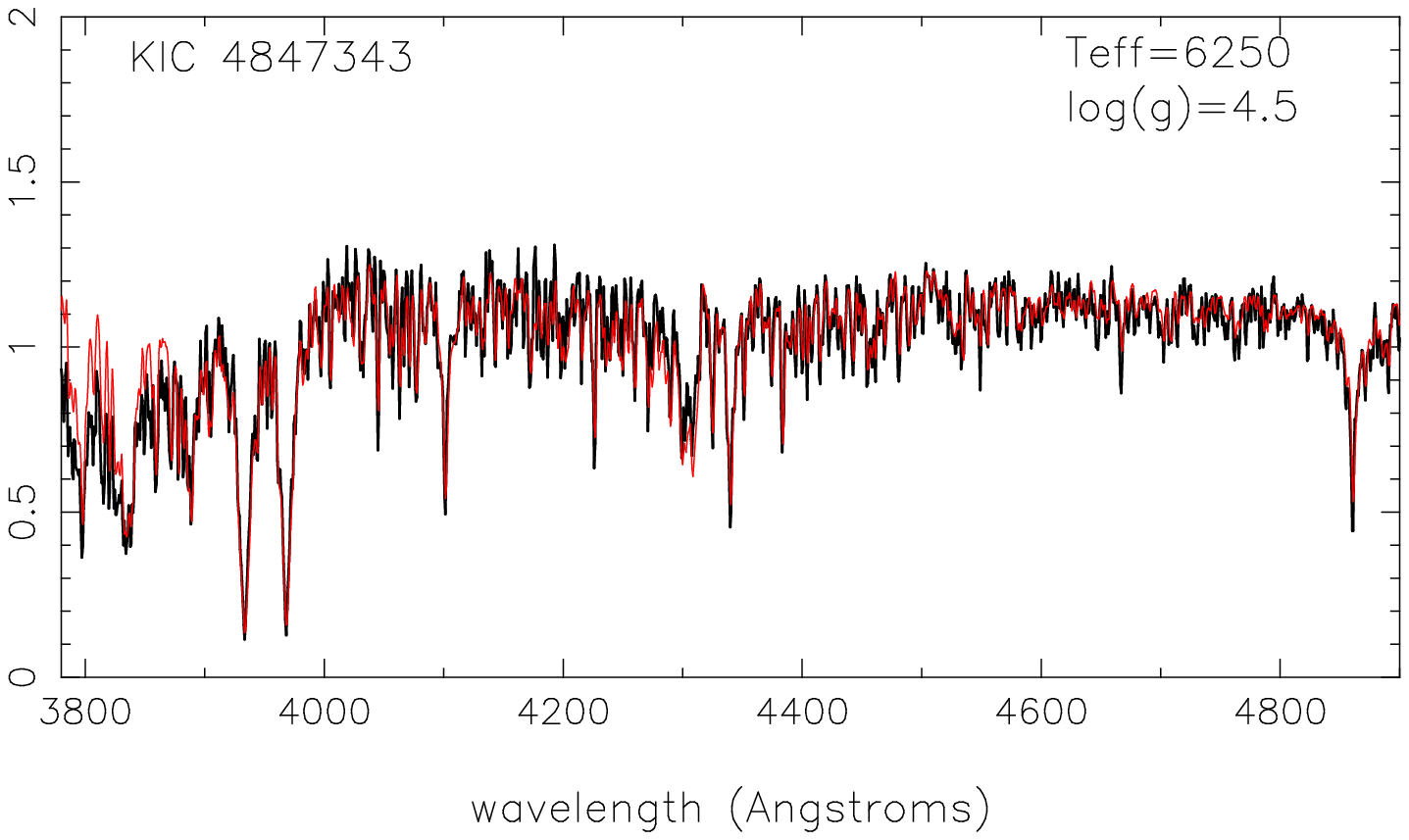}
\includegraphics[scale=.42,angle=270]{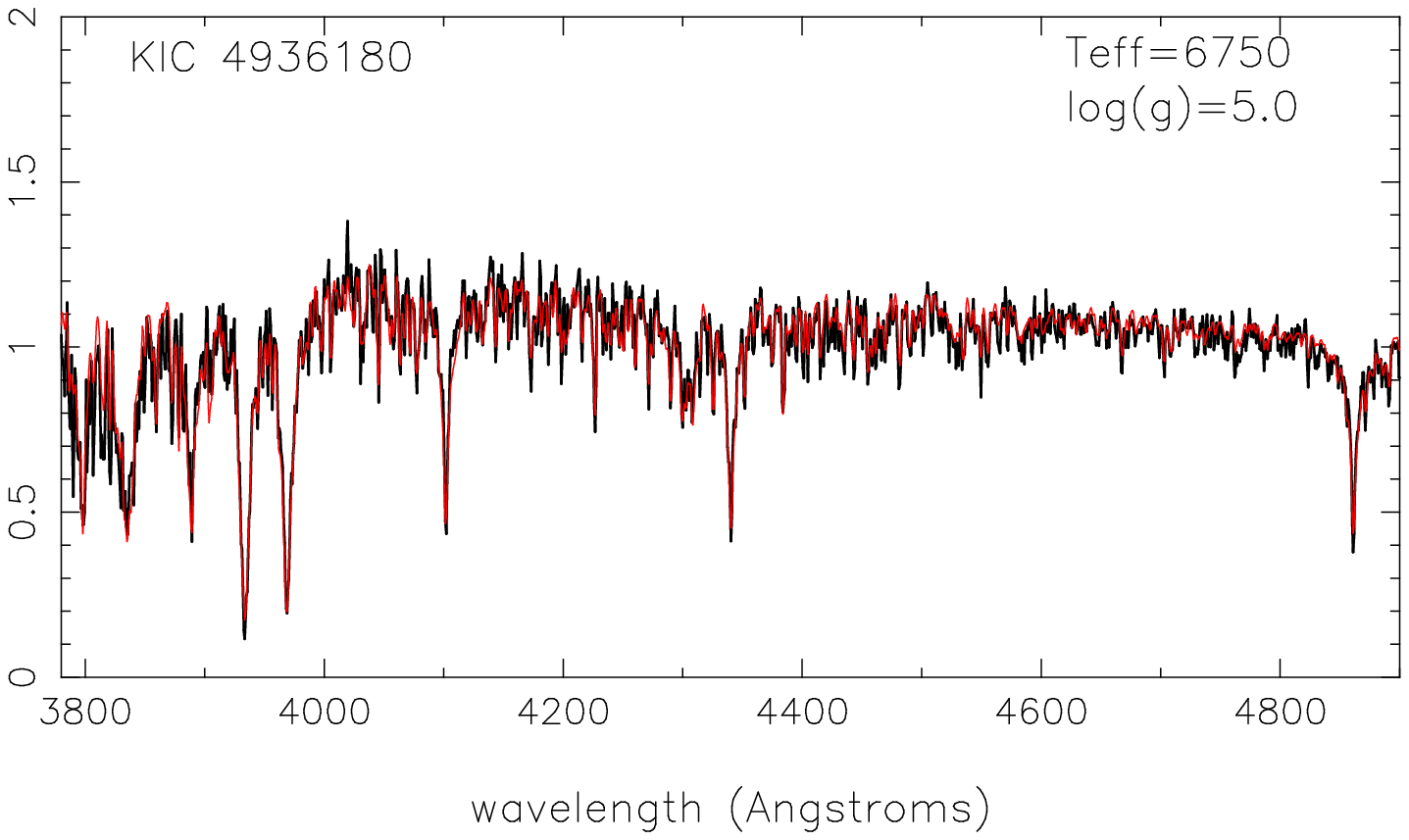}
\includegraphics[scale=.42,angle=270]{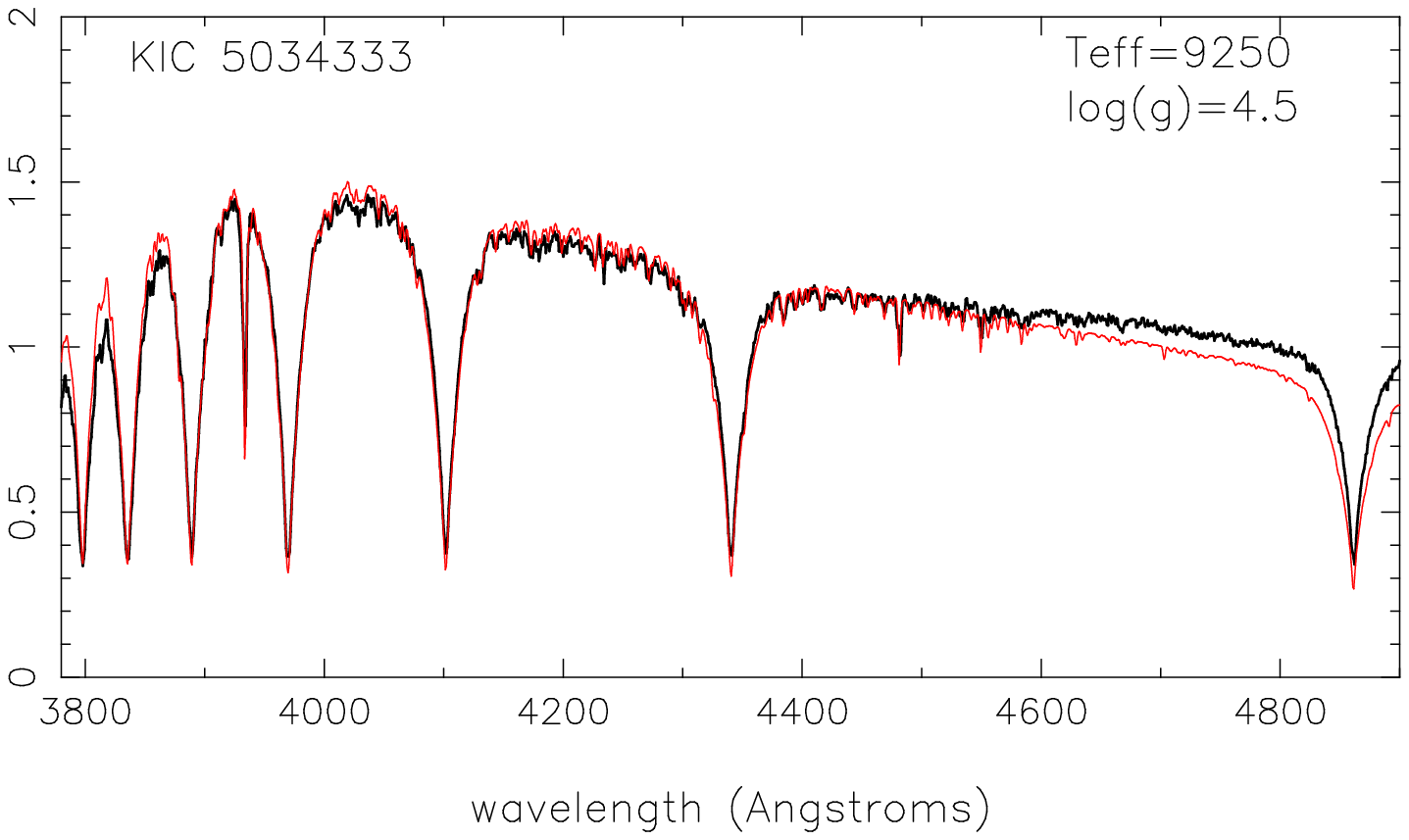}
\includegraphics[scale=.42,angle=270]{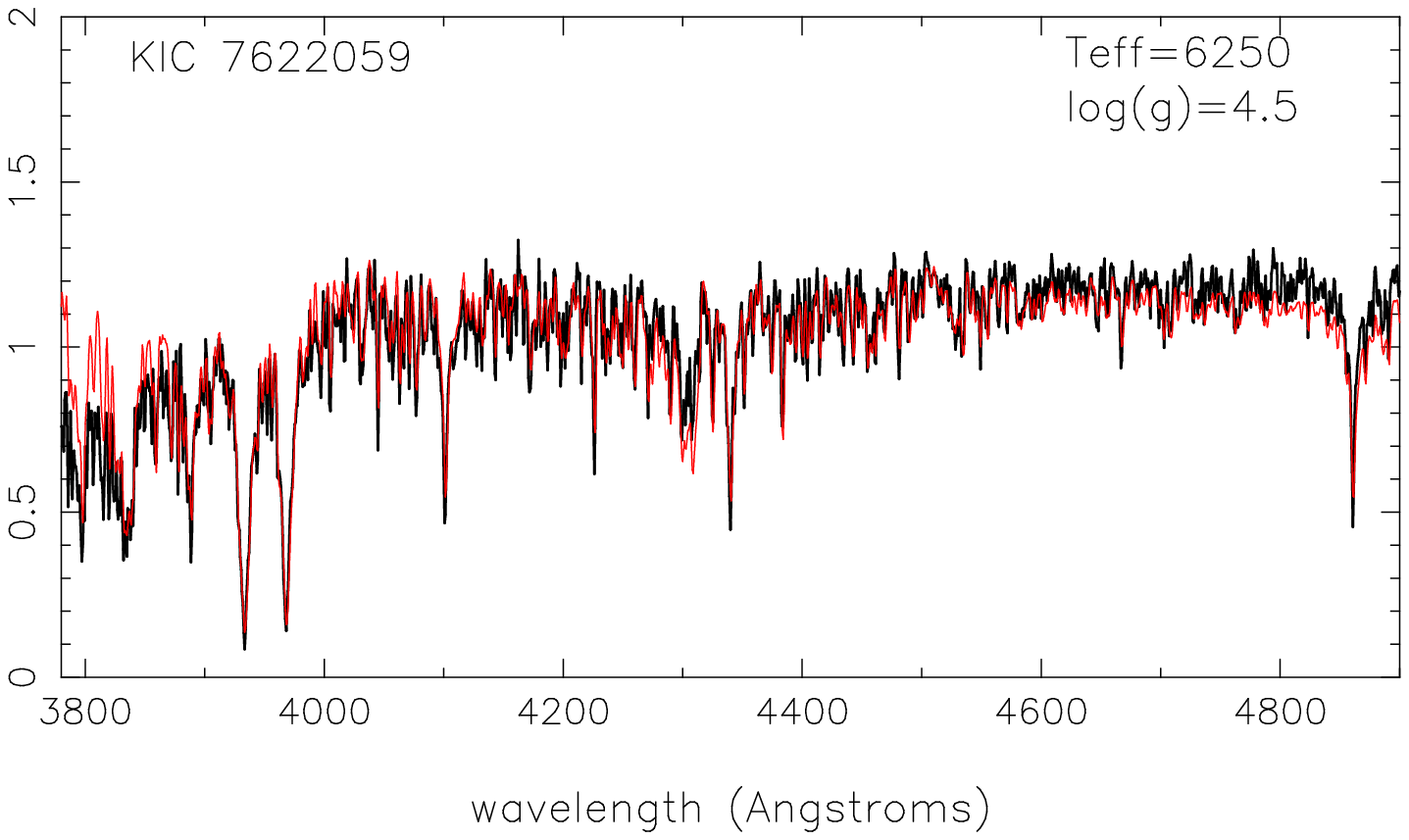}
\includegraphics[scale=.42,angle=270]{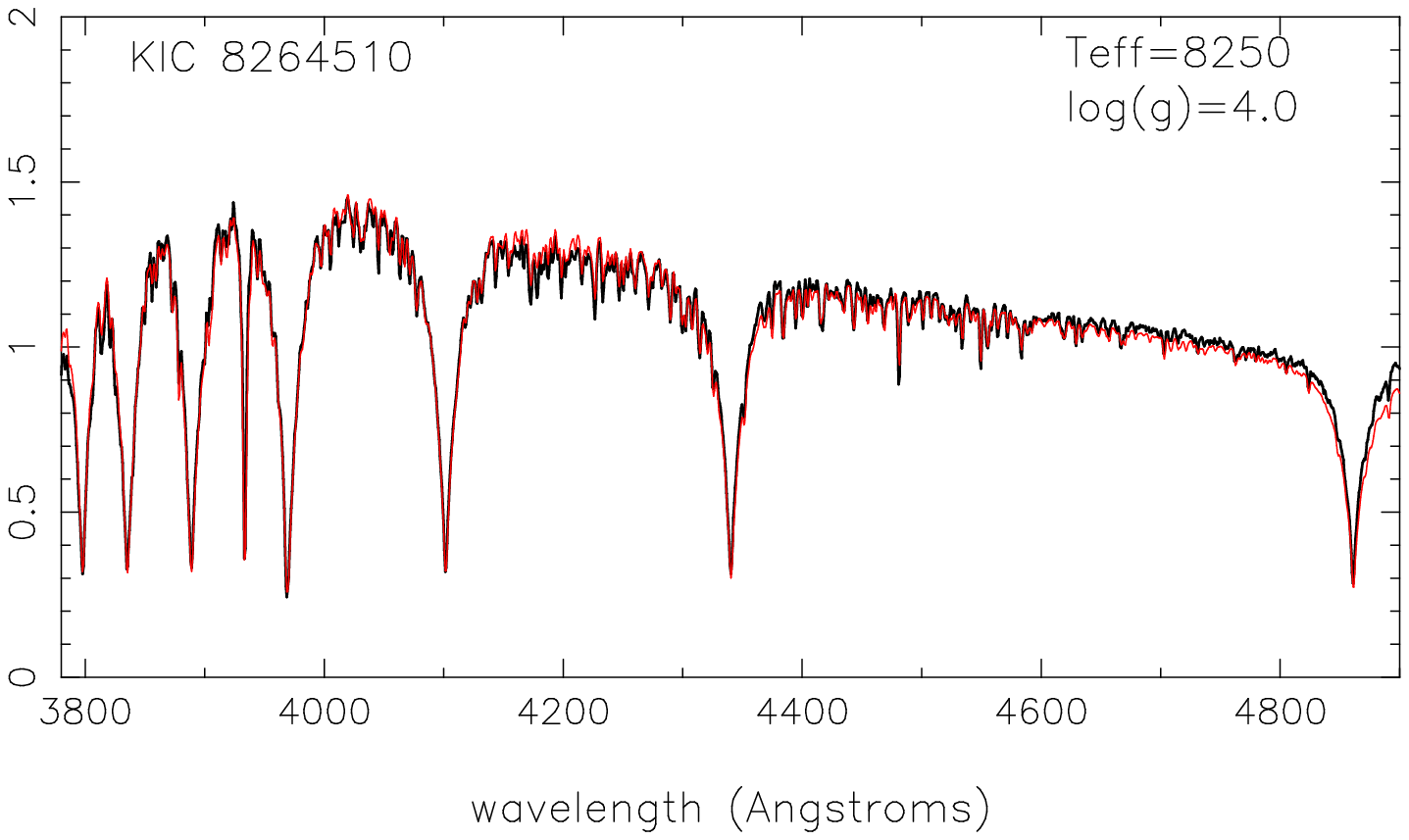}
\includegraphics[scale=.42,angle=270]{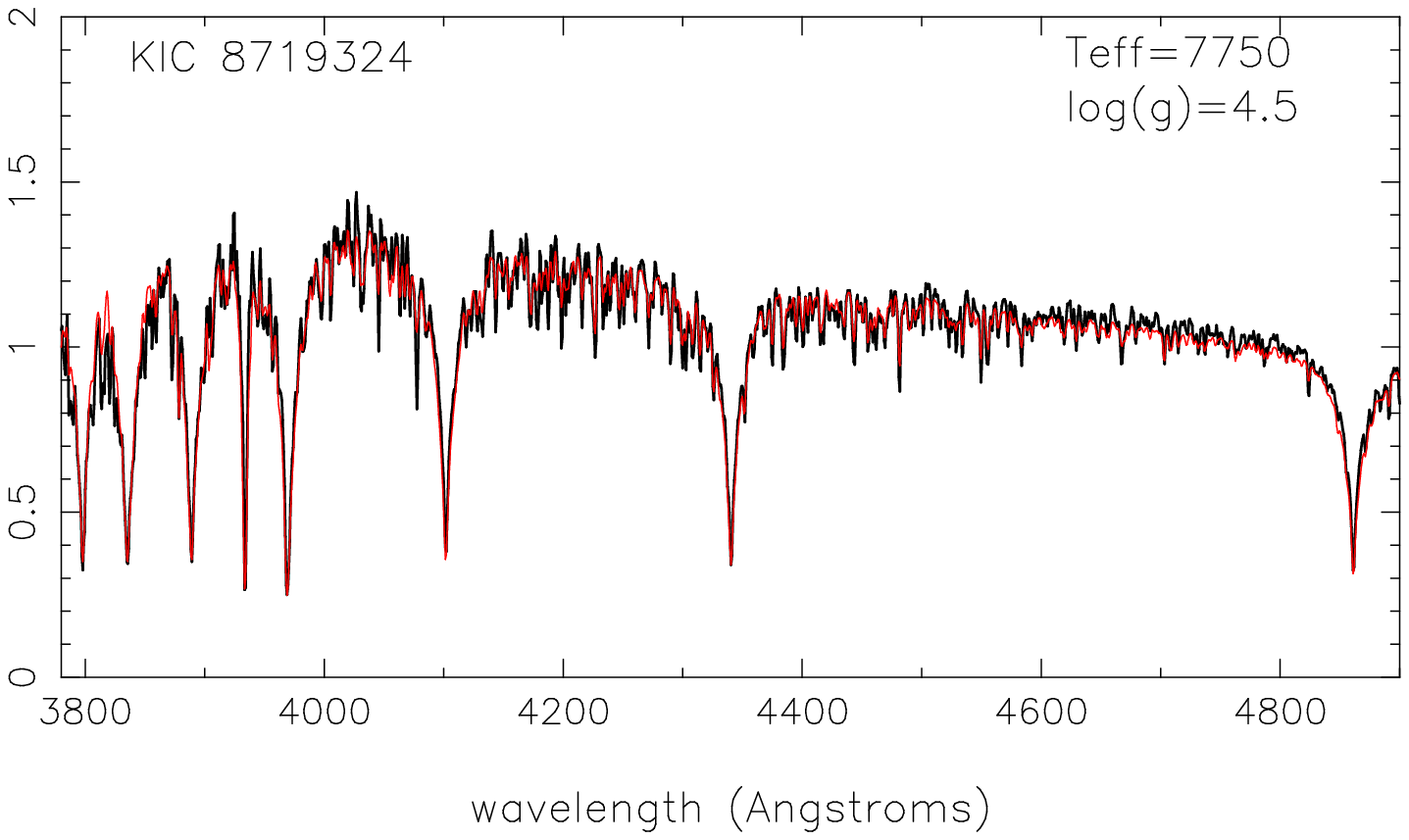}
\includegraphics[scale=.42,angle=270]{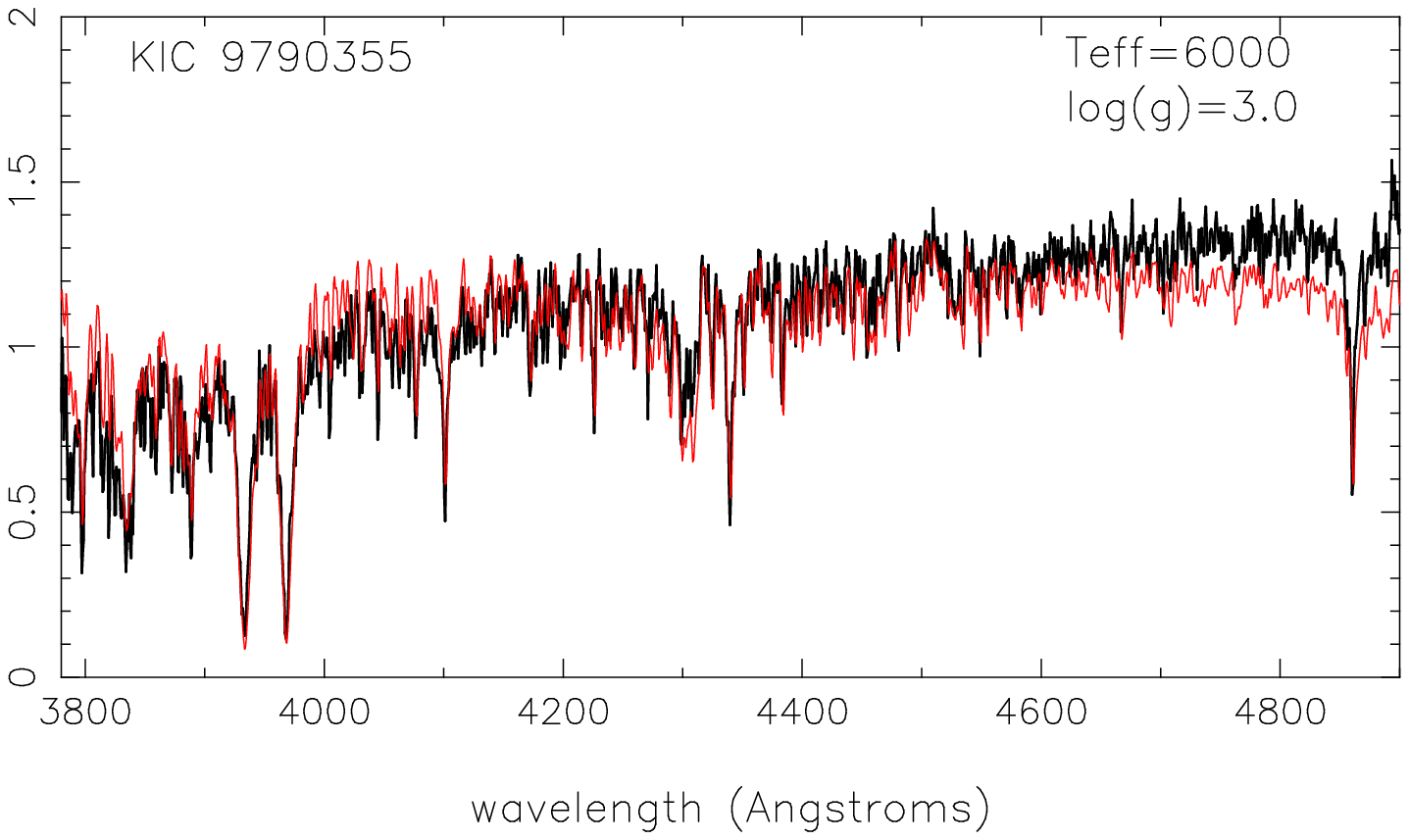}
\includegraphics[scale=.42,angle=270]{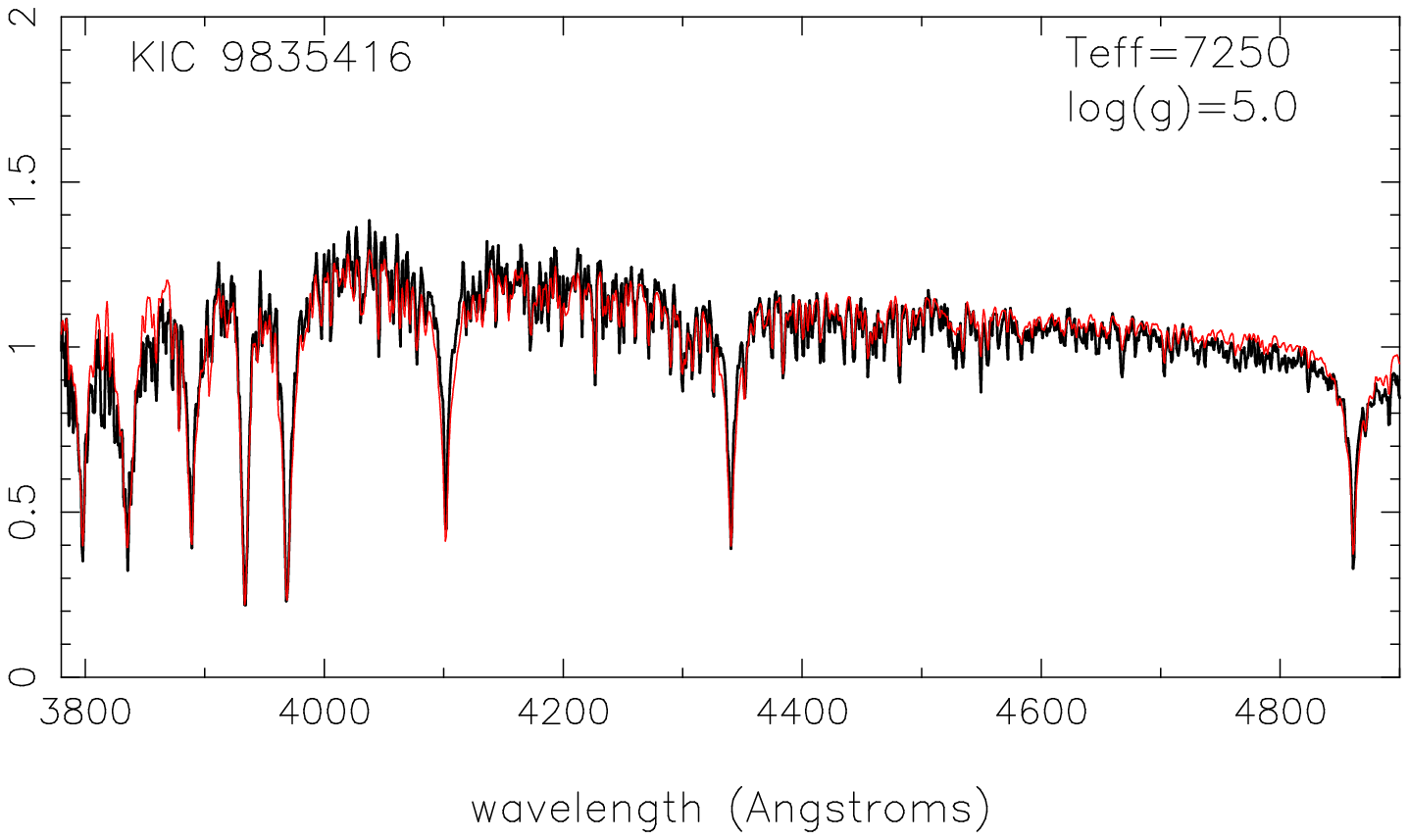}
\includegraphics[scale=.42,angle=270]{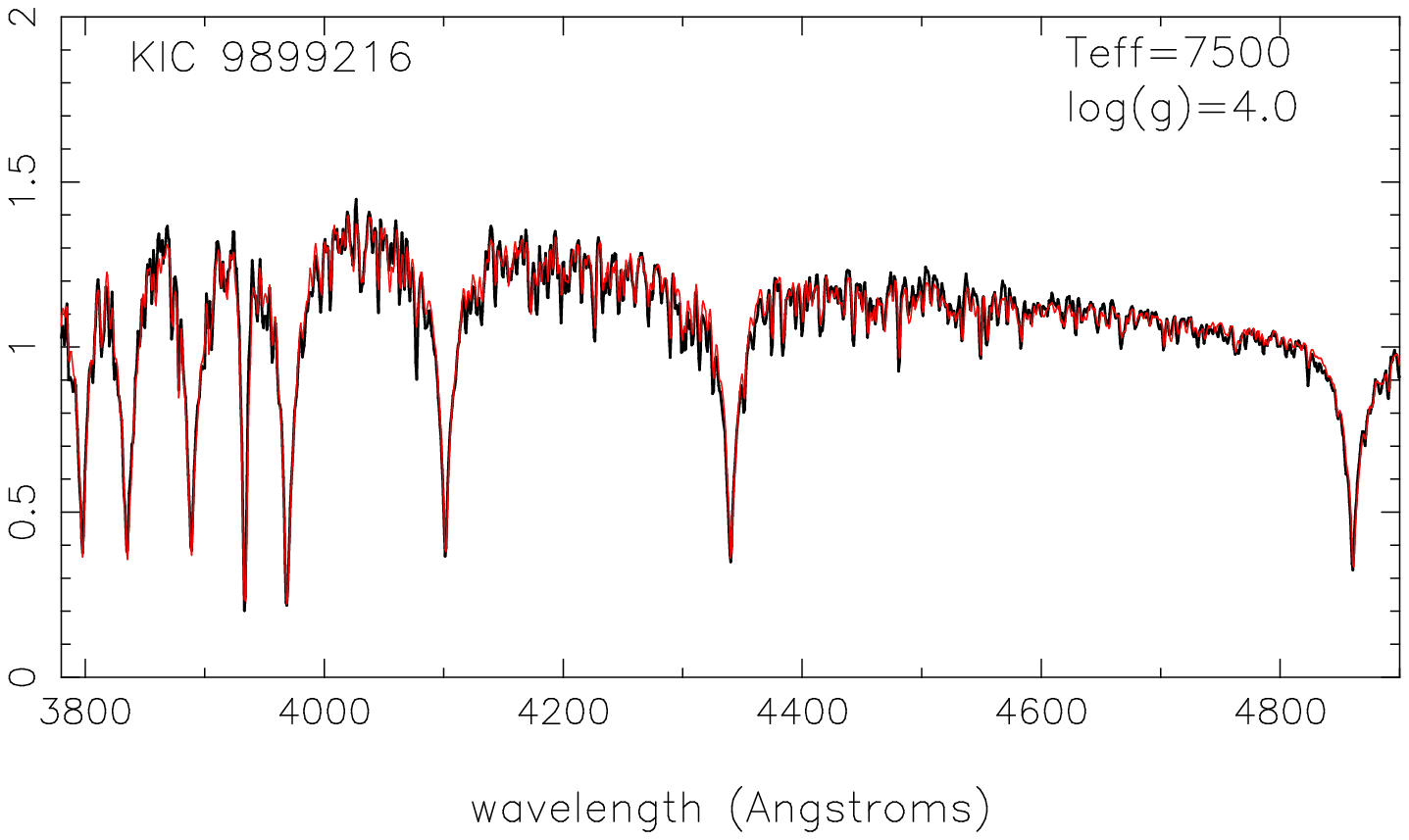}
\includegraphics[scale=.42,angle=270]{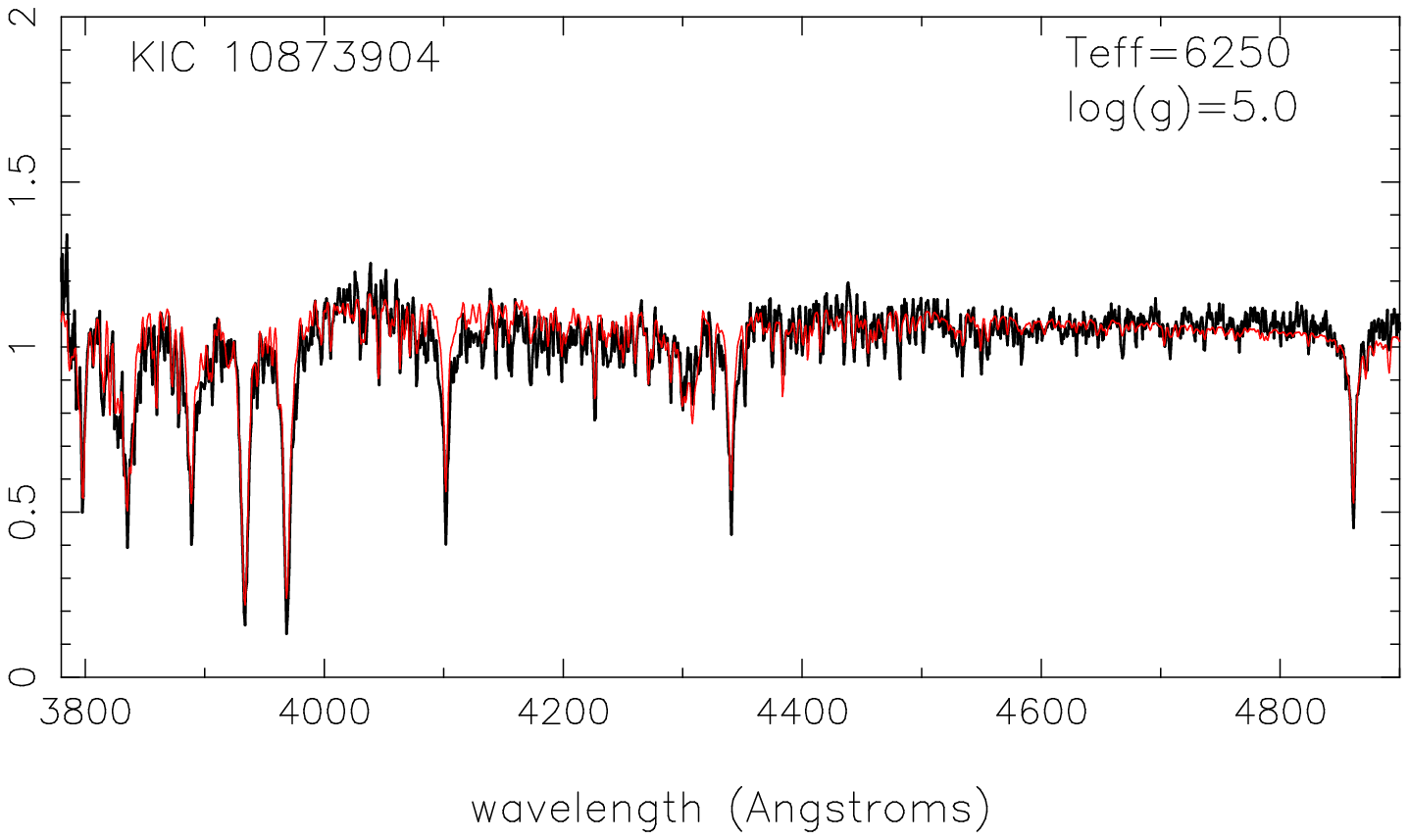}
\includegraphics[scale=.42,angle=270]{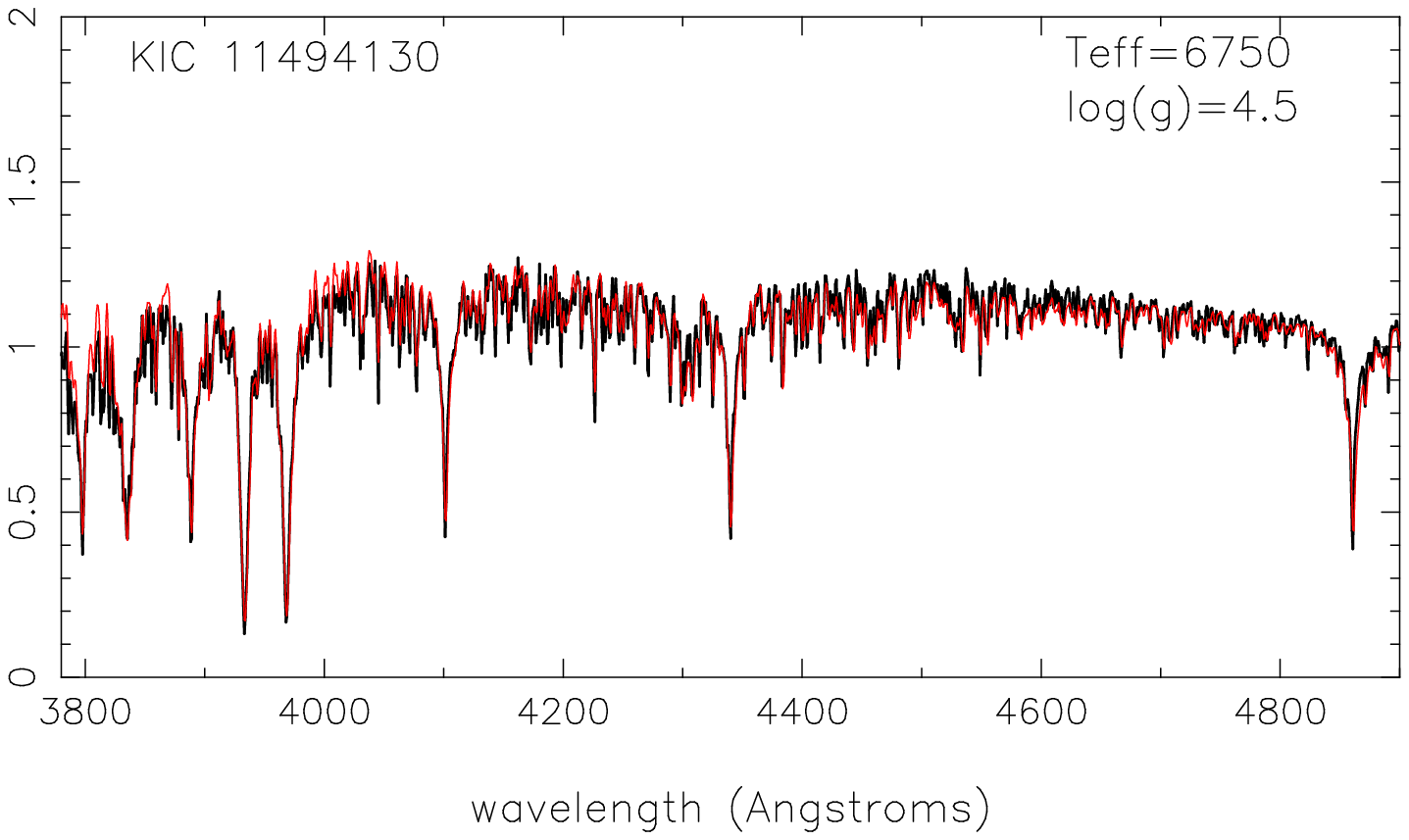}
\includegraphics[scale=.42,angle=270]{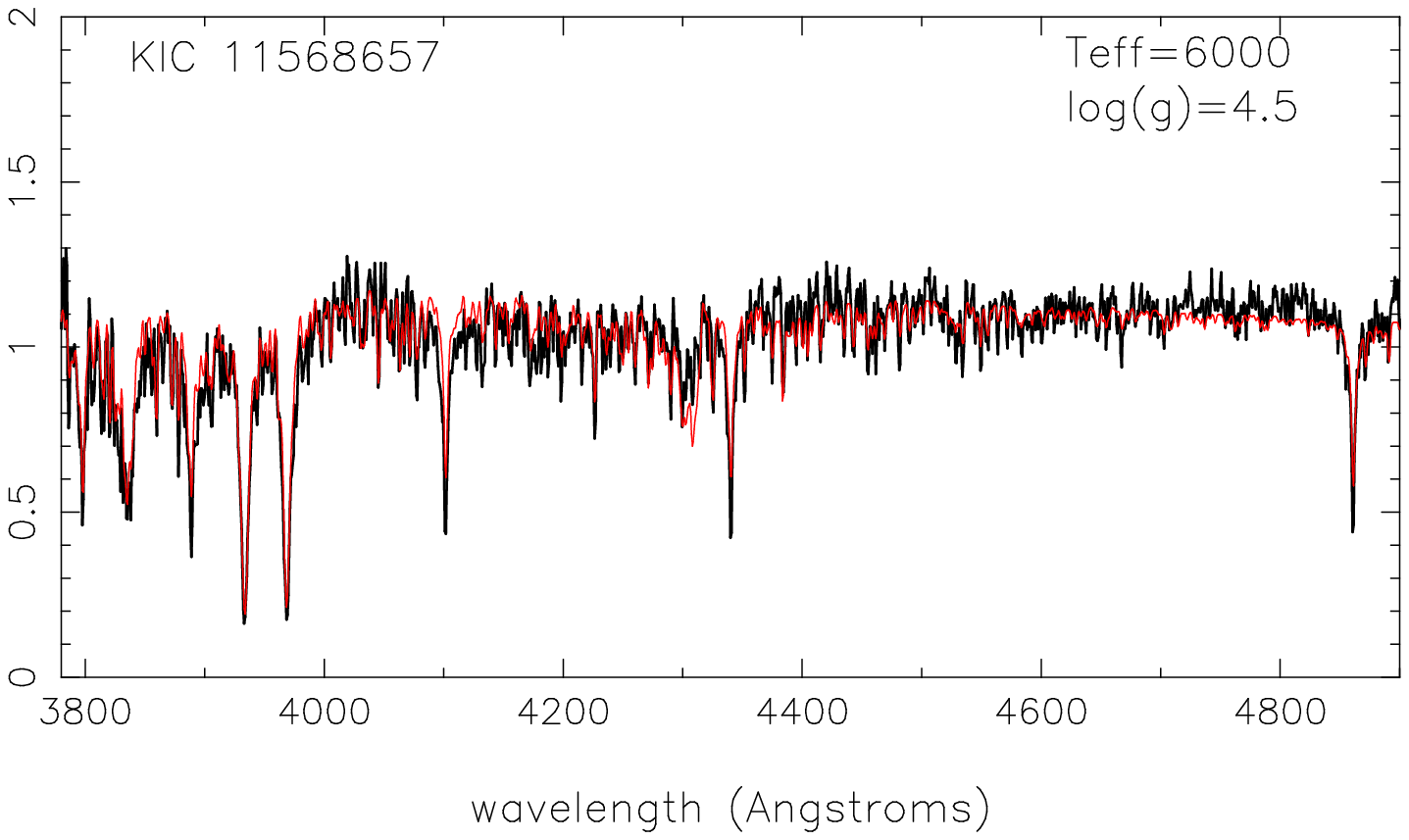}
\includegraphics[scale=.42,angle=270]{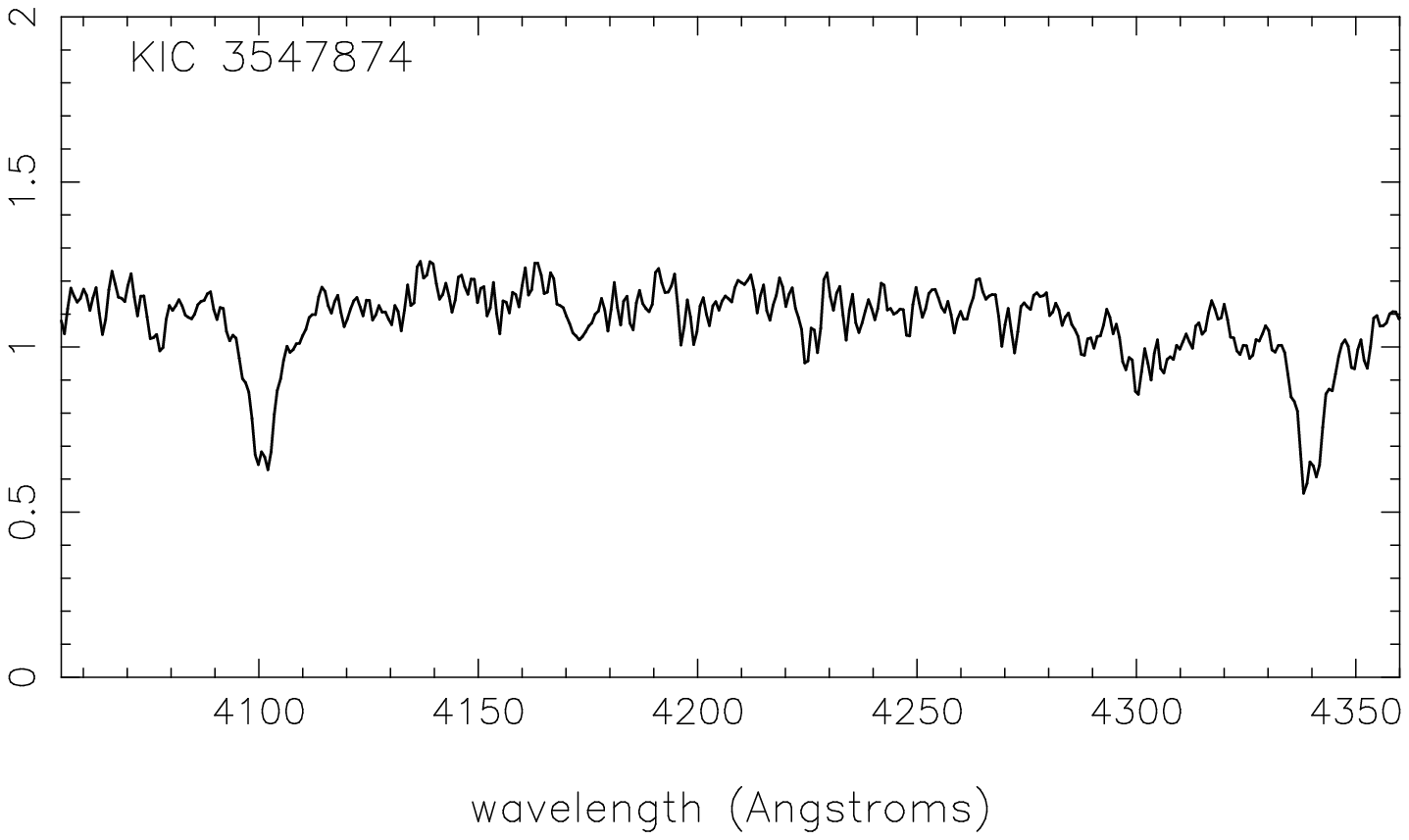}

\caption{\label{spectra}Spectrum of each object in black. Model fits to the spectra are over-plotted in red. Physical parameters of the model are specified in the upper right hand corner of the plot. The last panel shows the double-lined spectrum of \alan\ obtained at an orbital phase of 0.986. }
\end{figure*}

We have modeled each spectrum using a library of synthetic model
atmosphere spectra from \citet{Munari2005}, which are based on the
model atmospheres of \citet{Castelli2003}.  The predicted
spectra are resampled to a resolution to match the observed spectra
and the best-fitting model, scaled in flux to best match the spectrum,
is found by minimizing the rms residuals in the fit over the
wavelength interval $3800-4700$\,\AA.  The grid of atmosphere models are spaced at discrete intervals on \teff\,(250\,K) and \logg\,(0.5
cgs), and so the best fits have values at these grid points with uncertainties at least as large as half of the grid point spacing.  See Figure~\ref{spectra} for individual spectra of each system over-plotted with the best fit model. The results are listed in Table~\ref{table}.

Our spectral fits assume that each target is dominated by the primary in the system.  In all but one case, we only see spectral lines for one star, even when taking observations near periastron. The exception is \alan\ which shows evidence of being a double lined binary like KOI-54 (see Figure~\ref{spectra}). The spectrum that reveals the double lines was taken at a phase concurrent with the heartbeat. From the separation in the spectral lines, the spectrum indicates that the primary has a radial velocity of 100$\pm$11~km\,s$^{-1}$ relative to a spectrum taken just before the heartbeat. 

Since our spectra are of moderate resolution, several systems may show evidence of double lines if spectra were taken at the correct phase with a higher resolution spectrograph. However, in order for the secondary to contribute a significant flux, it would have to have a similar temperature as the primary (or be a giant star). A blend of two stars with similar temperatures will not cause a large error in our fitted temperatures and gravities.  Note that \carl\ and \thor\ both had spectra taken when the radial velocity shift was much larger than the spectral resolution and therefore cannot be double line binaries.  Both would have shown evidence of double lines if the companion was of a similar temperature.     

\subsection{Radial Velocity Measurements}
\label{s:rv}
\begin{figure*}[ht]

\includegraphics[scale=.32,angle=270]{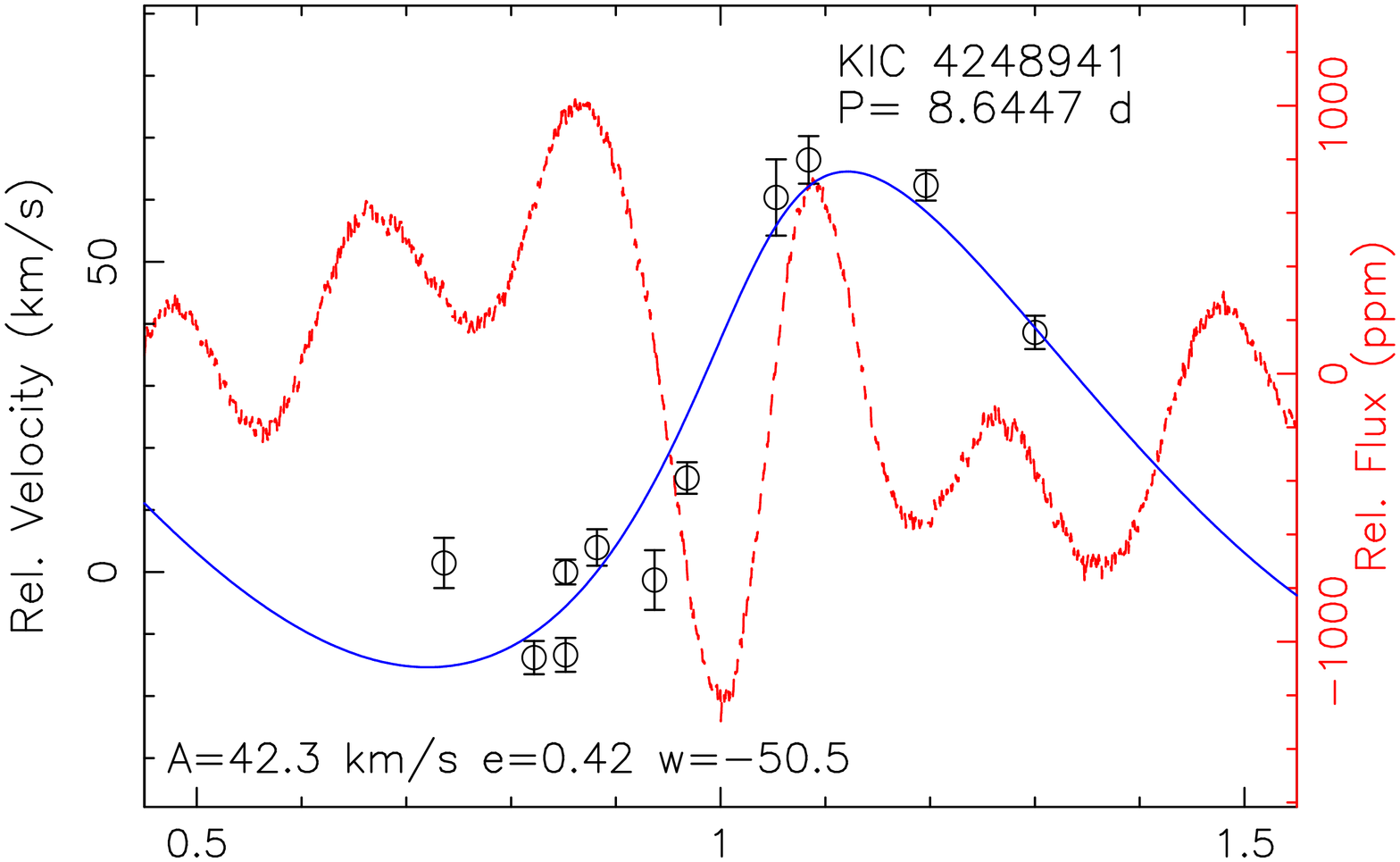}\hspace{1em}
\includegraphics[scale=.32,angle=270]{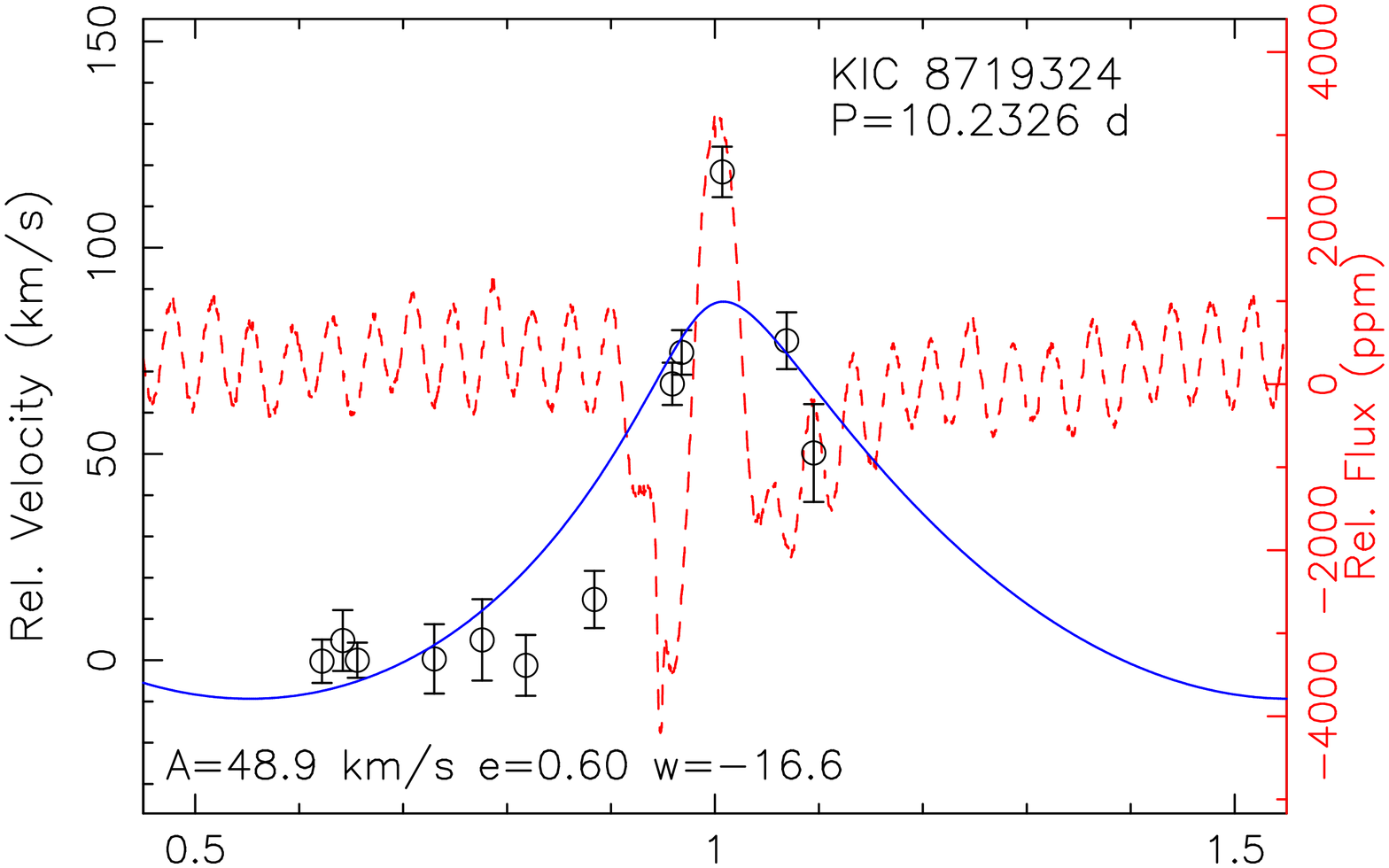}\\
\includegraphics[scale=.32,angle=270]{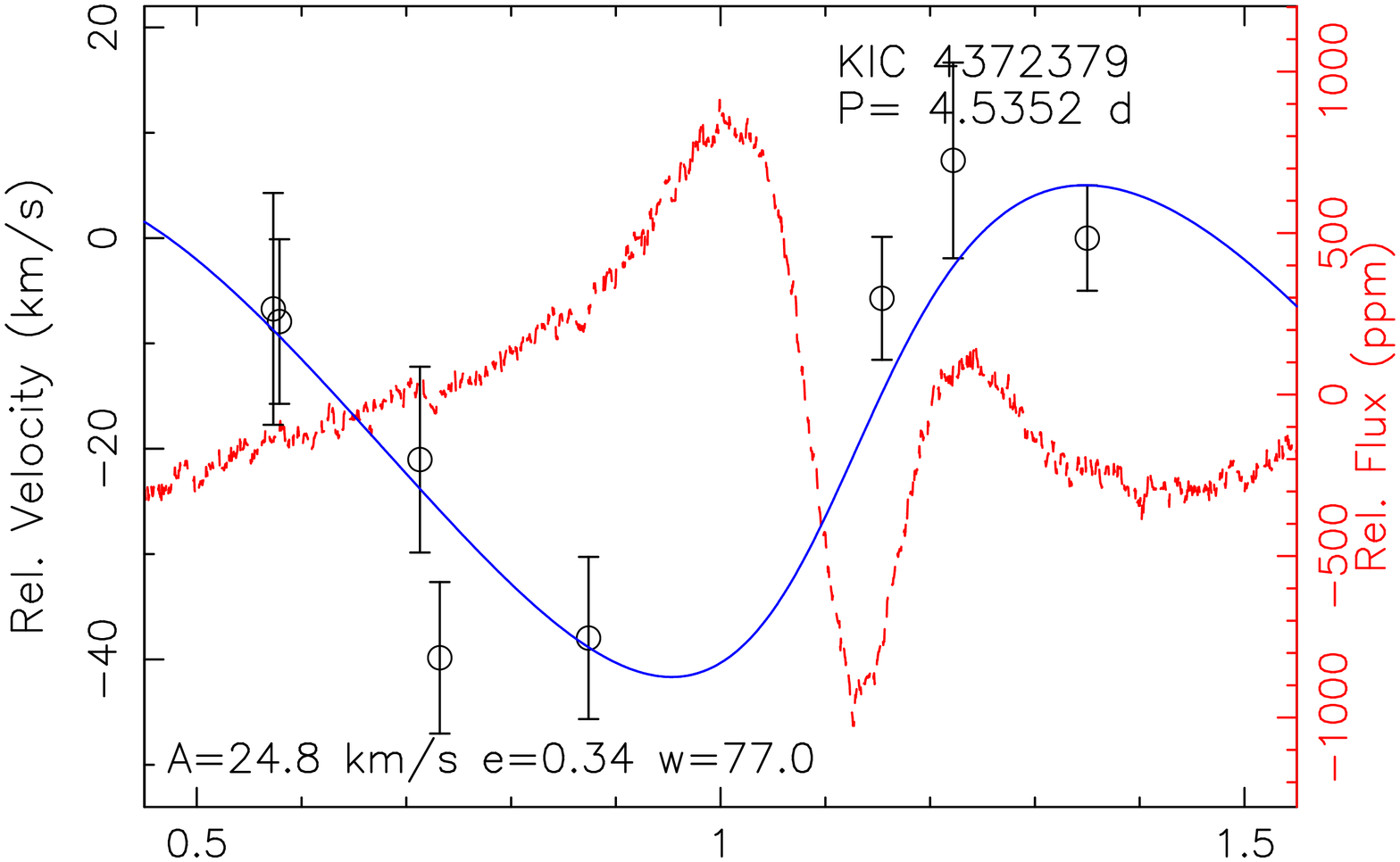}\hspace{1em}
\includegraphics[scale=.32,angle=270]{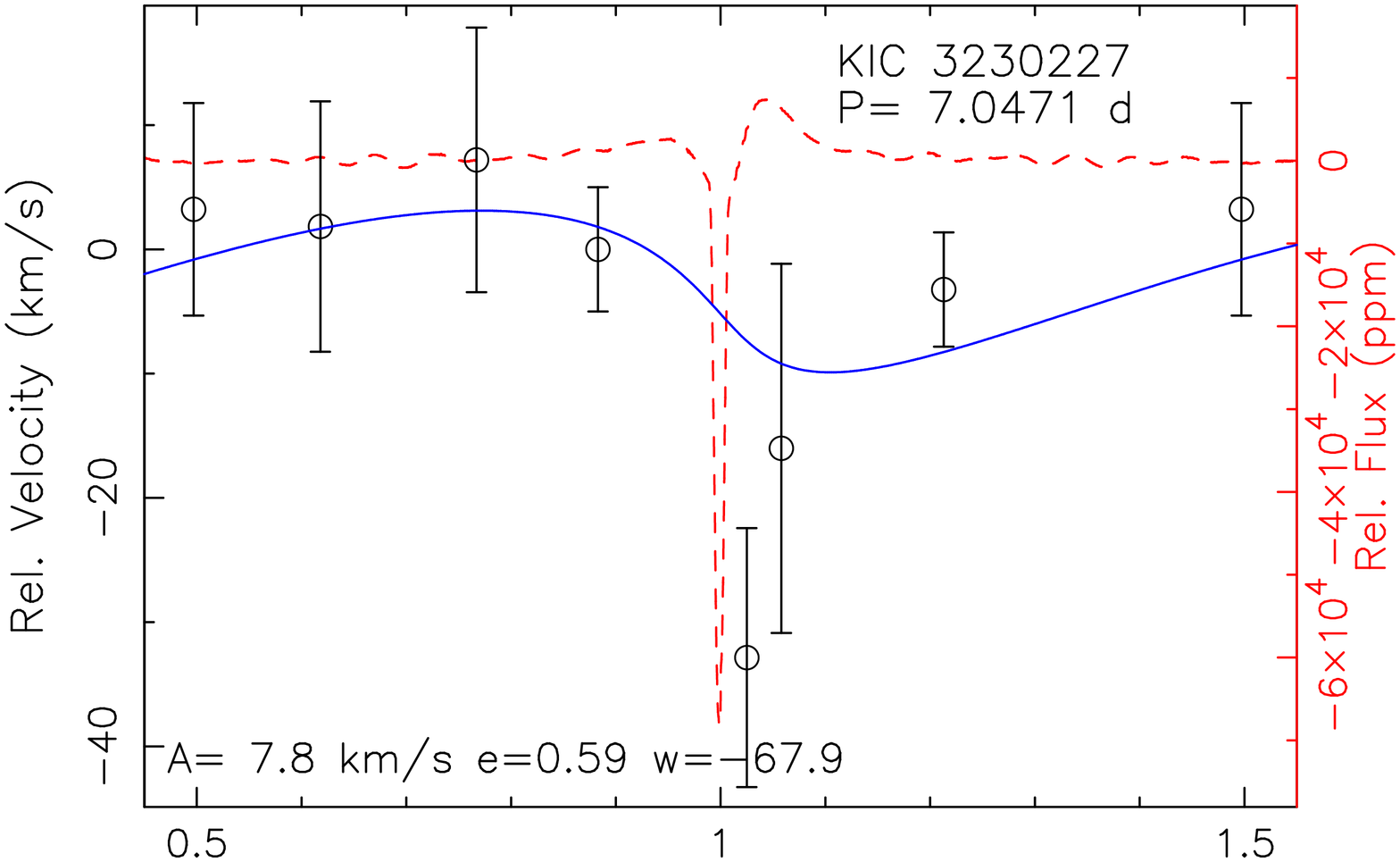}

\caption{\label{rv}Radial velocity measurements (circles with error bars) plotted against orbital phase. The folded light curves, are presented as dashed lines in red with units on the right side of each graph.  For four systems, we fit the radial velocity points while constraining the eccentricity, angle of periastron and time of periastron passage to the values measured in \S\ref{s:fittingrv} (blue line). The amplitude ($A$), eccentricity ($e$) and angle of periastron ($\omega$) of the radial velocity fit are listed on the plot.}
\end{figure*}

For five systems, we obtained more than one spectrum and were able to measure the relative velocities at different orbital phases. We measured the radial velocity between the spectral lines by performing a cross-correlation against one spectrum.  We show relative radial velocity measurements for these systems phased with the folded light curve in Figure~\ref{rv}.  All radial velocities have been corrected for the Earth's motion around the sun. The error bars are estimated from the full width half maximum of a Gaussian fit to the peak of the cross-correlation function. Note that \alan\ is a double-lined binary and so our cross-correlation technique is too simple to accurately measure the radial velocities, especially when the lines are not well resolved.

\subsection{Fitting Radial Velocity Measurements}
\label{s:fittingrv}

For four systems we have a time series of radial velocity measurements that show a distinct non-sinusoidal shape, indicative of an eccentric binary system; see Figure~\ref{rv}. Unfortunately we do not have enough data for any of these systems to constrain the orbital parameters of the system; however we can show that the radial velocity measurements are consistent with the orbital parameters determined from the fits to the folded light curve.  We fit the radial velocity curve constraining the eccentricity, angle of periastron, and time of periastron passage to those listed in Table~\ref{table}.  The velocity semi-amplitude was a free parameter of the fit. See the blue line in Figure~\ref{rv} for the results of this fit.  The parameters of the fit are labeled on the figure for each system. 

For \ardi, \thor , \lars\ and \carl\ the tidal distortion model is mostly consistent with the radial velocity fits. Those remaining inconsistencies are in part caused by using orbital parameters determined by fitting an incomplete model to the light curves.  However, another possible reason for the inconsistency is that the radial velocity measurements may include the unmodelled tidally-induced distortions and oscillations on the surface of the star. \citet{Willems2002} discuss these velocity fields and show that for high inclinations, the amplitude of these velocities may be as high as 15 km\,s$^{-1}$ for one pulsation mode on a B-type star with an eccentricity of 0.5. Despite being cooler, significant pulsation velocity fields may exist on our heartbeat stars.  Factors that increase the size of the tidal force, such as eccentricity and the mass of the secondary, can also increase the size of this effect.  Note, this would not have been a dominant effect for KOI-54 because of the low inclination of that system.  Whether we are seeing these surface velocity fields is unclear without better time-resolved radial velocity measurements, but such considerations must be made when fitting the data on these complicated, tidally dynamic systems.


\section{Discussion}
\label{s:discussion}

Our collection of heartbeat stars were discovered serendipitously while searching for planets in the \kepler\ archive.  \kepler\ targets were intentionally selected to look for ``Sun-like" stars, and so primarily G and K-type stars were selected.  For stars brighter than 15.5\,mag, 97 per\,cent of the selected stars are cooler than 6500\,K, with the distribution peaking near 5500\,K \citep{Batalha2010}.  The temperatures of the observed stars in our group of binaries range from 6000 to 9250\,K.  While we certainly have not found all eccentric systems with dynamic tidal distortions in the exoplanet data, we might expect our collection to be dominated by G-type stars if heartbeat stars were common across all main sequence stars. 

\begin{figure}[t]
\includegraphics[scale=0.56,angle=270]{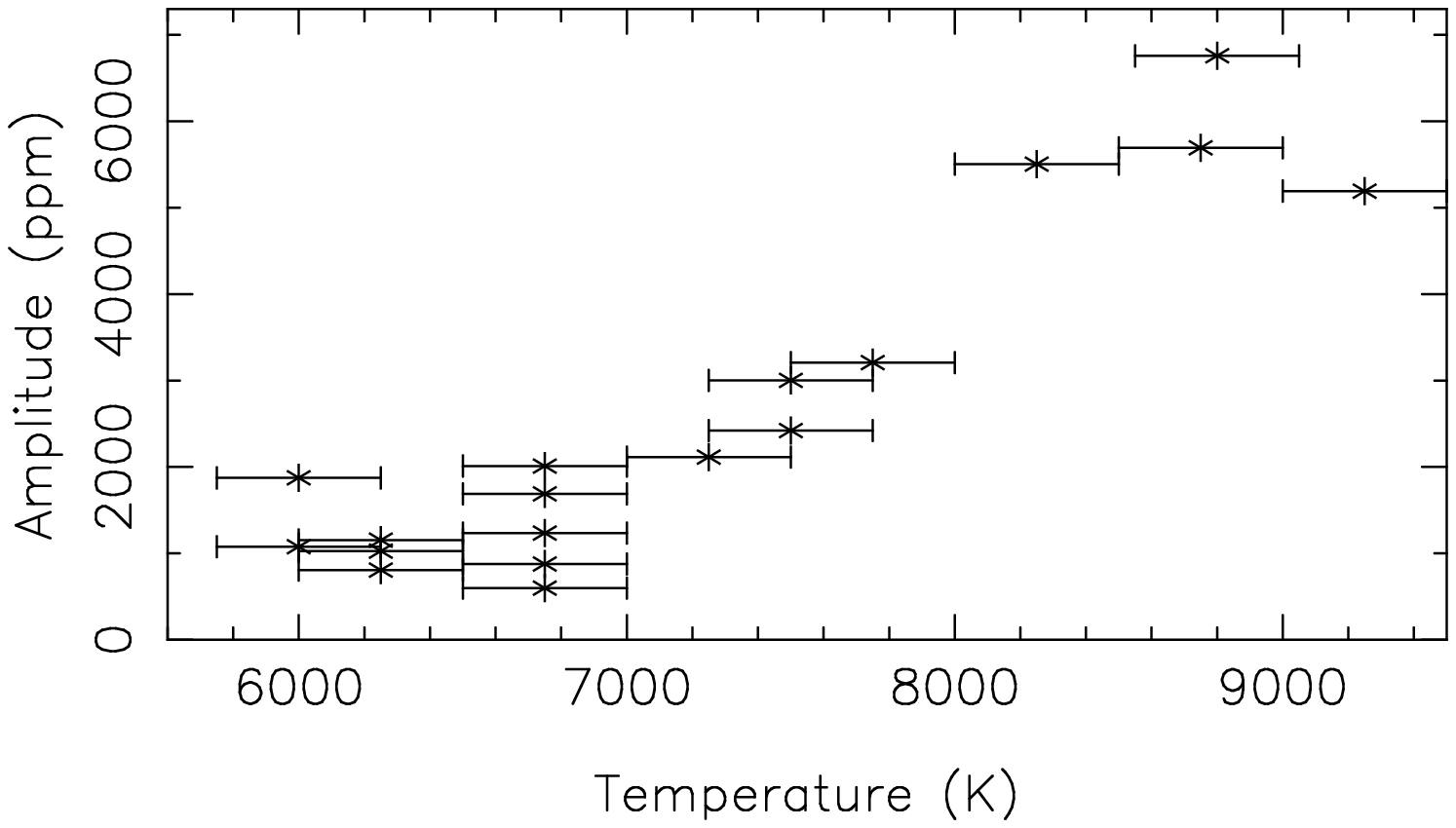}
\caption{\label{amplitude} The observed amplitude of the heartbeat plotted against the measured temperature of the primary star.  The error in the amplitude is approximately the size of the point.\newline }
\end{figure}

The dominance of hot stars in our sample is likely the result of the size of the tides on the primary star.  The fractional amplitude of the tidal distortions depends on the star's surface gravity as well as the size of the tidal force from the secondary \citep{Pfahl2008}. An A5V star has a lower \logg\ and will have approximately twice the amplitude of a G5V star undergoing the same tidal force, making it easier to detect. We may be seeing this trend in Figure~\ref{amplitude} where we plot amplitude against effective temperature.  We measured the amplitude as the absolute value of the largest deviation from zero level in the folded light curves (without including the amplitude of the obvious eclipsing events).  We have not adjusted this plot for effects that can affect the strength of the external tidal forces (such as eccentricity and the mass of the companion relative to the primary). We do note that no obvious trend between temperature and eccentricity exists in this collection of heartbeat stars (see Table~\ref{table}).  As we continue to search the \kepler\ data set and can fold longer data sets, we will likely find a larger population of cooler main sequence heartbeat stars with low amplitudes.

\subsection{Circularization}

From the perspective of stellar evolution, highly eccentric short-period binary systems are a mystery. Highly eccentric systems with small periastron separations are expected to tend towards circular orbits at a rapid rate as their tidal interactions are efficient at dissipating energy (tidal interactions scale as ($R_\star$/a)$^{-6}$). Thus the lack of short-period, highly-eccentric stars in Figure~\ref{eccentricity} is not unexpected. 

There are currently two theories that describe the circularization of binary orbits: the equilibrium tide model \citep{Hut1980} and the dynamical tide model \citep{Zahn1975}. \citet{Eggleton1998} states that the former is most likely applicable to stars that have relatively small radii with respect to their periastron distance and the latter for systems where the stellar surfaces are in close proximity at periastron. The equilibrium tide model attributes orbital circularization to the torque caused by the misalignment of the tidal bulges on the stellar surface with respect to the star's instantaneous equipotential shape. This model suggests that in certain cases, when the ratio of the spin to the orbital angular momentum is large,  the orbital evolution of the system can be erratic; eventually leading to circularization, collision or even escape.

For stars that interact more dramatically during periastron (as likely the case for these highly eccentric heartbeat stars), \citet{Zahn1975} has shown that the dynamical tidal forces generate oscillation in the surface layers of the star that are damped through tidal interactions; a trend confirmed by \citet{Khaliullin2011}. This process is highly effective at energy loss and results in a circularization time scale much shorter than the lifetime of hot main sequence stars \citep[on the order of 10$^7$\,y,][]{Hut1980}. 

One way to maintain eccentric binaries in hot main sequence stars is by including a third body.  Advances on the work of \citet{Mayor1987} by \citet{Tokovinin2006} recently determined that in binary systems with periods less than 3\,d the frequency of tertiary components is as large as 96\% (dropping to 34\% for binary systems with periods $>$\,12\,d). A tertiary component can both remove angular momentum from and perturb the inner binary system, allowing it to maintain a high eccentricity despite the strong dynamic tidal forces. Thus the existence of highly eccentric binary systems, such as those in our sample, is possibly a consequence of the presence of a third body.




\subsection{Induced Pulsations}
The tidal forces also induce pulsations on these stars.  We do no analysis of these obvious features in the light curve, but we do note their presence.  A cursory analysis of the Fourier transform and folded light curves show that a majority have pulsations that are harmonics of the orbital period and several also show modes independent of the orbital period.  The pulsations with a period at an integer harmonic of the orbital period are likely driven by the variable tidal forces \citep{Welsh2011,Willems2003,Kumar1995}.  Other modes could be caused by one star being a normally pulsating star, or, the tidal forces are exciting additional modes through nonlinear couplings with resonant modes \citep{Fuller2011,Wu2001}. 

Since the tidally-induced modes are harmonics of the orbital frequency, the only way to accurately measure the pulsations is to first correctly fit and remove the orbital effects \citep{Welsh2011}. Because of the careful modeling effort required for each individual star, we leave the extraction and analysis of the tidally-induced pulsations for subsequent papers. 

\section{Conclusions}
With the new level of precision and coverage afforded by the \kepler\ time series data, we have discovered \nheartbeat\ eccentric, tidally distorted, binary stars we have called heartbeat stars. KOI-54 is a long period, highly eccentric, example of this class. The variety of light curve shapes arises primarily from variable tidal distortions being observed from different inclinations and periastron angles. Radial velocity measurements, as well as fits to the light curves, support this model.  As \kepler\ obtains more data, we will likely find more examples of stars in this class with lower amplitudes.

Detailed modeling of these systems will be challenging. The tidal forces are changing throughout the orbit causing large tidal distortions and induced pulsations on possibly both stars. These effects can cause non-sinusoidal variations in both the light curve and the radial velocity measurements.  Plus, large orbital velocities can cause Doppler boosting to add to the flux variations while the secondary can cause significant irradiation variations. When more complete radial velocity measurements are obtained, the parameters of the companion can be constrained and a better model can be developed for each star. With detailed models, these stars offer unique laboratories in which higher order effects can be studied, effects often easily ignored in stellar structure calculations.

This is the first opportunity to observationally study dynamic tidal interactions on a class of stars.  Our first analysis of each system has already shown some of the potential of these systems to constrain theories regarding energy transfer in tidally dynamic systems.  From our fits to these heartbeat systems, we confirm that highly eccentric binaries with very short periods must be rare, indicating that tidal friction may have circularized these systems.  Further investigation may confirm the theories of \citet{Zahn1975} and \citet{Khaliullin2011} regarding the time scales of circularization.  Additionally, the strength and period of the tidally-induced pulsations varies across these systems. With a class of such systems, we can begin to decipher how this driving mechanism excites certain oscillation modes in a diverse sample of stars.

\begin{acknowledgements}
Funding for the \kepler\ mission is provided by the NASA Science Mission directorate. We thank the larger \kepler\ team for their support and hard work. Kitt Peak National Observatory is operated by the Association of Universities for Research in Astronomy (AURA) under cooperative agreement with the National Science Foundation. Some of the data presented in this paper were obtained from the Multi-mission Archive at the Space Telescope Science Institute (MAST). STScI is operated by the Association of Universities for Research in Astronomy, Inc., under NASA contract NAS5-26555. Support for MAST for non-HST data is provided by the NASA Office of Space Science via grant NNX09AF08G and by other grants and contracts.\\

\end{acknowledgements}

\bibliography{heartbeat}

\end{document}